\documentclass[aps,pra,superscriptaddress,twocolumn]{revtex4-2}
\usepackage{epsfig,amssymb, amsmath,fancyhdr,upgreek,color}
\newcommand{\bomega}{{\boldsymbol \omega}}
\newcommand{\dbomega}{{\boldsymbol{\delta\omega}}}

\begin{document}
\title{Entanglement source and quantum memory analysis for zero added-loss multiplexing}
\author{Jeffrey H. Shapiro}
\email{jhs@mit.edu}
\affiliation{Research Laboratory of Electronics, Massachusetts Institute of Technology, Cambridge, Massachusetts 02139 USA}
\author{Michael G. Raymer}
\email{raymer@uoregon.edu}
\affiliation{Department of Physics, Oregon Center for Optical, Molecular, and Quantum Science, University of Oregon, Eugene, Oregon 97403 USA} 
\author{Clark Embleton}
\email{cemblet2@uoregon.edu}
\affiliation{Department of Physics, Oregon Center for Optical, Molecular, and Quantum Science, University of Oregon, Eugene, Oregon 97403 USA}
\author{Franco N. C. Wong}
\email{ncw@mit.edu}
\affiliation{Research Laboratory of Electronics, Massachusetts Institute of Technology, Cambridge, Massachusetts 02139 USA}
\author{Brian J. Smith}
\email{bjsmith@uoregon.edu}
\affiliation{Department of Physics, Oregon Center for Optical, Molecular, and Quantum Science, University of Oregon, Eugene, Oregon 97403 USA}
\date{9/24/24}

\begin{abstract}
High-rate, high-fidelity entanglement distribution is essential to the creation of a quantum internet, but recent achievements in fiber (248\,km at 9/s rate) and satellite-based (1200\,km at 1.1/s rate) entanglement distribution fall far short of what is needed.  Chen~\emph{et al}.~[Phys. Rev. Appl. {\bf 19}, 054209 (2023)] proposed a means for dramatically increasing entanglement-distribution rates via a scheme they call zero added-loss multiplexing (ZALM).  ZALM's quantum transmitter employs a pair of Sagnac-configured spontaneous parametric downconverters (SPDCs), channelization via dense wavelength-division multiplexing (DWDM) filtering, and partial Bell-state measurements (BSMs) to realize a heralded source of frequency-multiplexed polarization-entangled biphotons.  Each biphoton is transmitted to Alice and Bob along with a classical message identifying its frequency channel and whether a $\psi^-$ singlet or a $\psi^+$ triplet was heralded.  Alice and Bob's quantum receivers then use DWDM filtering and temporal-mode conversion to interface their received biphotons to intra-cavity color-center quantum memories.  This paper delves deeply into ZALM's SPDCs, partial-BSMs, and Duan-Kimble loading of Alice and Bob's quantum memories.  Its principal results---the density operators for the SPDC sources and the quantum memories---allow heralding probability, heralding efficiency, and fidelity to be evaluated for both the polarization-entangled biphotons and the loaded quantum memories, thus enabling exploration of the parameter space for optimizing ZALM's performance.  Even without a comprehensive optimization analysis, the paper's examples already demonstrate two critical features of the ZALM architecture:  (1) the necessity of achieving a near-separable channelized biphoton wave function to ensure the biphoton sent to Alice and Bob is of high purity; and (2) the premium placed on Alice and Bob's temporal-mode converters' enabling narrowband push-pull memory loading to ensure the arriving biphoton's state is faithfully transferred to the intra-cavity color centers.

\end{abstract}

\maketitle

\section{Introduction \label{Intro}}
Entanglement, the quintessential manifestation of quantum physics,~\cite{Einstein1935} enables a host of new technological capabilities.    Among them is quantum teleportation~\cite{Bennett1993}, a communication primitive on which a quantum internet can be built~\cite{Lloyd2004,Kimble2008,Wehner2018,Azuma2023}.  Consequently, long-distance high-rate entanglement distribution---over optical fiber or via free-space propagation---is eagerly being sought.  A 248\,km distance record was established in 2022 for a fiber link~\cite{Neumann2022}, but it provided a minuscule 9/s rate owing to the 79\,dB link loss.  Quantum repeaters~\cite{Azuma2023,Sangouard2011,Muralidharan2015} can multiply the distance over which entanglement can be distributed at high rate over optical fiber, but they have yet to reach an advanced state of development.  Alternatively, a low earth orbit (LEO) satellite-based system has distributed entanglement between ground stations separated by 1200\,km~\cite{Yin2017}.  Moreover, a LEO satellite's motion, relative to the ground, affords a distribution rate largely independent of the distance between ground stations, cf.\@ the quantum key distribution results from the same satellite~\cite{Liao2017}.  However, the satellite-based entanglement-distribution rate (1.1/s) at 1200\,km ground-station separation is still far below that needed for a useful quantum internet, and the satellite system's secret-key rate (1.1\,kbit/s) at that separation is likewise far below that needed for one-time pad transmission of large data files.  

Recently, a potential means for dramatically increasing entanglement-distribution rates---over fiber or via satellite---was proposed by Chen~\emph{et al.}~\cite{Chen2023}.  Their proposal uses spontaneous parametric downconverters (SPDCs) and partial Bell-state measurements (BSMs)~\cite{Braunstein1995} to herald the generation of polarization-entangled biphotons across a large number of frequency-multiplexed signal channels that collectively span the downconverters' $\sim$10\,THz phase-matching bandwidth.   
Thus, multi-pair generation in any particular frequency channel can be kept low while the overall generation rate improves as the pump power and the number of channels increase. Furthermore, as explained in Ref~\cite{Chen2023}, by transmitting the heralded biphotons to Alice and Bob's terminals---along with  classical messages identifying their heralded frequency channels and whether the heralded polarization-entangled Bell state is a $\psi^-$ singlet or a $\psi^+$ triplet---Alice and Bob can use quantum frequency conversion~\cite{Kumar1990,Huang1992} and time-lens bandwidth compression~\cite{Karpinski2017} to efficiently couple the received biphotons to pairs of intra-cavity color-center quantum memories.   

Chen~\emph{et al}.\@ dubbed their approach zero-added-loss multiplexing (ZALM), because it avoids the switching losses incurred in previously proposed multiplexing schemes~\cite{Mower2011,Dhara2022}.  Like Ref.~\cite{Lloyd2004}'s fiber-based entanglement-distribution architecture, and Ref.~\cite{Yin2017}'s satellite-based entanglement-distribution system, ZALM uses a source-in-the-middle configuration, which is known~\cite{Jones2016} to offer an efficiency---hence a distribution rate---advantage over its meet-in-the-middle and sender-receiver alternatives in both fiber and free-space operation.  The free-space advantage is further increased when atmospheric turbulence is concentrated near Alice and Bob's terminals, as it is if the ZALM source is on a LEO satellite.    

Reference~\cite{Chen2023} contains a broad assessment of ZALM performance, concentrating on the  entanglement-distribution rate and its tradeoff with color-center entangled-state fidelity, and it includes simulation results for ZALM's mode conversion, i.e., the combination of quantum frequency conversion and bandwidth compression needed to efficiently couple arriving biphotons to the intra-cavity color centers, as well as  other implementation considerations. 

The present paper takes a more narrow view, viz., a deep dive into the SPDCs and the partial BSMs that yield heralded polarization-entangled biphotons across a set of frequency channels, and the ensuing color-center entangled-state fidelity between Alice and Bob's Duan-Kimble~\cite{Duan2004} intra-cavity quantum memories. Two major results are the density operators for the heralded biphoton produced by the source and for the loaded quantum memories.  These operators quantify the need to input near-separable channelized biphoton states from two SPDC sources to the partial BSMs in order to get a high-purity heralded biphoton, and thereby a high color-center entangled-state fidelity.  Moreover, this need applies even for ideal partial-BSM apparatus, and for intra-cavity color-center quantum memories with perfect---unit magnitude and $\pi$-rad phase difference---state-dependent reflectivities in their Duan-Kimble loading protocol.  More importantly, our density operators lend themselves to numerical evaluations that enable accounting for nonidealities in the ZALM source and for memories with realistic state-dependent reflectivities.  In this regard it is important to note that our memory analysis employs a rigorously derived and improved ``push-pull'' version of the Duan-Kimble protocol, which we obtained in our paper's companion work, Raymer~\emph{et al}.~\cite{Raymer2024}.

The remainder of the paper is organized as follows.  We begin, in Sec.~\ref{ZALM}, with an overview of ZALM, focusing on the dual-Sagnac SPDC source, partial BSM, and Duan-Kimble quantum memories.   In Section~\ref{Sagnac}, we characterize a channelized, pulse-pumped Sagnac-configured SPDC source~\cite{Wong2006} of unheralded polarization-entangled biphotons, starting from the perturbation theory result for  arbitrary pump spectrum and phase-matching function, and then specializing to a transform-limited Gaussian pump pulse and Gaussian phase-matching function.  Section~\ref{Sagnac} also presents the channelized biphoton wave function's singular-value decomposition (SVD), which is useful for theoretical calculations and 
importantly offers a practical route to numerical evaluation of memory-loading performance in broadband conditions.  Next, in Sec.~\ref{partial-BSM}, we derive the density operator of the heralded polarization-entangled biphoton produced by the channelized partial BSM on the idler photons from a dual-Sagnac source.  There we evaluate the heralding probability, heralding efficiency, purity, and error probability of the general heralded biphoton and report numerical results for the all-Gaussian special case.   In Sec.~\ref{memory}, we derive the intra-cavity color-center density operator for the Duan-Kimble quantum memories operated in push-pull mode.  This section reports the fidelity for ideal (lossless, $\pi$-rad phase shift) state-dependent memory-cavity reflectivities and how that fidelity may be realized---at the expense of a reduction in the entanglement-distribution rate---in narrowband operation with the push-pull protocol.  The main text concludes, in Sec.~\ref{Discuss}, with a brief treatment of the tradeoff between ZALM's entanglement-distribution rate and its inter-channel interference, followed by discussion of some possibilities for extending our analysis to include equipment nonidealities, alternative SPDC sources, alternative memory protocols, etc.  Appendix~\ref{AppendA} and Appendix~\ref{AppendB} then supply derivation details for Secs.~\ref{partial-BSM} and \ref{memory}, respectively.     

\section{ZALM Overview \label{ZALM}}
\begin{figure}[hbt]
    \centering
    \includegraphics[width=0.3\textwidth]{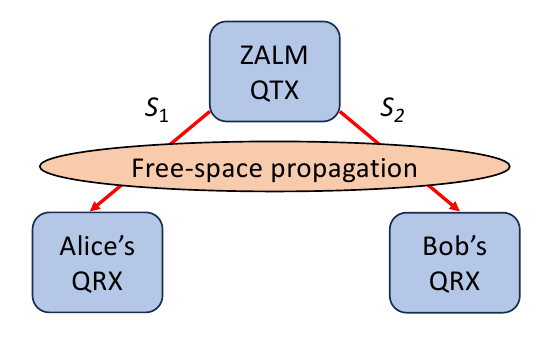}
    \caption{High-level cartoon of free-space entanglement distribution via zero added-loss multiplexing (ZALM). QTX:  quantum transmitter.  QRX:  quantum receiver. \label{ZALM-boxes}}    
\end{figure}

\begin{figure}[hbt]
    \centering
    \includegraphics[width=0.49\textwidth]{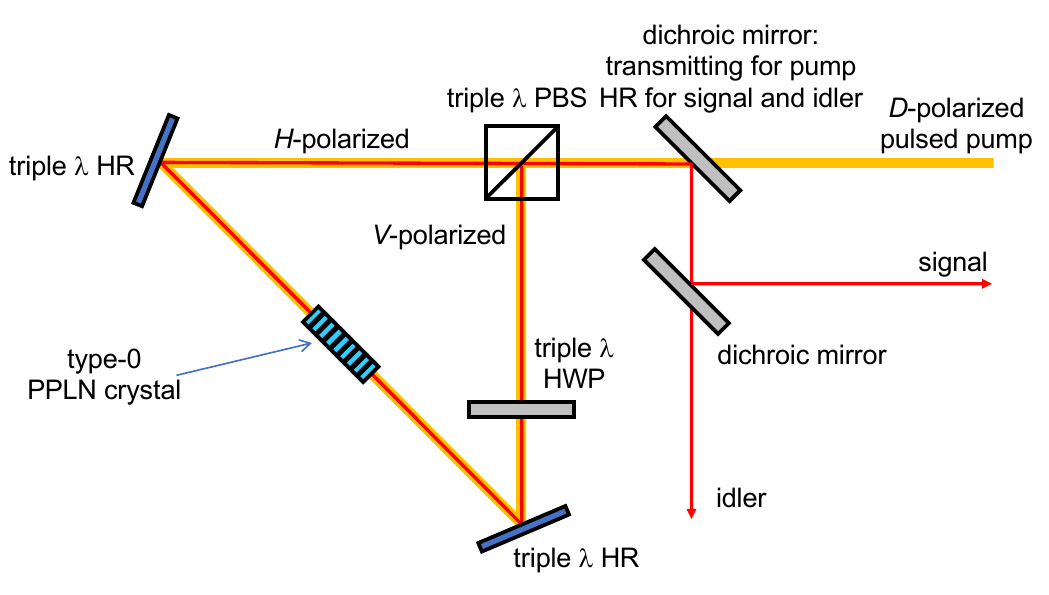}
    \caption{Schematic of a Sagnac-configured SPDC source~\cite{Wong2006} of singlet-state signal-idler biphotons suitable for use in ZALM.  A periodically-poled lithium niobate (PPLN) crystal~\cite{footnote1} is bidirectionally pulse-pumped for type-0 nondegenerate phase matching.  $D$, $H$, and $V$: diagonal, horizontal, and vertical polarizations.  HR:  high reflector. $\lambda$:  wavelength.  PBS:  polarizing beam splitter.  HWP:  half-wave plate. \label{Sagnac-source_fig}}    
\end{figure}

\begin{figure}[hbt]
    \centering
   \includegraphics[width=0.49\textwidth]{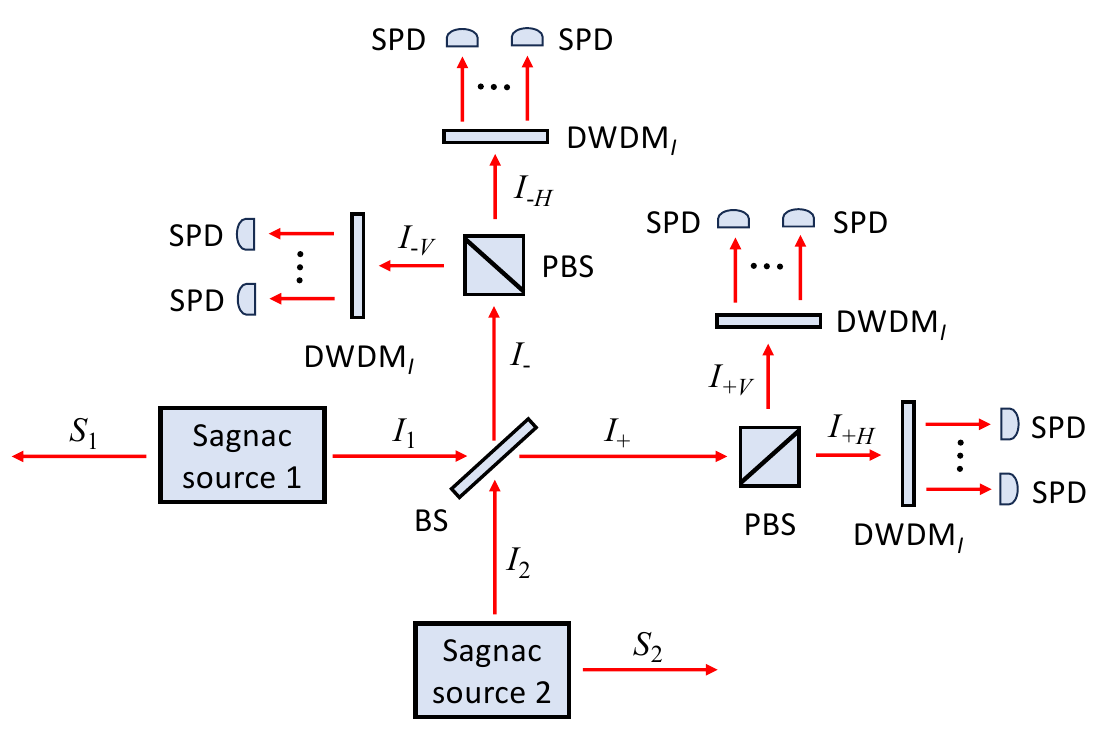}
    \caption{Schematic of ZALM's partial Bell-state measurement for heralding polarization-entangled photon pairs.  $S_k$ and $I_k$ for $k = 1,2$:  signal ($S$) and idler ($I$) beams from the $k$th dual-Sagnac source.  BS: 50--50 beam splitter. $I_\pm$:  the idler-beam outputs from the 50--50 beam splitter.  PBS:  polarizing beam splitter.  $I_{\pm P}$ for $P = H, V$:  horizontally ($H$) and vertically ($V$) polarized outputs from the PBS illuminated by $I_\pm$.  DWDM$_I$:  idler-beam dense wavelength-division multiplexing filter.  SPD:  single-photon detector.
 \label{partial-BSM_fig} }
\end{figure}

Figure~\ref{ZALM-boxes} is a high-level cartoon of free-space entanglement distribution via ZALM.  The ZALM quantum transmitter (QTX) contains two coherently pulse-pumped Sagnac-configured SPDC sources, which we assume to be identical, each arranged to emit polarization-entangled singlet states, as shown in  Fig.~\ref{Sagnac-source_fig} and analyzed in Sec.~\ref{Sagnac}.  Their idler beams are routed to the spectrally channelized partial-BSM setup shown in Fig.~\ref{partial-BSM_fig} and analyzed in Sec.~\ref{partial-BSM}.  Specifically, these idlers are combined on a 50--50 beam splitter whose outputs have their horizontal ($H$) and vertical ($V$) polarization components separated by polarizing beam splitters (PBSs). The PBSs' outputs are then channelized by dense wavelength-division multiplexing (DWDM) filters whose channelized outputs illuminate single-photon detectors (SPDs).  

A particular pump pulse results in an $n$th-channel partial-BSM success when its combined idlers register one $H$ and one $V$ detection at the $n$th channel's SPDs; see Figure~\ref{partial-BSM_fig}.  When these detections come from the $I_{+H}$ and $I_{-V}$ or the $I_{+V}$ and $I_{-H}$ SPDs, they herald a $\psi^-$ polarization-singlet state, commonly referred to as the $HV$$-$$VH$ state.  On the other hand, when they come from the $I_{+H}$ and $I_{+V}$ or the $I_{-H}$ and $I_{-V}$ SPDs, they herald a $\psi^+$ polarization-triplet state, commonly called the $HV$+$VH$ state.  

The heralded polarization-entangled signal-signal ($S_1$-$S_2$) biphotons are then sent to Alice and Bob's quantum receivers (QRXs) via free-space propagation, accompanied by classical messages specifying the heralded states and their frequency channels.   
(For fiber connections to Alice and Bob's QRXs, polarization entanglement would be converted to time-bin entanglement at the ZALM QTX before transmission.  See Sec.~\ref{Discuss} for a discussion of how our analysis can be applied to time-bin entanglement distribution over fiber.)  

\begin{figure}[hbt]
    \centering
    \includegraphics[width=0.37\textwidth]{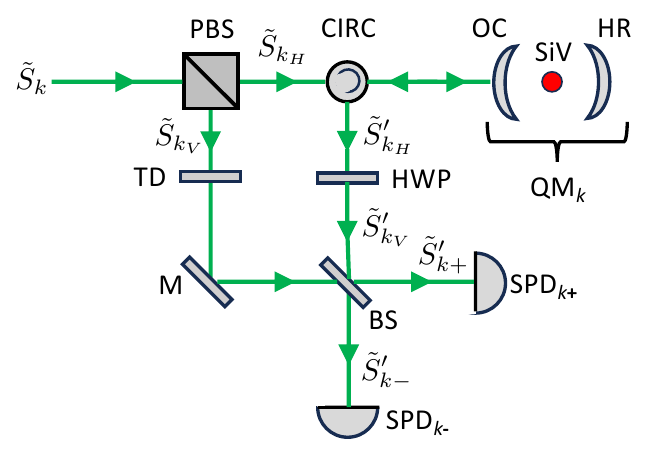}
    \caption{Notional schematic for Duan-Kimble loading of Alice and Bob's intra-cavity color-center quantum memories, assuming the partial BSM heralds an $n$th-channel biphoton. $\tilde{S}_k$:  the frequency-converted, bandwidth-compressed $n$th-channel signal photon arriving at Alice's memory ($k =1$) and Bob's memory ($k=2$).  PBS:  polarizing beam splitter.  $\tilde{S}_{k_P}$:  the horizontally ($P=H$) and vertically ($P=V$) polarized components of $\tilde{S}_k$.  CIRC:  optical circulator.  QM:  quantum memory.  OC and HR:  optical coupler and high-reflector of a single-ended Fabry-P\'{e}rot (FP) cavity~\cite{footnote2}.  SiV:  silicon-vacancy color center~\cite{footnote3} contained in the FP cavity.  $\tilde{S}'_{k_H}$: $H$-polarized signal photon after reflection from the QM.  TD:  $T$-sec-long time delay.  M: mirror.  HWP:  half-wave plate.  $\tilde{S}'_{k_V}$:  $V$-polarized signal photon obtained by passing $\tilde{S}'_{k_H}$ through the HWP.  BS:  50--50 beam splitter.  $\tilde{S}'_{K\pm}$:  outputs from the 50--50 beam splitter.  SPD$_\pm$:  single-photon detectors on the beam splitter's $\pm$ output ports. \label{DuanKimble}}    
\end{figure}

The propagation-surviving signal photons arriving at Alice and Bob's QRXs are DWDM channelized and, when the $n$th channel has been heralded by the ZALM QTX, Alice and Bob's $n$th-channel DWDM outputs are frequency converted and bandwidth compressed to prepare them for loading into Alice and Bob's intra-cavity color-center quantum memories.  Following these transformations---which we take to be ideal, so as to focus our attention on the fundamental limits of ZALM---the $n$th-channel signal at each QRX is separated into its horizontally and vertically polarized components by a PBS, with the former undergoing a state-dependent reflectivity from the memory cavity and the latter being time delayed, as shown in Fig.~\ref{DuanKimble} and analyzed in Sec.~\ref{memory}.   After rotation to vertical polarization, the cavity-reflected signal is interfered with the time-delayed version of the PBS's vertically-polarized output on Fig.~\ref{DuanKimble}'s 50--50 beam splitter~\cite{footnote1a}.    Entanglement of Alice and Bob's memories is heralded when each QRX detects a photon from a particular QTX herald.  As shown in Sec.~\ref{memory}, a successful herald loads a $\psi^-$ singlet state into their memories with, in general, its entangled-state fidelity depending on which combination of detectors registered counts.   

\begin{figure}[hbt]
    \centering
    \includegraphics[width=0.28\textwidth]{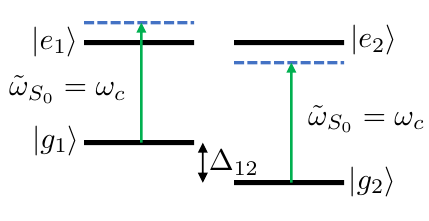}
    \caption{Schematic of the SiV memory's four-state model with push-pull operation.  $\tilde{\omega}_{S_0}$:  center frequency of the frequency-shifted, bandwidth-compressed signal photons.  $\omega_c$:  resonance frequency of the SiV-containing cavity.  $|g_1\rangle$ and $|g_2\rangle$:  ground states of the two independent SiV transitions in which the qubit will be stored.  $|e_1\rangle$ and $|e_2\rangle$:  excited states of the two independent SiV transitions.  $\Delta_{12}$: blue detuning of transition~2 from transition~1.  In push-pull operation, $\tilde{\omega}_{S_0}=\omega_c$ is blue (red) detuned by $\Delta_{12}/2$ from the first (second) transition. \label{transitions}}    
\end{figure}

The original version of the Duan-Kimble quantum memory~\cite{Duan2004}, and related  experiments with SiV-based 
memories~\cite{Nguyen2019,Bersin2024,Knaut2024}, used an on-off protocol.   In a simplified four-state model, in which the qubit is stored in two ground states each of which couples to a different excited state, on-off operation has the optical frequency \emph{on} resonance with the cavity and one of the ground-to-excited-state transitions and far \emph{off} resonance for the other transition.  In Raymer~\emph{et al}.~\cite{Raymer2024} we introduced the push-pull protocol, in which the optical frequency is on resonance with the cavity but midway between the two ground-to-excited-state transitions, as shown in Fig.~\ref{transitions}, and our analysis there showed that push-pull operation offers several advantages over on-off operation. Hence we adopt the push-pull scheme for this study, although our memory analysis can be modified to apply to on-off operation.   

In summary, ZALM's goal is to spectrally multiplex heralded biphotons at its QTX thereby increasing the entanglement-distribution rate to remote locations while minimizing the occurrence of multiple biphotons from a given pump pulse in a given DWDM channel. 
ZALM's advantageous scaling, in comparison with its entanglement-distribution competitors, was demonstrated theoretically in Chen~\emph{et al}.~\cite{Chen2023}.

We are now prepared to dive into analyzing ZALM's source, partial-BSM, and memory loading.  Before doing so, however, it is worth warding off a possible misunderstanding about heralding.  Aside from losses, our analyses will assume ideal equipment, e.g., the detectors will not have dark counts, the beam splitters (polarizing and 50--50) will be perfect, etc.  Nevertheless, we will see that a partial BSM which heralds a $\psi^\mp$ state need \emph{not} result in a $\psi^\mp$ state being sent to Alice and Bob.  Indeed, there is, in general, a nonzero probability that a $\psi^\mp$ herald from the partial-BSM results in a $\psi^\pm$ state being transmitted to Alice and Bob, thus degrading the ZALM scheme's performance.  The culprit here is the ZALM QTX's having Sagnac-configured SPDC sources that emit \emph{nonseparable} channelized biphotons.   To the best of our knowledge, our work is the first to explicitly treat that error probability and its effect on heralded-biphoton purity and consequent color-center entangled-state fidelity. See Ou~\emph{et al}.~\cite{Ou1999}, however, for an early treatment of quantum interference between the outputs from a pair of SPDC sources.

\section{Sagnac-Configured SPDC Source \label{Sagnac}}
Here we begin our paper's journey---from biphoton source, through partial BSM, to loaded intra-cavity color-center quantum memories---by analyzing the Sagnac-configured SPDC source shown in Fig.~\ref{Sagnac-source_fig}, focusing on its ensuing behavior when its signal and idler are channelized by complementary DWDM filters. 
 
\subsection{Perturbation theory for the SPDC source \label{singlePerturbation}}
In Fig.~\ref{Sagnac-source_fig}, a horizontally ($H$) polarized pulse with center frequency $\omega_P$ pumps a type-0 phase matched PPLN crystal in both the counter-clockwise (ccw) and clockwise (cw) directions.  Operation is nondegenerate in frequency, with the SPDC's ccw-propagating and cw-propagating signal and idler having center frequencies $\omega_{S_0}$ and $\omega_{I_0}$ satisfying $\omega_{S_0} + \omega_{I_0} = \omega_P$ and $|\omega_{S_0}-\omega_{I_0}| \gg \Omega_{\rm PM}$, where $\Omega_{\rm PM}$ is the crystal's phase-matching bandwidth.  Type-0 phase matching results in the ccw- and cw-propagating signal and idler being $H$ polarized as they exit the crystal. Consequently, Fig.~\ref{Sagnac-source_fig}'s half-wave plate (HWP), triple-wavelength PBS and dichroic mirrors are able to separate the pump, signal, and idler after they exit the Sagnac loop. 

Using standard perturbation theory~\cite{Chen2023,Wong2006}, justified by the inefficient nature of spontaneous parametric downconversion, a pump pulse will, to first order, result in the following ccw-propagating and cw-propagating unnormalized (vacuum plus biphoton) states entering Fig.~\ref{Sagnac-source_fig}'s PBS:
\begin{align}
&|\tilde{\psi}\rangle_{SI}^{\rm ccw} = \sqrt{P_0}\,|{\bf 0}\rangle_{SI} \nonumber \\[.05in]
&\,\,- \sqrt{P_1}\int\!\frac{{\rm d}\omega_S}{2\pi}\int\!\frac{{\rm d}\omega_I}{2\pi}\,\Psi_{SI}(\omega_S,\omega_I)|\omega_S\rangle_{S_V}|\omega_I\rangle_{I_V},\\[.05in]
&|\tilde{\psi}\rangle_{SI}^{\rm cw} = \sqrt{P_0}\,|{\bf 0}\rangle_{SI} \nonumber \\[.05in]
&\,\, + \sqrt{P_1}\int\!\frac{{\rm d}\omega_S}{2\pi}\int\!\frac{{\rm d}\omega_I}{2\pi}\,\Psi_{SI}(\omega_S,\omega_I)|\omega_S\rangle_{S_H}|\omega_I\rangle_{I_H},
\end{align}
where $P_0 \sim 1$ is the probability that the ccw (cw) pump pulse does not produce a ccw (cw) biphoton, $P_1\ll P_0$ is the probability that it does, and the sign difference between the two biphotons is due to the HWP.  In these equations: $|{\bf 0}\rangle_{SI}$ is the signal-idler vacuum state; $\Psi_{SI}(\omega_S,\omega_I)$ is the normalized,
\begin{equation}
\int\!\frac{{\rm d}\omega_S}{2\pi}\int\!\frac{{\rm d}\omega_I}{2\pi}\,|\Psi_{SI}(\omega_S,\omega_I)|^2 = 1,
\end{equation}
frequency-domain biphoton wave function; the $|\omega_K\rangle_{K_P}$ for $K = S,I$ and $P = H,V$, denote horizontally ($P=H$) and vertically ($P=V$) polarized, frequency-$\omega_K$, single-photon states~\cite{footnote4} of the signal ($K=S$) and idler ($K=I$); and integrals without limits are taken to be from $-\infty$ to $\infty$ because $\Omega_{\rm PM} \ll \omega_{K_0}$.  So, because the ccw and cw pumps are coherent with each other, the preceding ccw and cw states give us the following unnormalized (vacuum plus biphoton) state leaving the Sagnac source:
\begin{equation}
|\tilde{\psi}\rangle_{SI} = P_0|\,{\bf 0}\rangle_{SI} + \sqrt{2P_0P_1}\,|\psi\rangle_{SI},
\end{equation}
where
\begin{equation}
|\psi\rangle_{SI} \equiv \int\!\frac{{\rm d}\omega_S}{2\pi}\!\int\!\frac{{\rm d}\omega_I}{2\pi}\,\Psi_{SI}(\omega_S,\omega_I)|\psi^-(\omega_S,\omega_I)\rangle_{SI},
\label{SagnacBiphoton}
\end{equation}
is the Sagnac source's post-selected, normalized biphoton state, with 
$|\psi^-(\omega_S,\omega_I)\rangle_{SI}$ being the polarization-entangled singlet state,
\begin{equation}
|\psi^-(\omega_S,\omega_I)\rangle_{SI} \equiv (|\omega_S\rangle_{S_H}|\omega_I\rangle_{I_V} - 
|\omega_S\rangle_{S_V}|\omega_I\rangle_{I_H})/\sqrt{2}.
\end{equation}

The biphoton wave function plays a pivotal role in what follows.  It is properly normalized and proportional to $\mathcal{E}_P(\omega_S+\omega_I)\Phi_{\rm PM}(\omega_S,\omega_I)$~\cite{Chen2023,Wong2006}, where $\mathcal{E}_P(\omega)$ is the spectrum of the pump pulse's positive-frequency field and $\Phi_{\rm PM}(\omega_S,\omega_I)$ is the PPLN crystal's phase-matching function.  The next subsection introduces the channelized version of $\Psi_{SI}(\omega_S,\omega_I)$, and uses its all-Gaussian special case to illustrate a fundamental tradeoff between heralding efficiency and purity.

\subsection{Channelized SPDC source \label{singleChannelized}}
Anticipating the partial-BSM's DWDM channelizing of the idlers from the ZALM QTX's pair of Sagnac SPDCs, and the complementary DWDM channelizing of the signals arriving at Alice and Bob's QRXs, let us examine their effects on the $|\psi\rangle_{SI}$ biphoton from Eq.~(\ref{SagnacBiphoton}).  To do so, we will employ idealized $N$-channel DWDM filters---DWDM$_S$ for the signal and DWDM$_I$ for the idler---with $\delta B$\,Hz bandwidth brickwall channels and $\Delta B$\,Hz channel spacing, where $\Delta B > \delta B$ provides inter-channel guard bands.  Their $n$th-channel frequency responses are
\begin{equation}
H_{S_n}(\omega_S) = \left\{\begin{array}{ll}
1, & \mbox{for $|\omega_S-\omega_{S_0}+2\pi n \Delta B|\le \pi\delta B$}\\[.05in]
0, & \mbox{otherwise},
\end{array}
\right.
\end{equation}
and
\begin{equation}
H_{I_n}(\omega_I) = \left\{\begin{array}{ll}
1, & \mbox{for $|\omega_I-\omega_{I_0}-2\pi n \Delta B|\le \pi\delta B$}\\[.05in]
0, & \mbox{otherwise},
\end{array}
\right.
\end{equation}
for $n = -(N-1)/2, -(N-3)/2,\ldots,(N-1)/2$, with $N$ being an odd integer.  

For a biphoton produced with monochromatic SPDC pumping, energy conservation dictates that an idler photon at frequency $\omega_I = \omega_{I_0} + \delta\omega$ is accompanied by a signal photon at frequency $\omega_S = \omega_{S_0}-\delta\omega$.  For pulsed pumping, however, this need not be the case, because there is a range of energy-conserving frequencies for the signal-photon companion of an idler at frequency $\omega_I = \omega_{I_0} + \delta\omega$.  

To illustrate this behavior---and for future use throughout the partial-BSM and memory-loading analyses---we introduce the all-Gaussian biphoton wave function~\cite{U'Ren2005,Zhang2014},
\begin{align}
\Psi_{SI}(\omega_S,\omega_I) &= \sqrt{8\pi \sigma_P/\Omega_{\rm PM}}\,e^{-(\Delta\omega_S+\Delta\omega_I)^2\sigma^2_P/16} \nonumber \\[.05in]
&\,\,\times e^{-4(\Delta\omega_S-\Delta\omega_I)^2/\Omega_{\rm PM}^2},
\label{allGauss}
\end{align}
with $\Delta\omega_K\equiv \omega_K-\omega_{K_0}$, for $K = S,I$, being the signal and idler's detunings from their respective center frequencies. 
The first exponential term in Eq.~(\ref{allGauss}) comes from $\mathcal{E}_P(\omega_S+\omega_I)$ for a transform-limited Gaussian pump pulse with time-domain complex envelope proportional to $e^{-4t^2/\sigma_P^2}$, and the second exponential term comes from $\Phi_{\rm PM}(\omega_S,\omega_I)$ for a PPLN crystal engineered to have a Gaussian phase-matching function~\cite{U'Ren2005,Dixon2013,Chen2017,Chen2019}.    

\begin{figure}[hbt]
    \centering
    \includegraphics[width=0.45\textwidth]{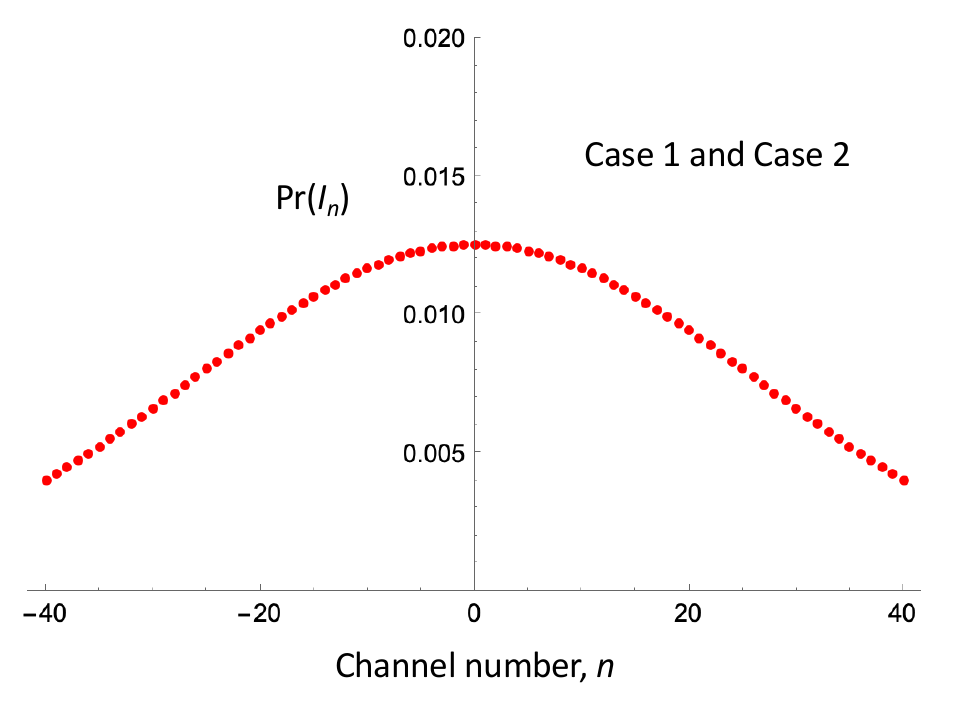}
    \caption{Heralding probability, $\Pr(I_n)$ from Eq.~(\ref{singleHerald}), plotted versus channel number assuming Eq.~(\ref{allGauss})'s all-Gaussian biphoton wave function with brickwall filtering and parameter values from Table~\ref{singleParams}.  The Case~1 and Case~2 curves are indistinguishable.  \label{singleHerald_fig}}    
\end{figure}

\begin{figure}[hbt]
    \centering
    \includegraphics[width=0.45\textwidth]{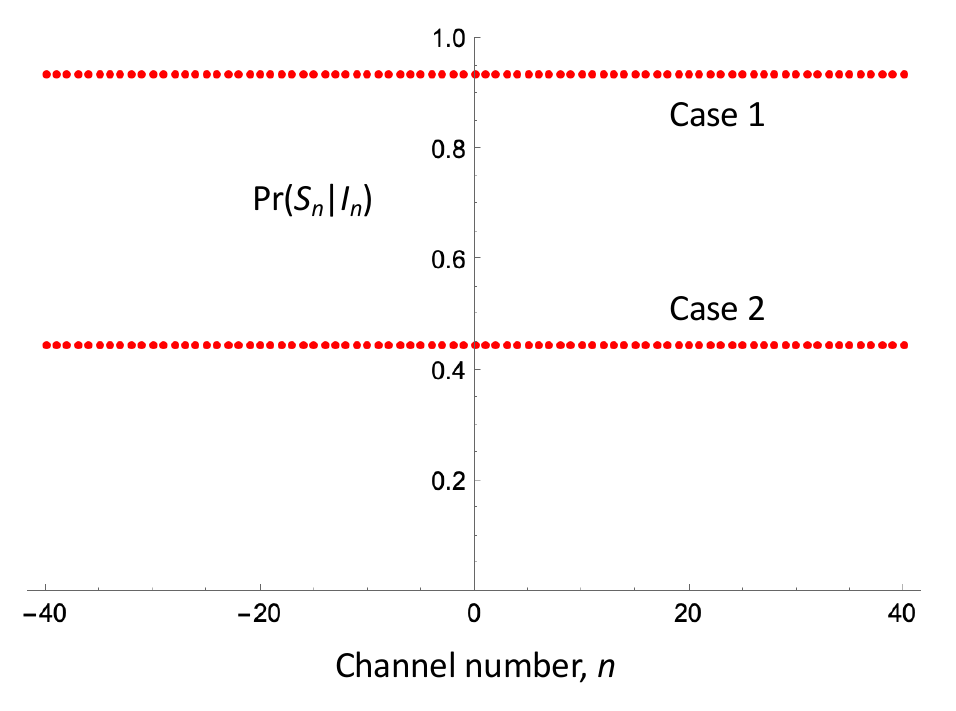}
    \caption{Heralding efficiency, $\Pr(S_n\mid I_n)$ from Eq.~(\ref{singleEfficiency}), plotted versus channel number assuming Eq.~(\ref{allGauss})'s all-Gaussian biphoton wave function with brickwall filtering and  parameter values from Table~\ref{singleParams}.  \label{singleEfficiency_fig}}    
\end{figure}

Using the all-Gaussian wave function with brickwall filtering and parameter values from Table~\ref{singleParams}, Fig.~\ref{singleHerald_fig} plots, versus channel number, the SPDC source's heralding probability, i.e., the probability that the idler photon will pass through the DWDM$_I$ filter's $n$th channel, 
\begin{equation}
\Pr(I_n) = \int\!\frac{{\rm d}\omega_S}{2\pi}\int\!\frac{{\rm d}\omega_I}{2\pi}\,|\Psi_{SI_n}(\omega_S,\omega_I)|^2,
\label{singleHerald}
\end{equation}
where $\Psi_{SI_n}(\omega_S,\omega_I) \equiv \Psi_{SI}(\omega_S,\omega_I)H_{I_n}(\omega_I)$. Figure~\ref{singleEfficiency_fig} then plots, versus channel number, the heralding efficiency, viz.,  
the probability that the signal photon will pass through the DWDM$_S$ filter's $n$th channel given that its idler companion passed through the DWDM$_I$ filter's $n$th channel,
\begin{equation}
\Pr(S_n\mid I_n) = \frac{\displaystyle \int\!\frac{{\rm d}\omega_S}{2\pi}\int\!\frac{{\rm d}\omega_I}{2\pi}\,|\Psi_{S_nI_n}(\omega_S,\omega_I)|^2}{\displaystyle \int\!\frac{{\rm d}\omega_S}{2\pi}\int\!\frac{{\rm d}\omega_I}{2\pi}\,|\Psi_{SI_n}(\omega_S,\omega_I)|^2},
\label{singleEfficiency}
\end{equation}
where $\Psi_{S_nI_n}(\omega_S,\omega_I) \equiv \Psi_{SI}(\omega_S,\omega_I)H_{S_n}(\omega_S)H_{I_n}(\omega_I)$ is the dual-Sagnac source's channelized biphoton wave function.

Figures~\ref{singleHerald_fig} and \ref{singleEfficiency_fig} reveal some interesting behaviors.  First, $\Pr(I_n)$ is virtually independent of the pump pulse's duration and rolls off, with increasing $|n|$, owing to the Gaussian nature of the phase-matching function.  Second, $\Pr(S_n\mid I_n)$ is constant across DWDM channels---hence largely independent of the phase-matching bandwidth---but it is strongly affected by the pump pulse's duration.  Indeed, as expected, when the pump pulse's bandwidth, $\sim$$1/\sigma_P$ is significantly less than the channel bandwidth, $\delta B$, as it is in Case~1, we get $\Pr(S_n\mid I_n) \approx 0.94$, but when $1/\sigma_P = 2.5\,\delta B$, as it is in Case~2, $\Pr(S_n\mid I_n)$ drops to 0.44.
\begin{table}[hbt]
\begin{tabular}{|c|c|c|c|}\hline\hline
 Parameter & Symbol & Case~1& Case~2 \\ \hline
 pump-pulse duration & $\sigma_P$ &160\,ps & 16\,ps \\[.05in]
 phase-matching bandwidth   & $\Omega_{\rm PM}/2\pi$ & 6.37\,THz & 6.37\,THz\\[.05in]
 channel bandwidth & $\delta B$ & 25\,GHz & 25\,GHz \\[.05in]
DWDM channel spacing & $\Delta B$ &30\,GHz &30\,GHz \\[.05in]
\# of DWDM channels & $N$ & 81 & 81 \\[.05in]
\hline\hline
 \end{tabular}
 \caption{Parameter values used in Figs.~\ref{singleHerald_fig}--\ref{crosstalk2_fig}.  $\Omega_{\rm PM}/2\pi = 6.37$\,THz corresponds to a $\sigma_{\rm cor} = 0.1$\,ps biphoton correlation time in Ref.~\cite{Zhang2014}'s time-domain wave function.
 \label{singleParams}}
 \end{table}  

It might seem that long pump pulses would be desirable, because they ensure high values of the Sagnac source's channelized heralding efficiency, i.e., $\Pr(S_n\mid I_n)$.  Heralding efficiency, however, is \emph{not} the only figure of merit worthy of consideration. Both ZALM's partial BSMs and its intra-cavity color-center quantum memories rely, for their proper functioning, on high-quality quantum interference, e.g., between $\tilde{S}'_{kV}$ and $\tilde{S}_{kV}$ in Fig.~\ref{DuanKimble} for the quantum memories.  High-quality quantum interference requires co-polarized photons of high purity.  To assess what it takes to achieve that high purity, consider the the unnormalized density operator of an $n$th-channel biphoton from a single Sagnac source, 
\begin{align}
\tilde{\rho}_{S_n, I_n} &= \int\!\frac{{\rm d}^2\bomega_S}{4\pi^2}
\int\!\frac{{\rm d}^2\bomega_I}{4\pi^2}\,\Psi_{S_nI_n}(\omega_S,\omega_I)\Psi_{S_nI_n}^*(\omega'_S,\omega_I')\nonumber \\[.05in]
&\,\, \times |\psi^-(\omega_S,\omega_I)\rangle_{SI}\,{}_{SI}\langle \psi^-(\omega_S',\omega_I')|,
\end{align}
where, for $K=S,I$, $\bomega_K \equiv (\omega_K,\omega'_K)$ and ${\rm d}^2\bomega_K \equiv {\rm d}\omega_K\,{\rm d}\omega_K'$. By tracing out the idler
we obtain an unnormalized version of the conditional density operator given by
\begin{align}
\tilde{\rho}_{S_n\mid I_n} &= \int\!\frac{{\rm d}^2\bomega_S}{4\pi^2}
\int\!\frac{{\rm d}\omega_I}{2\pi}\,\Psi_{S_nI_n}(\omega_S,\omega_I)\Psi_{S_nI_n}^*(\omega'_S,\omega_I)\nonumber \\[.05in]
&\,\, \times \left[|\omega_S\rangle_{S_H}\,{}_{S_H}\langle \omega'_S| + |\omega_S\rangle_{S_V}\,{}_{S_V}\langle \omega'_S|\right]/2,
\label{SnInUnpol}
\end{align}
whose normalization constant is
\begin{align}
{\rm Tr}(\tilde{\rho}_{S_n\mid I_n}) &= \int\!\frac{{\rm d}\omega_S}{2\pi}\int\!\frac{{\rm d}\omega_I}{2\pi}\,|\Psi_{S_nI_n}(\omega_S,\omega_I)|^2  \nonumber \\[.05in]
& = \Pr(S_n,I_n),
\label{PrSnIn}
\end{align}
where $\Pr(S_n,I_n)$ is the joint probability that the signal and idler photons both pass through their DWDM filter's $n$th channels. 

Note that the quantity defined as
\begin{equation}
\hat{I}(\bomega_S) \equiv \left[|\omega_S\rangle_{S_H}\,{}_{S_H}\langle \omega'_S| + |\omega_S\rangle_{S_V}\,{}_{S_V}\langle \omega'_S|\right]/2
\end{equation} 
indicates that Eq.~(\ref{SnInUnpol}) is a randomly-polarized state, as expected from the $S_n$-$I_n$ biphoton's having maximal polarization entanglement.   Because of that random polarization, we have that
\begin{equation}
\hat{I}(\bomega_S)\equiv \left[|\omega_S\rangle_{S_P}\,{}_{S_P}\langle \omega'_S| + |\omega_S\rangle_{S_{P_\perp}}\,{}_{S_{P_\perp}}\langle \omega'_S|\right]/2,
\end{equation} 
for $(P,P_\perp)$ being an \emph{arbitrary} polarization basis, and so our quest for high-quality quantum interference in ZALM's partial BSMs and quantum memories can focus on the purity of $S_{n_P}$.  

Projecting onto Eq.~(\ref{SnInUnpol})'s $P$-polarized component, we get
\begin{align} 
\tilde{\rho}_{S_{n_P}\mid I_n} &= \int\!\frac{{\rm d}^2\bomega_S}{4\pi^2}
\int\!\frac{{\rm d}\omega_I}{2\pi}\,\Psi_{S_nI_n}(\omega_S,\omega_I)\Psi_{S_nI_n}^*(\omega'_S,\omega_I)
\nonumber \\[.05in]
&\,\, \times\,
|\omega_S\rangle_{S_P}\,{}_{S_P}\langle \omega'_S|/2,
\end{align}
whose normalization constant is $\Pr(S_n,I_n)/2$. The square of the normalized version of this density operator is therefore
\begin{eqnarray}
\lefteqn{\hspace{-.1in}\hat{\rho}^2_{S_n\mid I_n}=} \nonumber \\[.05in] 
&& \hspace{-.27in}\frac{\displaystyle 
\int\!\frac{{\rm d}^2\bomega_S}{4\pi^2}\int\!\frac{{\rm d}\tilde{\omega}_S}{2\pi}\,\Phi_n(\omega_S,\omega'_S)\Phi_n(\omega_S',\tilde{\omega}_S)  |\omega_S\rangle_{S_P}\,{}_{S_P}\langle \tilde{\omega}_S|}
{\displaystyle \left(\int\!\frac{{\rm d}\omega_S}{2\pi}\int\!\frac{{\rm d}\omega_I}{2\pi}\,|\Psi_{S_nI_n}(\omega_S,\omega_I)|^2\right)^2}\!,
\end{eqnarray}
with
\begin{equation}
\Phi_n(\omega_S,\omega'_S) \equiv \int\!\frac{{\rm d}\omega_I}{2\pi}\,\Psi_{S_n,I_n}(\omega_S,\omega_I) \Psi^*_{S_nI_n}(\omega'_S,\omega_I),
\label{PhiNdefn}
\end{equation}
from which we get that the purity of $S_{n_P}$ is 
\begin{eqnarray}
\lefteqn{{\rm Tr}(\hat{\rho}^2_{S_{n_P}\mid I_n}) =}\nonumber \\[.05in]
&&\hspace{-.15in}\frac{\displaystyle \int\!\frac{{\rm d}^2\bomega_S}{4\pi^2}\,\left| \int\!\frac{{\rm d}\omega_I}{2\pi}\,\Psi_{S_nI_n}(\omega_S,\omega_I)\Psi^*_{S_nI_n}(\omega'_S,\omega_I)\right|^2}{\displaystyle \left(\int\!\frac{{\rm d}\omega_S}{2\pi}\int\!\frac{{\rm d}\omega_I}{2\pi}\,|\Psi_{S_nI_n}(\omega_S,\omega_I)|^2\right)^2}.
\label{singlePurity}
\end{eqnarray}

Applying the Schwarz inequality to Eq.~(\ref{singlePurity})'s numerator verifies that ${\rm Tr}(\hat{\rho}^2_{S_{n_P}\mid I_n}) \le 1$,  and it is 
seen that equality occurs when the channelized biphoton wave function is separable, i.e., $\Psi_{S_nI_n}(\omega_S,\omega_I) = \Psi_{S_n}(\omega_S)\Psi_{I_n}(\omega_I)$.  In the next subsection we prove that separability is the only way to obtain 100\% purity. 

\begin{figure}[hbt]
    \centering
    \includegraphics[width=0.45\textwidth]{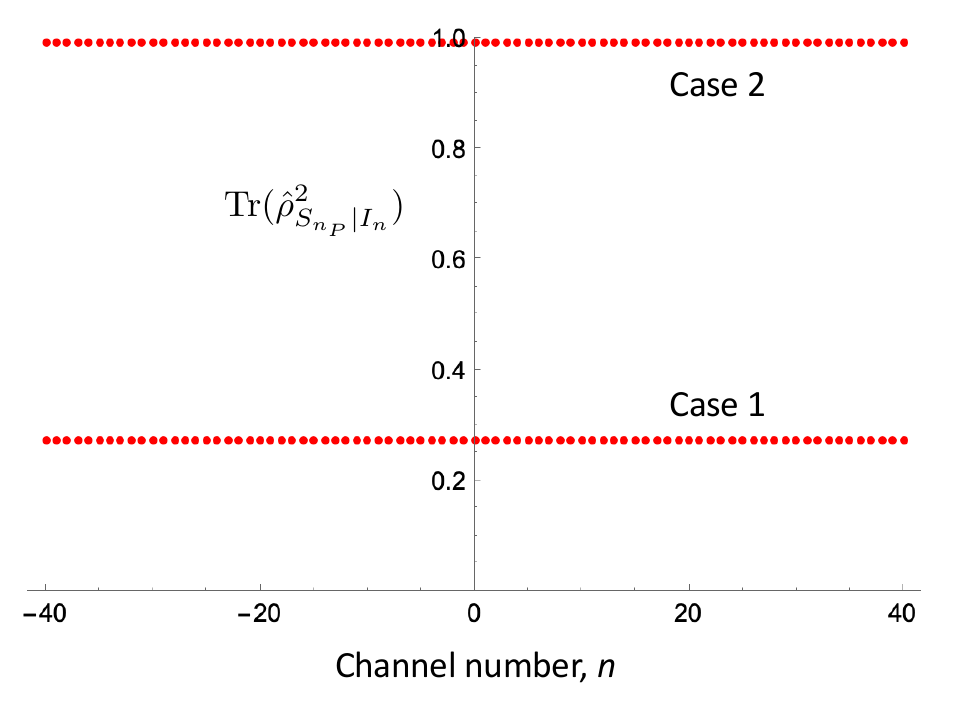}
    \caption{Single source's single-polarization purity, ${\rm Tr}(\hat{\rho}_{S_{n_P}\mid I_n}^2)$ from Eq.~(\ref{singlePurity}), versus channel number for the $n$th-channel signal photon resulting from an $n$th-channel idler herald assuming Eq.~(\ref{allGauss})'s all-Gaussian biphoton wave function with brickwall filtering and parameter values from Table~\ref{singleParams}. \label{singlePurity_fig}}   
\end{figure}

Figure~\ref{singlePurity_fig} plots the single source's single-polarization purity, ${\rm Tr}(\hat{\rho}_{S_{n_P}\mid I_n}^2)$, versus channel number assuming Eq.~(\ref{allGauss})'s all-Gaussian biphoton wave function and brickwall filtering with parameter values from Table~\ref{singleParams}.  Like what we found for the heralding efficiency, we see that the phase-matching bandwidth has no effect on the purity while the pump pulse's bandwidth does impact the purity.  Unlike what we found for the heralding efficiency, however, Case~2 is now preferable to Case~1, because the former's low time-bandwidth product,  $\sigma_P\delta B =  0.4$, ensures a nearly single temporal mode (100\% purity) signal photon, whereas the latter's $\sigma_P\delta B = 4$ yields a low-purity, ${\rm Tr}(\hat{\rho}_{S_{n_P}\mid I_n}^2) = 0.27$, signal photon.  

\subsection{Singular-value decomposition of $\Psi_{S_nI_n}(\omega_S,\omega_I)$ \label{subsectionSVD}}
The channelized biphoton wave function, $\Psi_{S_nI_n}(\omega_S,\omega_I)$,  has a singular-value decomposition that we will write in the form
\begin{equation}
\Psi_{S_nI_n}(\omega_S,\omega_I) = \sum_{\ell =1}^\infty \lambda_\ell\, \phi_\ell(\omega_S)\psi_\ell(\omega_I),
\label{SVD}
\end{equation}
where $\lambda_1 \ge \lambda _2 \ge \cdots \ge \lambda_\ell \ge \cdots \ge 0$ are its singular values, and the $\{\phi_\ell(\omega_S)\}$ and $\{\psi_\ell(\omega_I)\}$ are, respectively, complete orthonormal (CON) function sets on the passbands of $H_{S_n}(\omega_S)$ and $H_{I_n}(\omega_I)$~\cite{footnote6}.  This SVD will play important roles going forward, beginning in this subsection where it is applied to Sec.~\ref{singleChannelized}'s formulas for $\Pr(S_n,I_n)$, $\Phi_n(\omega_S,\omega'_S)$, and ${\rm Tr}(\hat{\rho}^2_{S_{n_P}\mid I_n})$.

Substituting Eq.~(\ref{SVD}) into Eqs.~(\ref{PrSnIn}), (\ref{PhiNdefn}), and (\ref{singlePurity}), we can use the CON natures of the mode sets $\{\phi_\ell(\omega_S)\}$ and $\{\psi_\ell(\omega_I)\}$ to obtain their SVD forms:  
\begin{align}
\Pr(S_n,I_n) &= \sum_{\ell = 1}^\infty\sum_{\ell' = 1}^\infty\lambda_\ell\lambda_{\ell'}\int\!\frac{{\rm d}\omega_S}{2\pi}\,\phi_\ell(\omega_S)\phi^*_{\ell'}(\omega'_S) \nonumber \\[.05in]
&\,\, \times \int\!\frac{{\rm d}\omega_I}{2\pi}\,\psi_\ell(\omega_I)\psi^*_{\ell'}(\omega_I) = \sum_{\ell =  1}^\infty\lambda_\ell^2,
\end{align}
\begin{align}
\Phi_n(\omega_S,\omega'_S) &= \sum_{\ell = 1}^\infty\sum_{\ell' = 1}^\infty\lambda_\ell\lambda_{\ell'}\,\phi_\ell(\omega_S)\phi^*_{\ell'}(\omega'_S) \nonumber \\[.05in]
&\,\, \times \int\!\frac{{\rm d}\omega_I}{2\pi}\,\psi_\ell(\omega_I)\psi^*_{\ell'}(\omega_I) \\[.05in]
&= \sum_{\ell =  1}^\infty\lambda_\ell^2\, \phi_\ell(\omega_S)\phi^*_\ell(\omega'_S),
\label{PhiSVD}
\end{align}
and
\begin{align}
{\rm Tr}(\hat{\rho}_{S_{n_P}\mid I_n}^2) &= \int\!\frac{{\rm d}^2\bomega_S}{4\pi^2}\,\left|\sum_{\ell = 1}^\infty \tilde{\lambda}_\ell^2\,\phi_\ell(\omega_S)\phi^*_\ell(\omega'_S)\right|^2 \\[.05in]
&= \sum_{\ell =1}^\infty\sum_{\ell'=1}^\infty\tilde{\lambda}^2_\ell\tilde{\lambda}^2_{\ell'}\int\!\frac{{\rm d}\omega_S}{2\pi}\,\phi_\ell(\omega_S)\phi^*_{\ell'}(\omega_S)\nonumber \\[.05in]
&\,\,\times \int\!\frac{{\rm d}\omega'_S}{2\pi}\,\phi^*_\ell(\omega'_S)\phi_{\ell'}(\omega'_S)  = \sum_{\ell =1}^\infty \tilde{\lambda}^4_\ell,
\label{singlePurity2}
\end{align}
where the $\{\tilde{\lambda}_\ell\}$, given by
$\tilde{\lambda}_\ell \equiv \lambda_\ell/\sqrt{\sum_{\ell = 1}^\infty\lambda^2_\ell}$, 
are $\Psi_{S_nI_n}(\omega_S,\omega_I)$'s normalized ($\sum_{\ell= 1}^\infty \tilde{\lambda}^2_\ell = 1$) singular values.  

Because of the $\{\tilde{\lambda}_\ell\}$'s foregoing normalization, it is obvious that ${\rm Tr}(\hat{\rho}_{S_{n_P}\mid I_n}^2) \le 1$, with equality if and only if $\tilde{\lambda}_1 = 1$ and $\tilde{\lambda}_\ell = 0$ for $\ell = 2,3,\ldots$ 
In other words, ${\rm Tr}(\hat{\rho}_{S_{n_P}\mid I_n}^2) = 1$ if and only if the channelized wave function is separable,
\begin{equation}
\Psi_{S_nI_n}(\omega_S,\omega_I) = \sqrt{\Pr(S_n,I_n)}\,\phi_1(\omega_S)\psi_1(\omega_I),
\end{equation}
as claimed 
in Sec.~\ref{singleChannelized}.  

\section{Partial Bell-State Measurement \label{partial-BSM}}
This section's principal goals are to derive: (1) the dual-Sagnac source's normalized heralding probability, i.e., the probability that an ideal (unit efficiency) $n$th-channel partial BSM will herald a $\psi^\mp$ biphoton~\cite{footnote7}; (2) the dual-Sagnac source's heralding efficiency, viz., the probability that the photons sent to Alice and Bob's QRXs by the ZALM QTX pass through those QRX's $n$th DWDM$_S$ channels, given that the QTX's partial BSM heralds an $n$th-channel $\psi^\mp$ state; (3) the normalized density operator, $\hat{\rho}_{S_{1_n},S_{2_n}\mid \psi^\mp_n}$,  of the biphoton sent to Alice and Bob, given that the $\{S_{1_n},S_{2_n} \mid \psi^\mp_n\}$ event has occurred; and (4)  the purity, ${\rm Tr}(\hat{\rho}^2_{S_{1_n},S_{2_n}\mid \psi^\mp_n})$, of the biphoton sent to Alice and Bob, given that the $\{S_{1_n},S_{2_n} \mid \psi^\mp_n\}$ event has occurred.  In pursuing these goals we assume ideal equipment, except for symmetric losses that we coalesce into identical efficiencies, $0 < \eta_{\rm qtx} < 1$, assigned to Fig.~\ref{partial-BSM_fig}'s SPDs but ignored for now as they are the normalization constant for the first goal and they do not affect the others' post-selected events.  \\ 

The partial-BSM's normalized heralding probability, which we denote $\Pr(\psi^\mp_n)$, is 
obtained as follows.  Assuming ideal (unit efficiency) equipment, we have that $\Pr(\psi^\mp_n)$ is the probability that both idler photons pass through the DWDM$_I$ filter's $n$th channel \emph{and} result in one $H$ and one $V$ detection from that channel's SPDs.  Because the two idlers are equally likely to be in any of the four polarization Bell states, we get
\begin{align}
\Pr(\psi^\mp_n) &= [\Pr(I_n)]^2/4 \\[.05in]
&= \frac{1}{4}\!\left(\int\!\frac{{\rm d}\omega_S}{2\pi}\int\!\frac{{\rm d}\omega_I}{2\pi}\,|\Psi_{SI_n}(\omega_S,\omega_I)|^2\right)^2.
\label{dualHeraldProb}
\end{align} 
 Finding the other quantities of interest will require greater effort.

\subsection{Derivations of $\Pr(S_{1_n},S_{2_n}\mid \psi^\mp_n)$ and $\hat{\rho}_{S_{1_n},S_{2_n}\mid \psi^\mp_n}$}
Assuming that neither of its coherently-pumped, identical, Sagnac SPDC sources makes a multi-pair emission,  ZALM's QTX produces, with probability $(2P_0P_1)^2$, the post-selected joint state $|\psi\rangle_{S_1I_1}\otimes |\psi\rangle_{S_2I_2}$,
where
\begin{equation}
|\psi\rangle_{S_kI_k} \equiv \int\!\frac{{\rm d}\omega_S}{2\pi}\!\int\!\frac{{\rm d}\omega_I}{2\pi}\,\Psi_{SI}(\omega_S,\omega_I)|\psi^-(\omega_S,\omega_I)\rangle_{S_kI_k},
\label{SkIk}
\end{equation}
for $k=1,2$, with
\begin{eqnarray}
\lefteqn{|\psi^-(\omega_S,\omega_I)\rangle_{S_kI_k} \equiv} \nonumber \\[.05in]
&& (|\omega_S\rangle_{S_{k_H}}|\omega_I\rangle_{I_{k_V}} - 
|\omega_S\rangle_{S_{k_V}}|\omega_I\rangle_{I_{k_H}})/\sqrt{2}.
\end{eqnarray}
Note that $P_1 \ll P_0$ suffices, to first order, for ignoring a  multi-pair emission from a single Sagnac-configured SPDC, but it does \emph{not} suffice for ignoring multi-pair emissions from \emph{both} sources.  That insufficiency arises because using the first-order expressions from Eq.~(\ref{SkIk}) in $|\psi\rangle_{S_1I_1}\otimes |\psi\rangle_{S_2I_2}$ yields a second-order expression.  Thus terms involving source~1 emitting two pairs while none are emitted by source~2 and vice versa should be included to get a joint $S_1$-$S_2$ state that is correct to second order.  

We will not include the unwanted second-order terms in our analysis, because they do not send photons to \emph{both} Alice \emph{and} Bob.  Hence, when background light and dark counts at Alice and Bob's QRXs can be ignored, as we shall assume, the unwanted second-order terms cannot lead to memory loads at both receivers and we can safely proceed without including them.

Let $|\tilde{\psi}\rangle_{{\bf S}_n{\bf I}_n} \equiv |\tilde{\psi}\rangle_{S_{1_n}I_{1_n}}\otimes|\tilde{\psi}\rangle_{S_{2_n}I_{2_n}}$ be the unnormalized state that results from restricting $|\psi\rangle_{S_1I_1}\otimes |\psi\rangle_{S_2I_2}$ to the $n$th DWDM$_S$ and DWDM$_I$ channels for the signal and idler, respectively.  Unnormalized density operators for $S_{1_n},S_{2_n},\psi^\mp$ events are found by tracing over their heralding detections, giving us
\begin{widetext}
\begin{align}
\tilde{\rho}_{S_{1_n}S_{2_n}, \psi^-_n} &\equiv
\int\!\frac{{\rm d} \omega_{I_+}}{2\pi}\int\!\frac{{\rm d}\omega_{I_-}}{2\pi}\,({}_{I_{+H}}\!\langle \omega_{I_+}|\,{}_{I_{-V}}\!\langle \omega_{I_-}|)|\tilde{\psi}\rangle_{{\bf S}_n{\bf I}_n}\,{}_{{\bf S}_n{\bf I}_n}\!\langle \tilde{\psi}|(|\omega_{I_-}\rangle_{I_{-V}}\,|\omega_{I_+}\rangle_{I_{+H}})\nonumber \\[.05in]
&\,\,+ \int\!\frac{{\rm d} \omega_{I_+}}{2\pi}\int\!\frac{{\rm d}\omega_{I_-}}{2\pi}\,({}_{I_{+V}}\!\langle \omega_{I_+}|\,{}_{I_{-H}}\!\langle \omega_{I_-}|)|\tilde{\psi}\rangle_{{\bf S}_n{\bf I}_n}\,{}_{{\bf S}_n{\bf I}_n}\!\langle \tilde{\psi}|(|\omega_{I_-}\rangle_{I_{-H}}\,|\omega_{I_+}\rangle_{I_{+V}}),\label{rho-psiminus}
\end{align}
and
\begin{align}
\tilde{\rho}_{S_{1_n}S_{2_n}\, \psi^+_n} &\equiv
\int\!\frac{{\rm d} \omega_{I_+}}{2\pi}\int\!\frac{{\rm d}\omega'_{I_+}}{2\pi}\,({}_{I_{+H}}\!\langle \omega_{I_+}|\,{}_{I_{+V}}\!\langle \omega'_{I_+}|)|\tilde{\psi}\rangle_{{\bf S}_n{\bf I}_n}\,{}_{{\bf S}_n{\bf I}_n}\!\langle \tilde{\psi}|(|\omega'_{I_+}\rangle_{I_{+V}}\,|\omega_{I_+}\rangle_{I_{+H}})\nonumber \\[.05in]
&\,\,+ \int\!\frac{{\rm d} \omega_{I_-}}{2\pi}\int\!\frac{{\rm d}\omega'_{I_-}}{2\pi}\,({}_{I_{-H}}\langle \omega_{I_-}|\,{}_{I_{-V}}\!\langle \omega'_{I_-}|)|\tilde{\psi}\rangle_{{\bf S}_n{\bf I}_n}\,{}_{{\bf S}_n{\bf I}_n}\!\langle \tilde{\psi}|(|\omega'_{I_-}\rangle_{I_{-V}}\,|\omega_{I_-}\rangle_{I_{-H}}).\label{rho-psiplus}
\end{align}
Now, using the beam-splitter relations
\begin{equation}
|\omega_I\rangle_{I_{1P}} = \frac{|\omega_I\rangle_{I_{+P}} + |\omega_I\rangle_{I_{-P}}}{\sqrt{2}}, \mbox{ and }
|\omega_I\rangle_{I_{2P}} = \frac{|\omega_I\rangle_{I_{+P}} - |\omega_I\rangle_{I_{-P}}}{\sqrt{2}},\mbox{ for $P = H,V$},
\label{beamSplitters}
\end{equation}
in $|\tilde{\psi}\rangle_{{\bf S}_n{\bf I}_n}$, and their adjoints in ${}_{{\bf S}_n{\bf I}_n}\langle\tilde{\psi}|$, Appendix~\ref{AppendA} shows that
\begin{align}
\tilde{\rho}_{S_{1_n},S_{2_n},\psi^\mp_n} &= \frac{1}{8}
\int\!\frac{{\rm d}^2\bomega_S}{4\pi^2}\int\!\frac{{\rm d}^2\bomega'_S}{4\pi^2}\,K^{(c)}_{S_{1_n}S_{2_n}}(\bomega_S;\bomega'_S) |\psi^\mp(\bomega_S)\rangle_{S_1S_2}\,{}_{S_1S_2}\langle \psi^\mp(\bomega'_S)| \nonumber \\[.05in]
&\,\,+\frac{1}{8}
\int\!\frac{{\rm d}^2\bomega_S}{4\pi^2}\int\!\frac{{\rm d}^2\bomega'_S}{4\pi^2}\,K^{(e)}_{S_{1_n}S_{2_n}}(\bomega_S;\bomega'_S) |\psi^\pm(\bomega_S)\rangle_{S_1S_2}\,{}_{S_1S_2}\langle \psi^\pm(\bomega'_S)|,
\label{tilde-rhominusplus}
\end{align}
\end{widetext}
where the kernels are
\begin{align}
K^{(c)}_{S_{1_n}S_{2_n}}(\bomega_S;\bomega'_S) &\equiv \Phi_n(\omega_{S_1},\omega'_{S_1})\Phi_n(\omega_{S_2},\omega'_{S_2}) \nonumber \\[.05in]
&\,\,+ \Phi_n(\omega_{S_1},\omega'_{S_2})\Phi_n(\omega_{S_2},\omega'_{S_1}),
\label{KcDefn}
\end{align}
and
\begin{align}
K^{(e)}_{S_{1_n}S_{2_n}}(\bomega_S;\bomega'_S) &\equiv \Phi_n(\omega_{S_1},\omega'_{S_1})\Phi_n(\omega_{S_2},\omega'_{S_2}) \nonumber \\[.05in]
&\,\,-\Phi_n(\omega_{S_1},\omega'_{S_2})\Phi_n(\omega_{S_2},\omega'_{S_1}),
\label{KeDefn}
\end{align}
with $\bomega_S$ and ${\rm d}^2\bomega_S$ here and henceforth denoting $(\omega_{S_1},\omega_{S_2})$ and ${\rm d}\omega_{S_1}{\rm d}\omega_{S_2}$.

In Eqs.~(\ref{KcDefn}) and (\ref{KeDefn}), the superscripts $(c)$ and $(e)$ denote ``correct'' and ``error'', respectively, because the $K^{(c)}_{S_{1_n}S_{2_n}}(\bomega_S;\bomega'_S)$ kernel is associated with the term for which a $\psi^\mp_n$ herald results in a $\psi^\mp_n$ biphoton being sent to Alice and Bob, while the $K^{(e)}_{S_{1_n}S_{2_n}}(\bomega_S;\bomega'_S)$ kernel is associated with the term for which a $\psi^\mp_n$ herald results in a $\psi^\pm_n$ biphoton being sent to Alice and Bob.

Tracing $\tilde{\rho}_{S_{1_n},S_{2_n}, \psi^\mp_n}$ in the Bell basis we obtain the normalization constant,
\begin{eqnarray}
\lefteqn{\hspace{-.15in}\Pr(S_{1_n},S_{2_n}, \psi^\mp_n) =} \nonumber \\[.05in]
&&\hspace{-.25in}\frac{1}{8}\int\!\frac{{\rm d}^2\bomega_S}{4\pi^2}\!\left[K^{(c)}_{S_{1_n}S_{2_n}}(\bomega_S;\bomega_S)+K^{(e)}_{S_{1_n}S_{2_n}}(\bomega_S;\bomega_S)\right] \\[.05in]
&& = \frac{1}{4}\!\left(\int\!\frac{{\rm d}\omega_S}{2\pi}\,\Phi_n(\omega_S,\omega_S)\right)^2 \\[.05in]
&& = \frac{1}{4}\!\left(\int\!\frac{{\rm d}\omega_S}{2\pi}\int\!\frac{{\rm d}\omega_I}{2\pi}\,|\Psi_{S_nI_n}(\omega_S,\omega_I)|^2\right)^2 \\[.05in]
&& = [\Pr(S_n,I_n)]^2/4,
\label{normalization2}
\end{eqnarray}
so that \vspace*{-.1in}
\begin{equation}
\hat{\rho}_{S_{1_n},S_{2_n}\mid\psi^\mp_n} = \frac{\tilde{\rho}_{S_{1_n}S_{2_n},\psi^\mp_n}}{[\Pr(S_n,I_n)]^2/4}.
\end{equation}

Equation~(\ref{normalization2}) makes perfect sense.  Given that the $I_+$ and $I_-$ photons pass through their DWDM$_I$ filters' $n$th-channel passbands, the joint state of the $I_{+n}$ and $I_{-n}$ photons is equally likely to be any of the four polarization Bell states.  Thus, assuming they have passed through those passbands, the  probability of their resulting in a $\psi^\mp_n$ herald is 1/4.  Meanwhile, assuming the ZALM QTX's Sagnac-configured SPDCs each produce a signal-idler biphoton, the probability that both signals and both idlers pass through their DWDM filter's $n$th-channel passbands is $[\Pr(S_n,I_n)]^2$.  Combining these two results we get Eq.~(\ref{normalization2}).  We now see that the partial-BSM's heralding efficiency satisfies
\begin{equation}
\Pr(S_{1_n},S_{2_n}\mid \psi^\mp_n) = \frac{\Pr(S_{1_n},S_{2_n},\psi^\mp_n)}{\Pr(\psi^\mp_n)} = 
\left[\frac{\Pr(S_n,I_n)}{\Pr(I_n)}\right]^2.
\label{dualEfficiency}
\end{equation}

Our next goal is to find the purity of the heralded biphoton state sent to Alice and Bob, ${\rm Tr}(\hat{\rho}^2_{S_{1_n},S_{2_n}\mid\psi^\mp_n})$.  En route to that goal it will be fruitful to take a detour, in the next subsection, to express $\hat{\rho}_{S_{1_n},S_{2_n}\mid\psi^\mp_n}$ in terms of the single Sagnac source's SVD.

\subsection{Single-Sagnac SVD representation of $\hat{\rho}_{S_{1_n},S_{2_n}\mid\psi^\mp_n}$\label{BSMsvd}}
Substituting Eq.~(\ref{PhiSVD}) into Eqs.~(\ref{KcDefn}) and (\ref{KeDefn}) gives us
\begin{eqnarray}
\lefteqn{K^{(c)}_{S_{1_n}S_{2_n}}(\bomega_S;\bomega'_S) = \sum_{\ell =1}^\infty\sum_{\ell'=1}^\infty \lambda^2_\ell \lambda_{\ell'}^2} \nonumber \\[.05in]
&\times& [\phi_\ell(\omega_{S_1})\phi^*_\ell(\omega'_{S_1})\phi_{\ell'}(\omega_{S_2})\phi^*_{\ell'}(\omega'_{S_2}) \nonumber \\[.05in]
&&\,\, +\, \phi_\ell(\omega_{S_1})\phi^*_\ell(\omega'_{S_2})\phi_{\ell'}(\omega_{S_2})\phi^*_{\ell'}(\omega'_{S_1})],
\label{KcSVD} 
\end{eqnarray}
and
\begin{eqnarray}
\lefteqn{K^{(e)}_{S_{1_n}S_{2_n}}(\bomega_S;\bomega'_S) = \sum_{\ell =1}^\infty\sum_{\ell'=1}^\infty \lambda^2_\ell \lambda_{\ell'}^2} \nonumber \\[.05in]
&\times& [\phi_\ell(\omega_{S_1})\phi^*_\ell(\omega'_{S_1})\phi_{\ell'}(\omega_{S_2})\phi^*_{\ell'}(\omega'_{S_2}) \nonumber \\[.05in]
&&\,\, -\, \phi_\ell(\omega_{S_1})\phi^*_\ell(\omega'_{S_2})\phi_{\ell'}(\omega_{S_2})\phi^*_{\ell'}(\omega'_{S_1})].
\label{KeSVD} 
\end{eqnarray}
It is immediately evident from these results that both $K^{(c)}_{S_{1_n}S_{2_n}}(\bomega_S;\bomega'_S)$ and $K^{(e)}_{S_{1_n}S_{2_n}}(\bomega_S;\bomega'_S)$ are Hermitian kernels, i.e.,
\begin{equation}
K^{(p)}_{S_{1_n}S_{2_n}}(\bomega'_S;\bomega_S) = K^{(p)*}_{S_{1_n}S_{2_n}}(\bomega_S;\bomega'_S), \mbox{ for $p = c,e$}.
\end{equation}
As a result, they have eigenvalue-eigenfunction decompositions,
\begin{equation}
K^{(c)}_{S_{1_n}S_{2_n}}(\bomega_S;\bomega'_S) = \sum_{m=1}^\infty\mu_m\,\xi_m(\bomega_S)\xi^*_m(\bomega'_S),
\label{eigenXi}
\end{equation} 
and 
\begin{equation}
K^{(e)}_{S_{1_n}S_{2_n}}(\bomega_S;\bomega'_S) = \sum_{m=1}^\infty\nu_m\,\zeta_m(\bomega_S)\zeta^*_m(\bomega'_S),
\label{eigenZeta}
\end{equation}
with real eigenvalues, $\{\mu_m\}$ and $\{\nu_m\}$, and orthonormal eigenfunctions, $\{\xi_m(\bomega_S)\}$ and $\{\zeta_m(\bomega_S)\}$, for $\omega_{S_1}$ and $\omega_{S_2}$ in the DWDM$_S$ filter's $n$th-channel passband~\cite{footnote8}.  If eigenfunctions with zero eigenvalues are included, the $\{\xi_m(\bomega_S)\}$ and $\{\zeta_m(\bomega_S)\}$ can be taken to be CON function sets on their domains.  

To find the preceding eigenvalues and eigenfunctions we need to solve the following Fredholm integral equations,
\begin{equation}
\int\!\frac{{\rm d}^2\bomega'_S}{4\pi^2}\,K^{(c)}_{S_{1_n}S_{2_n}}(\bomega_S;\bomega'_S)\xi_m(\bomega'_S) = \mu_m\, \xi_m(\bomega_S),
\label{FredholmC}
\end{equation}
and
\begin{equation}
\int\!\frac{{\rm d}^2\bomega'_S}{4\pi^2}\,K^{(e)}_{S_{1_n}S_{2_n}}(\bomega_S;\bomega'_S)\zeta_m(\bomega'_S) = \nu_m\, \zeta_m(\bomega_S),
\label{FredholmE}
\end{equation}
for $\omega_{S_1},\omega_{S_2}$ in the DWDM$_S$ filter's $n$th-channel passband.  Ordinarily, solving these equations would be a formidable task.  However, because of Eqs.~(\ref{KcSVD}) and (\ref{KeSVD}), we note that  numerical solutions to Eqs.~(\ref{FredholmC}) and (\ref{FredholmE}) can be readily obtained via available numerical techniques for finding the SVD of an arbitrary $\Psi_{S_nI_n}(\omega_S,\omega_I)$.  Let us see how this comes about.

We start by changing the scalar ($m$) indices in Eqs.~(\ref{FredholmC}) and (\ref{FredholmE}) to vector indices ${\bf m} \equiv (m_1,m_2)$ and choosing the $\{\xi_{\bf  m}(\bomega_S)\}$ and $\{\zeta_{\bf  m}(\bomega_S)\}$ to be 
\begin{widetext}
\begin{equation}
\xi_{\bf m}(\bomega_S) = \left\{\begin{array}{cl}
\phi_m(\omega_{S_1})\phi_m(\omega_{S_2}), & \mbox{for $m_1=m_2 = m = 1,2,\ldots,$}\\[.05in]
[\phi_{m_1}(\omega_{S_1})\phi_{m_2}(\omega_{S_2})+ \phi_{m_1}(\omega_{S_2})\phi_{m_2}(\omega_{S_1})]/\sqrt{2}, & \mbox{for $m_1>m_2 = 1,2,\ldots, $}
\end{array}\right.
\label{xiBFm}
\end{equation}
and
\begin{equation}
\zeta_{\bf m} (\bomega_S) = [\phi_{m_1}(\omega_{S_1})\phi_{m_2}(\omega_{S_2})- \phi_{m_1}(\omega_{S_2})\phi_{m_2}(\omega_{S_1})]/\sqrt{2},  \mbox{ for $m_1>m_2 = 1,2,\ldots $}
\label{zetaBFm}
\end{equation}
\end{widetext}
Because $\{\phi_{m_1}(\omega_{S_1})\phi_{m_2}(\omega_{S_2}) : m_1,m_2 = 1,2,\ldots\}$ are CON on $\omega_{S_1},\omega_{S_2}$ in the DWDM$_S$ filter's $n$th-channel passband, it follows that the $\{\xi_{\bf m}(\bomega_S)\}$ and $\{\zeta_{\bf m}(\bomega_S)\}$ are, collectively, a CON function set on the DWDM$_S$ filter's $n$th-channel passband.  Furthermore, it is easily verified, see Appendix~\ref{AppendA}, that the $\{\xi_{\bf m}(\bomega_S)\}$ and $\{\zeta_{\bf m}(\bomega_S)\}$ are, respectively, solutions to Eqs.~(\ref{FredholmC}) and (\ref{FredholmE}) with eigenvalues
\begin{equation}
\mu_{\bf m} = \left\{\begin{array}{cl}
2\lambda^4_m, & \mbox{for $m_1 = m_2= m =1,2,\ldots, $}\\[.05in]
2\lambda^2_{m_1}\lambda^2_{m_2}, & \mbox{for $m_1>m_2 = 1,2,\ldots ,$}
\end{array}\right.
\label{muBFm}
\end{equation}
and 
\begin{equation}
\nu_{\bf m} = 
2\lambda^2_{m_1}\lambda^2_{m_2}, \mbox{ for $m_1>m_2 = 1,2,\ldots $}
\label{nuBFm}
\end{equation}
Finally, because the $\{\mu_{\bf m}\}$ and $\{\nu_{\bf m}\}$ are all non-negative, checking for consistency with Eq.~(\ref{normalization2}) viz., verifying that~\cite{footnote9}
\begin{align}
\frac{1}{8}&\int\!\frac{{\rm d}^2\bomega_S}{4\pi^2}\!\left[K^{(c)}_{S_{1_n}S_{2_n}}(\bomega_S;\bomega_S)+K^{(e)}_{S_{1_n}S_{2_n}}(\bomega_S;\bomega_S)\right] \nonumber \\[.05in]
&= \sum_{\bf m}\frac{\mu_{\bf m}+\nu_{\bf m}}{8} = [\Pr(S_n,I_n)]^2/4,
\end{align}
will ensure that Eqs.~(\ref{xiBFm}) and (\ref{zetaBFm}) include \emph{all} the eigenfunctions with non-zero eigenvalues.  Using Eqs.~(\ref{muBFm}) and (\ref{nuBFm}) we have that
\begin{align}
\sum_{\bf m}&\frac{\mu_{\bf m}+\nu_{\bf m}}{8} = \frac{1}{4}\sum_{m=1}^\infty \lambda_m^4 +
\frac{1}{2}\sum_{m_1=2}^\infty\sum_{m_2=1}^{m_1-1}\lambda^2_{m_1}\lambda^2_{m_2} \nonumber \\[.05in]
&= \frac{1}{4}\!\left(\sum_{m=1}^\infty\lambda^2_m\right)^2 = [\Pr(S_n,I_n)]^2/4, 
\end{align}
and the verification is complete.

We can now rewrite the state  of the heralded biphoton sent to Alice and Bob, $\hat{\rho}_{S_{1_n},S_{2_n}\mid \psi^\mp_n}$, as
\begin{align}
\hat{\rho}&_{S_{1_n},S_{2_n}\mid \psi^\mp_n} = \sum_{\bf m}\tilde{\mu}_{\bf m}\int\!\frac{{\rm d}^2\bomega_S}{4\pi^2}\int\!\frac{{\rm d}^2\bomega'_S}{4\pi^2}\,\xi_{\bf m}(\bomega_S)\nonumber \\[.05in]
&\,\,\,\,\times |\psi^\mp(\bomega_S)\rangle_{S_1S_2}\,{}_{S_1S_2}\langle \psi^\mp(\bomega'_S)|
\xi^*_{\bf m}(\bomega'_S) \nonumber \\[.05in]
& +  \sum_{\bf m}\tilde{\nu}_{\bf m}\int\!\frac{{\rm d}^2\bomega_S}{4\pi^2}\int\!\frac{{\rm d}^2\bomega'_S}{4\pi^2}\,\zeta_{\bf m}(\bomega_S)\nonumber \\[.05in]
&\,\,\,\,\times |\psi^\pm(\bomega_S)\rangle_{S_1S_2}\,{}_{S_1S_2}\langle \psi^\pm(\bomega'_S)|
\zeta^*_{\bf m}(\bomega'_S),
\label{heraldedDensOp}
\end{align}
where 
\begin{equation}
\tilde{\mu}_{\bf  m} \equiv \frac{\mu_{\bf m}}{\sum_{\bf m}(\mu_{\bf m} +\nu_{\bf m})},
\end{equation}
and 
\begin{equation}
\tilde{\nu}_{\bf  m} \equiv \frac{\nu_{\bf m}}{\sum_{\bf m}(\mu_{\bf m} +\nu_{\bf m})},
\end{equation}
are its normalized eigenvalues.  We note that Eqs.~(\ref{xiBFm})--(\ref{nuBFm}) allow us to instantiate Eq.~(\ref{heraldedDensOp}) for an arbitrary single-source biphoton wave function using straightforward numerical evaluation of Eq.~(\ref{SVD})'s single-source temporal modes and singular values.

\subsection{Purity, fidelity, and error probability}
Subsection~\ref{BSMsvd}'s results allow us to make short work of finding the partial-BSM's purity, fidelity, and error probability.  For its purity we start from
\begin{align}
\hat{\rho}&^2_{S_{1_n},S_{2_n}\mid \psi^\mp_n} = \sum_{\bf m}\tilde{\mu}^2_{\bf m}\int\!\frac{{\rm d}^2\bomega_S}{4\pi^2}\int\!\frac{{\rm d}^2\bomega'_S}{4\pi^2}\,\xi_{\bf m}(\bomega_S)\nonumber \\[.05in]
&\,\,\,\,\times |\psi^\mp(\bomega_S)\rangle_{S_1S_2}\,{}_{S_1S_2}\langle \psi^\mp(\bomega'_S)|
\xi^*_{\bf m}(\bomega'_S) \nonumber \\[.05in]
& +  \sum_{\bf m}\tilde{\nu^2}_{\bf m}\int\!\frac{{\rm d}^2\bomega_S}{4\pi^2}\int\!\frac{{\rm d}^2\bomega'_S}{4\pi^2}\,\zeta_{\bf m}(\bomega_S)\nonumber \\[.05in]
&\,\,\,\,\times |\psi^\pm(\bomega_S)\rangle_{S_1S_2}\,{}_{S_1S_2}\!\langle \psi^\pm(\bomega'_S)|
\zeta^*_{\bf m}(\bomega'_S),
\end{align}
which immediately gives us
\begin{align}
{\rm Tr}(\hat{\rho}&^2_{S_{1_n},S_{2_n}\mid\psi^\mp_n}) = \sum_{\bf m}(\tilde{\mu}^2_{\bf m} + \tilde{\nu}^2_{\bf m})
\\[.05in]
&= \sum_{\bf m}\tilde{\lambda}^8_{\bf m} + 2\sum_{m_1=2}^\infty\sum_{m_2 = 1}^{m_1-1}\tilde{\lambda}^4_{m_1}\tilde{\lambda}^4_{m_2} \\[.05in]
&= \left(\sum_{m=1}^\infty \tilde{\lambda}^4_m\right)^2 = \left[{\rm Tr}(\hat{\rho}^2_{S_{n_P}\mid I_n})\right]^2,
\label{SVDbsmPurity}
\end{align}
where the last equality shows, as could easily be expected, that the purity of the ZALM transmission to Alice and Bob is proportional to the product of its two Sagnac sources' purities.  Note that although the Sagnac source's SVD was a valuable tool for deriving Eq.~(\ref{SVDbsmPurity}), we can evaluate ${\rm Tr}[(\hat{\rho}_{S_{1_n},S_{2_n}\mid\psi^\mp_n})^2]$ directly from the channelized biphoton wave function by squaring the single-source result from Eq.~(\ref{singlePurity}).    

Turning now to the partial-BSM's fidelity, i.e., the probability, $\Pr(c_n)$, that a $\psi^\mp_n$ herald will result in a $\psi^\mp_n$ biphoton being sent to Alice and Bob, we have that 
\begin{align}
\Pr(c_n) &= \int\!\frac{{\rm d}^2\bomega_S}{4\pi^2}\,{}_{S_{1_n}S_{2_n}}\langle \psi^\mp(\bomega_S)|\nonumber 
\\[.05in]
&\hat{\rho}_{S_{1_n},S_{2_n}\mid \psi^\mp_n}|\psi^\mp_n(\bomega_S)\rangle_{S_{1_n}S_{2_n}}\\[.05in]
&= \sum_{\bf m}\tilde{\mu}_{\bf m} = \sum_{m=1}^\infty\tilde{\lambda}^4_m + \sum_{m_1=2}^\infty\sum_{m_2 = 1}^{m_1-1}\tilde{\lambda}^2_{m_1}\tilde{\lambda}^2_{m_2} \\[.05in]
&= \frac{1}{2}\!\left(1+\sum_{m=1}^\infty\tilde{\lambda}^4_m\right),
\label{PrCn}
\end{align}
where the last equality uses $\sum_{m=1}^\infty \tilde{\lambda}^2_m = 1$.  Here too the single-source SVD has been a useful tool, but $\Pr(c_n)$ can be found directly from the channelized biphoton wave function because
\begin{equation}
\Pr(c_n) = \frac{1 + \sqrt{{\rm Tr}(\hat{\rho}^2_{S_{1_n},S_{2_n}\mid\psi^\mp_n})}}{2}.
\end{equation}

To complete this subsection we compute the error probability, $\Pr(e_n)$.  Because $\hat{\rho}_{S_{1_n}S_{2_n}\mid \psi^\mp_n}$ is diagonal in the $\psi^\mp_n$ and $\psi^\pm_n$ Bell states, $\Pr(e_n)$ is the probability that a $\psi^\mp_n$ herald leads to a $\psi^\pm_n$ biphoton being sent to Alice and Bob.  As we must have $\Pr(c_n) + \Pr(e_n) = 1$, we get
\begin{equation}
\Pr(e_n) = \frac{1 - \sqrt{{\rm Tr}(\hat{\rho}^2_{S_{1_n},S_{2_n}\mid\psi^\mp_n})}}{2},
\label{SVDbsmErrorProb}
\end{equation}
which can be found directly from the channelized biphoton wave function without performing an SVD.  

\subsection{Partial-BSM performance for the all-Gaussian special case}
\begin{figure}[hbt]
    \centering
    \includegraphics[width=0.45\textwidth]{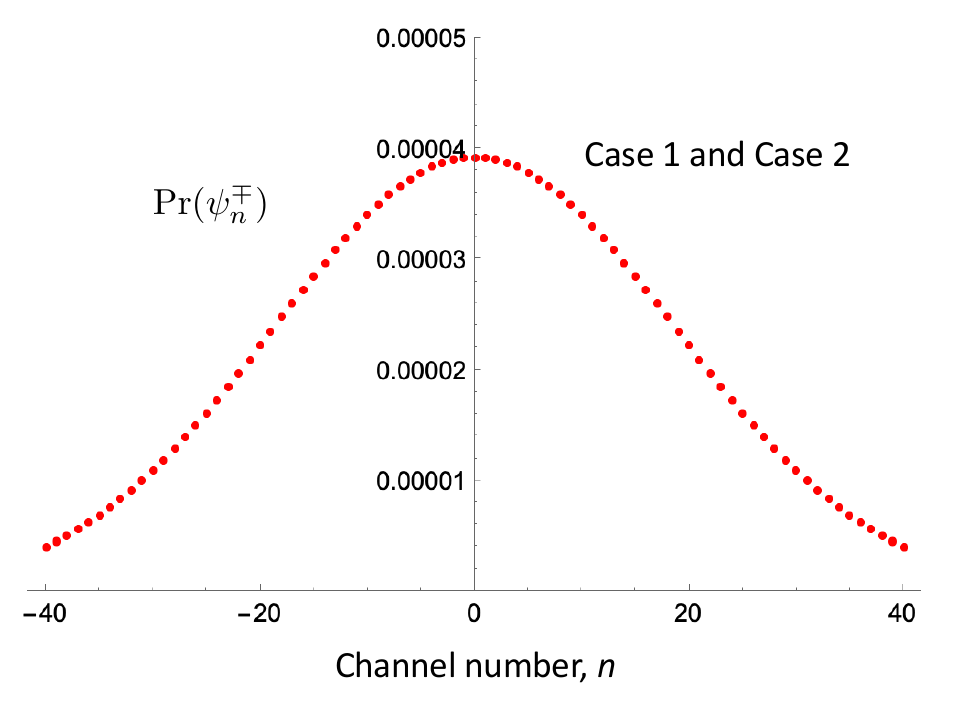}
    \caption{Partial-BSM normalized heralding probability, $\Pr(\psi^\mp_n)$ from Eq.~(\ref{dualHeraldProb}), versus channel number assuming Eq.~(\ref{allGauss})'s all-Gaussian biphoton wave function with brickwall filtering and  parameter values from Table~\ref{singleParams}.  \label{BSMherald_fig}}    
\end{figure}

\begin{figure}[hbt]
    \centering
    \includegraphics[width=0.45\textwidth]{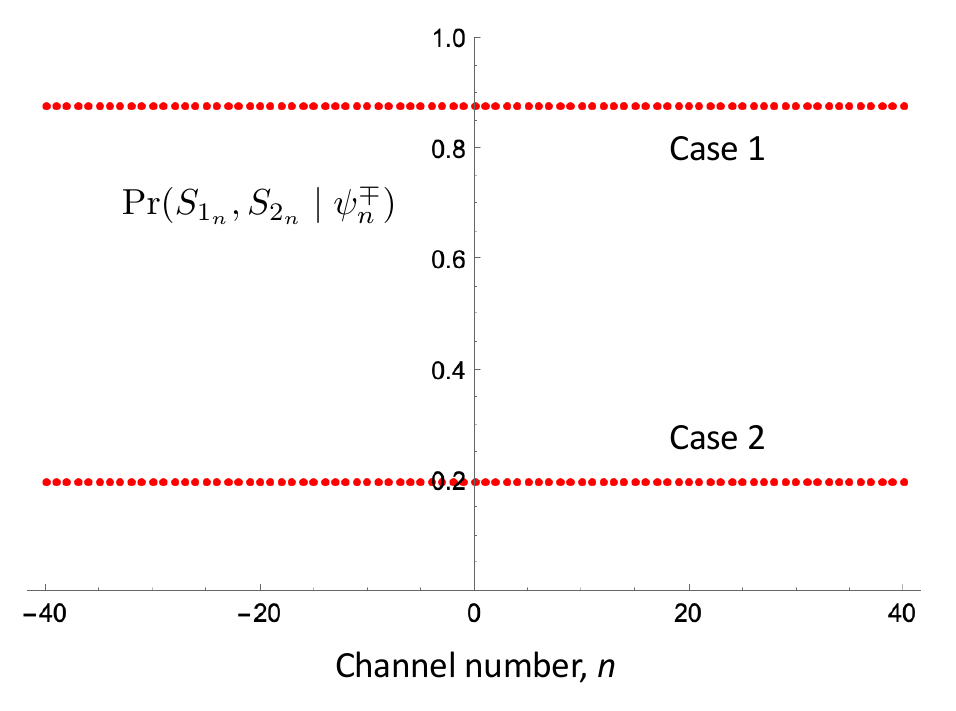}
    \caption{Partial-BSM heralding efficiency, $\Pr(S_{1_n},S_{2_n}\mid \psi^\mp_n)$ from Eq.~(\ref{dualEfficiency}), versus channel number assuming Eq.~(\ref{allGauss})'s all-Gaussian biphoton wave function with brickwall filtering and parameter values from Table~\ref{singleParams}.  \label{BSMefficiency_fig}}    
\end{figure}

\begin{figure}[hbt]
    \centering
    \includegraphics[width=0.45\textwidth]{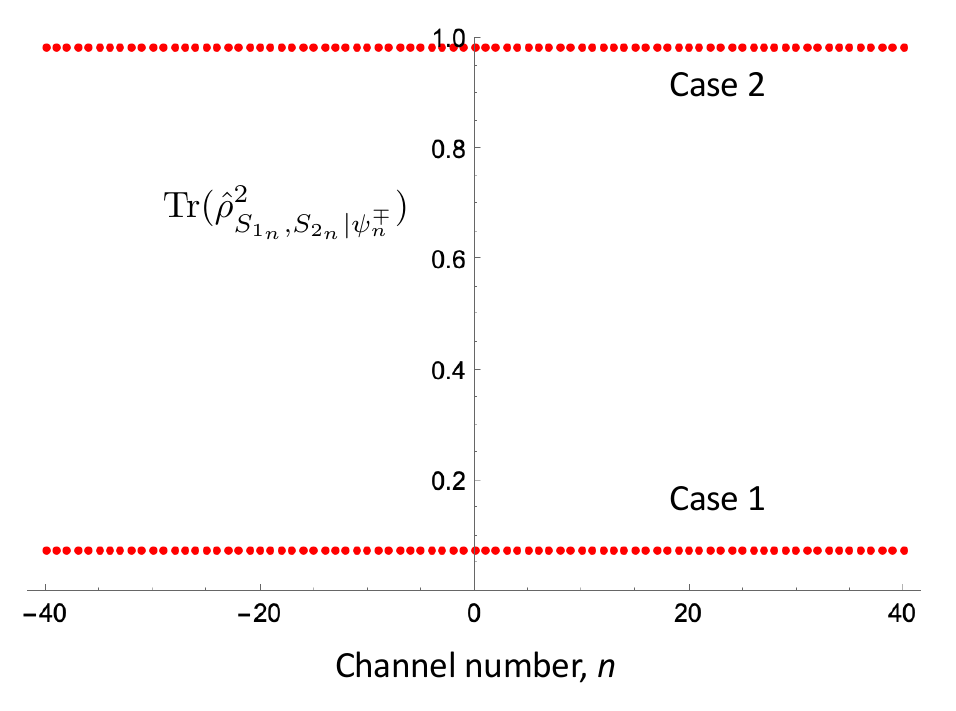}
    \caption{Partial-BSM purity, ${\rm Tr}(\hat{\rho}^2_{S_{1_n},S_{2_n}\mid\psi^\mp_n})$ from Eq.~(\ref{SVDbsmPurity}), versus channel number assuming Eq.~(\ref{allGauss})'s all-Gaussian biphoton wave function with brickwall filtering and parameter values from Table~\ref{singleParams}.  \label{BSMpurity_fig}}    
\end{figure}

\begin{figure}[hbt]
    \centering
    \includegraphics[width=0.45\textwidth]{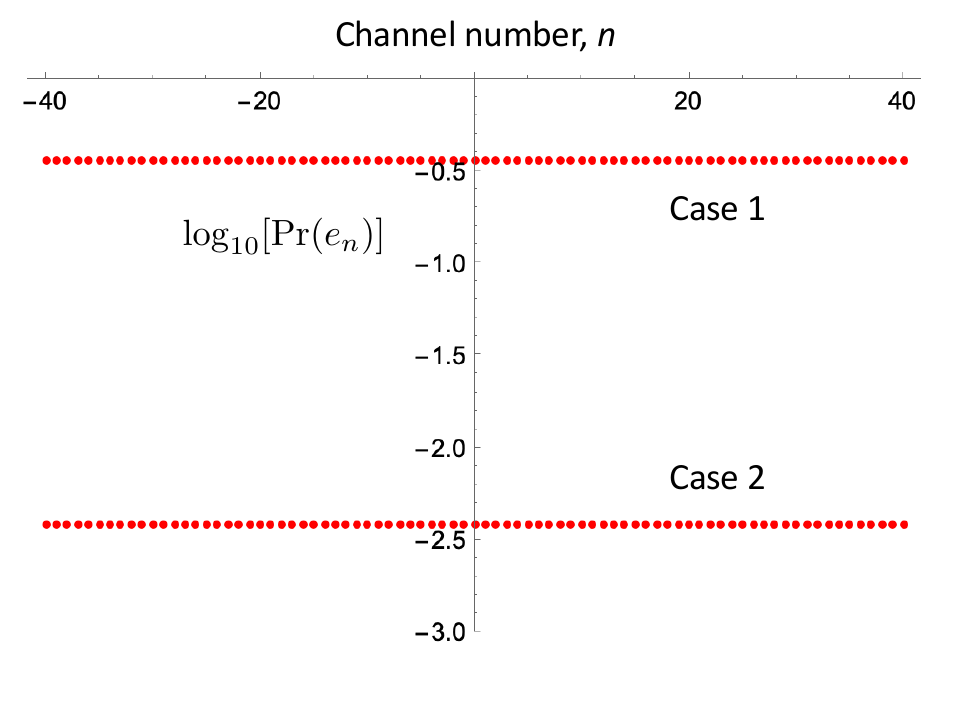}
    \caption{Partial-BSM error probability, $\Pr(e_n)$ from Eq.~(\ref{SVDbsmErrorProb}), versus channel number assuming Eq.~(\ref{allGauss})'s all-Gaussian biphoton wave function with brickwall filtering and parameter values from Table~\ref{singleParams}.  \label{BSMerrorprob_fig}}   
\end{figure}

Figures~\ref{BSMherald_fig}--\ref{BSMerrorprob_fig} plot, versus channel number, the $n$th-channel partial-BSM's normalized heralding probability, $\Pr(\psi^\mp_n)$, its  heralding efficiency, $\Pr(S_{1_n},S_{2_n}\mid \psi^\mp_n)$, its purity, ${\rm Tr}(\hat{\rho}^2_{S_{1_n},S_{2_n}\mid\psi^\mp_n})$, and its error probability, $\Pr(e_n)$, which equals $1-\Pr(c_n)$ where $\Pr(c_n)$ is the partial-BSM's fidelity. These figures assume the all-Gaussian biphoton wave function from Eq.~(\ref{allGauss}) and brickwall filtering with parameter values from Table~\ref{singleParams}.  

The behaviors seen in Figs.~\ref{BSMherald_fig}--\ref{BSMpurity_fig} are readily understood from what we saw for the single Sagnac source.  In particular, like their single-source counterparts, the partial-BSM's normalized heralding probability and its heralding efficiency are respectively, insensitive to the pump-pulse duration and the phase-matching bandwidth.  They also exhibit the tradeoff between heralding probability and heralding efficiency, albeit exacerbated by their requiring two photons to pass through DWDM filters instead of just one.  Comparing Fig.~\ref{BSMpurity_fig} with its single-source counterpart from Fig.~\ref{singlePurity_fig}, we see that Case~2 has lost very little purity in going from the single source's 99.2\% signal-photon purity to the partial-BSM's 98.5\%-purity heralded biphoton.   In Case~1, however, squaring has a more dramatic effect, with the partial-BSM's 7.50\% purity being well below the 27.4\% value for the single source's signal-photon purity.  

Error probability is a new metric that arises because the partial-BSM's $\psi^\mp_n$ herald may result in a $\psi^\pm_n$ state being sent to Alice and Bob.  Figure~\ref{BSMerrorprob_fig} shows that Case~2 does well on error probability, $\Pr(e_n) = 3.87\times 10^{-3}$, because of its high purity, while Case~1 suffers with $\Pr(e_n) =  0.363$ across all 81 channels, owing to its low purity.

\section{Entanglement Loading of Duan-Kimble Quantum Memories\label{memory}}
We now move on to our final major task, deriving the density operators, $\hat{\rho}^{({\bf s})}_{M_1,M_2\mid \psi^\mp_n}$ for ${\bf s} \equiv (s_1,s_2) = (\pm,\pm)$, of Alice and Bob's intra-cavity color-center quantum memories, given they were illuminated by an $n$th-channel heralded biphoton from the ZALM QTX and Fig.~\ref{DuanKimble}'s SPD$_{1_{s_1}}$ and SPD$_{2_{s_2}}$ detected photons after reflection from the cavities and interference at the 50--50 beam splitters.   

In our derivation we shall ignore all nonidealities in free-space propagation from the ZALM QTX to Alice and Bob's QRXs except for an identical transmissivity, $0 < \eta_{\rm prop} < 1$, associated with each propagation path.  Likewise, we shall ignore all nonidealities in Alice and Bob's mode converters, i.e., their frequency converters and bandwidth compressors, except for symmetrically-distributed losses that we combine into an efficiency, $0 < \eta_{\rm mc} < 1$.  The modeling situation is slightly different, however, for Fig.~\ref{DuanKimble}'s quantum memories, where we choose the SiV color center as the prototypical quantum memory and use the push-pull protocol's state-dependent reflectivities from our companion paper~\cite{Raymer2024}, which explicitly include leakage from the memory cavity's HR mirror and spontaneous emission from the SiV color center.  The rest of the Duan-Kimble memories' properties will be assumed ideal, except for symmetrically-distributed exo-cavity losses that we combine into identical efficiency factors, $0 < \eta_{\rm mem} < 1$.  The composite receiver efficiency, $\eta_{\rm qrx}\equiv \eta_{\rm mc}\eta_{\rm mem}$, is then assigned to the SPDs in Alice and Bob's QRXs.  Note that neither $\eta_{\rm prop}$ nor $\eta_{\rm qrx}$ appear in this section's density operator, as it applies to a post-selected event.  That said, $\eta_{\rm qtx}$, $\eta_{\rm prop}$, and $\eta_{\rm qrx}$ will show up in Sec.~\ref{Discuss}, where we discuss ZALM's entanglement-distribution rate and its tradeoff with inter-channel interference.

\subsection{Mode-converted density operator, $\hat{\rho}_{\tilde{S}_{1_n},\tilde{S}_{2_n}\mid \psi^\mp_n}$}

For a ZALM QTX that uses PPLN SPDCs, the $\psi^\mp_n$-heralded biphoton sent to Alice and Bob has signal-photon components with $\omega_{S_0}/2\pi - n\Delta B \sim 200$\,THz   center frequencies.  After they both successfully pass through Alice and Bob's DWDM$_S$ filters' $n$th-channel passbands, they have bandwidths $\delta B\sim 25$\,GHz in our Table~\ref{singleParams}'s Cases~1 and 2.  In contrast, Alice and Bob's SiV quantum memories have a center frequency $\tilde{\omega}_{S_0}/2\pi \sim 400$\,THz and bandwidth $\delta\tilde{B} \sim 600\,$MHz, quantifying the task to be performed by the QRX's mode converters.  To derive the mode-converted density operator, $\hat{\rho}_{\tilde{S}_{1_n},\tilde{S}_{2_n}\mid \psi^\mp_n}$, we make use of results from Sec.~\ref{partial-BSM}, which indicate that it suffices for us to derive the mode-converted form of $\Phi_n(\omega_{S_1},\omega_{S_2})$ and then use Eqs.~(\ref{tilde-rhominusplus})--(\ref{KeDefn}) to get the mode-converted density operator.  It is important to note that the ZALM QRX's mode converter may need to do more than just frequency conversion and bandwidth compression, i.e., the frequency-converted, bandwidth-compressed biphoton wave function may require shaping to optimize memory loading.  For now we will ignore that possibility, but we will comment on it later.

Ideal frequency conversion from center frequency $\omega_{S_0}-2\pi n\Delta B$ to center frequency $\tilde{\omega}_{S_0}$ changes $\Phi_n(\omega_{S_1},\omega_{S_2})$ to $\Phi_n(\omega_{S_1} +\delta\omega_S ,\omega_{S_2} + \delta\omega_S)$, where $\delta\omega_S \equiv \tilde{\omega}_{S_0}-\omega_{S_0}+2\pi n\Delta B$.  After bandwidth compression by $\beta \equiv \delta B/\delta\tilde{B}$, we get the following $\tilde{\Phi}_n(\omega_{S_1},\omega_{S_2})$ to use in lieu of $\Phi_n(\omega_{S_1},\omega_{S_2})$ in obtaining $\hat{\rho}_{\tilde{S}_{1_n},\tilde{S}_{2_n}\mid \psi^\mp_n}$ from Eqs.~(\ref{tilde-rhominusplus})--(\ref{KeDefn}),
\begin{equation}
\tilde{\Phi}_n(\omega_{S_1},\omega_{S_2}) = 
\beta^2\Phi_n[\beta(\omega_{S_1}+\delta\omega_S),\beta(\omega_{S_2}+\delta\omega_S)].
\end{equation}
Note that ideal mode conversion transforms $K^{(c)}_{S_{1_n}S_{2_n}}(\bomega_S;\bomega'_S)$ and $K^{(e)}_{S_{1_n}S_{2_n}}(\bomega_S;\bomega'_S)$'s eigenfunctions from $\xi_{\bf m}(\bomega_S)$ to 
$\tilde{\xi}_{\bf m}(\bomega_S)$ and from $\zeta_{\bf m}(\bomega_S)$ to 
$\tilde{\zeta}_{\bf m}(\bomega_S)$ as follows:
\begin{equation}
\tilde{\xi}_{\bf m}(\bomega_S) = \beta^2\xi_{\bf m}[\beta(\bomega_S + \dbomega_S)],
\end{equation}
and
\begin{equation}
\tilde{\zeta}_{\bf m}(\bomega_S) = \beta^2\zeta_{\bf m}[\beta(\bomega_S + \dbomega_S)],
\end{equation}
where $\dbomega_S \equiv (\delta\omega_S,\delta\omega_S)$.  
It does \emph{not}, however, change those kernel's normalized eigenvalues, $\{\tilde{\mu}_{\bf m}\}$ and $\{\tilde{\nu}_{\bf m}\}$.    Thus, as expected for ideal frequency conversion and bandwidth compression, the state's purity, fidelity, and error probability are unaffected and, as exemplified in Figs.~\ref{BSMpurity_fig} and \ref{BSMerrorprob_fig}, independent of channel number.  

\subsection{Push-pull Duan-Kimble entanglement loading}

Our approach to determining the density operator, $\hat{\rho}^{({\bf s})}_{M_1,M_2\mid \psi^\mp_n}$, for Alice ($M_1$) and Bob's ($M_2$) loaded quantum memories is straightforward.  We have that $\hat{\rho}_{\tilde{S}_{1_n},\tilde{S}_{2_n}\mid \psi^\mp_n}$'s eigenkets are
\begin{align}
|\tilde{\xi}^\mp_{\bf m}\rangle_{\tilde{S}_1\tilde{S}_2} \equiv \int\!\frac{{\rm d}^2\bomega_S}{4\pi^2}\,\tilde{\xi}_{\bf m}(\bomega_S)|\psi^\mp(\bomega_S)\rangle_{\tilde{S}_1\tilde{S}_2}
\label{tildeXiState}
\end{align}
and
\begin{align}
|\tilde{\zeta}^\mp_{\bf m}\rangle_{\tilde{S}_1\tilde{S}_2} \equiv \int\!\frac{{\rm d}^2\bomega_S}{4\pi^2}\,\tilde{\zeta}_{\bf m}(\bomega_S)|\psi^\pm(\bomega_S)\rangle_{\tilde{S}_1\tilde{S_2}},
\label{tildeZetaState}
\end{align}
where dependence on the channel number $n$ is implicit in $\tilde{\xi}_{\bf m}(\bomega_S)$ and $\tilde{\zeta}_{\bf m}(\bomega_S)$. From $|\tilde{\xi}^\mp_{\bf m}\rangle_{\tilde{S}_1\tilde{S}_2}$ and $|\tilde{\zeta}^\mp_{\bf m}\rangle_{\tilde{S}_1\tilde{S}_2}$ we will determine the unnormalized density operators, $\tilde{\rho}^{(\bf s)}_{M_1M_2\mid \tilde{\xi}^\mp_{\bf m}}$, and $\tilde{\rho}^{(\bf s)}_{M_1,M_2\mid \tilde{\zeta}^\mp_{\bf m}}$ of Alice and Bob's memories after these eigenkets have undergone the loading process sketched in Fig~\ref{DuanKimble} with SPD$_{1_{s_1}}$ and SPD$_{2_{s_2}}$ having detected photons. Then we get the quantum memories' unnormalized density operators via
\begin{equation}
\tilde{\rho}^{({\bf s})}_{M_1,M_2 \mid \psi^\mp_n} = \sum_{\bf m}\left[ \mu_{\bf m}\,\tilde{\rho}^{(\bf s)}_{M_1M_2\mid \tilde{\xi}^\mp_{\bf m}} + \nu_{\bf m}\,\tilde{\rho}^{(\bf s)}_{M_1,M_2\mid \tilde{\zeta}^\mp_{\bf m}}\right], 
\label{rhoBFs}
\end{equation}
whose trace provides the normalization constant needed to get the normalized density operators, $\hat{\rho}^{({\bf s})}_{M_1,M_2 \mid \psi^\mp_n}$.

Section~\ref{ZALM} has already provided a high-level description of the loading process.  The rest of this subsection is devoted to the details, specifically finding general expressions for the $\{\tilde{\rho}^{(\bf s)}_{M_1M_2\mid \tilde{\xi}^\mp_{\bf m}}\}$ and $\{\tilde{\rho}^{(\bf s)}_{M_1,M_2\mid \tilde{\zeta}^\mp_{\bf m}}\}$ and then specializing them to  the push-pull memory-loading operation.

Suppose that $|\tilde{\xi}_{\bf m}^\mp\rangle_{S_{1_n}S_{2_n}}$ has reached Alice and Bob's quantum memories, which have been prepared in the equal superpositions of their ground states, so that
\begin{equation}
|\tilde{\xi}_{{\bf m}_{\rm in}}^\mp\rangle_{\tilde{\bf S}{\bf M}} \equiv |\tilde{\xi}_{\bf m}^\mp\rangle_{\tilde{S}_1\tilde{S}_2}\otimes|\psi\rangle_{\bf M},
\end{equation}
with $|\psi\rangle_{\bf M} \equiv (|g_1\rangle_{M_1} + |g_2\rangle_{M_1})(|g_1\rangle_{M_2} + |g_2\rangle)_{M_2})/2$, is the initial biphoton-memories product state. After  the state-dependent reflections from the memory cavities, and $\tilde{S}'_{1_H}$ and $\tilde{S}'_{2_H}$'s passage through Fig.~\ref{DuanKimble}'s HWP, $|\tilde{\xi}_{{\bf m}_{\rm in}}^\mp\rangle_{{\bf S}{\bf M}} $ becomes the  entangled biphoton-memories state
\begin{align}
|\tilde{\xi}_{{\bf m}_{\rm out}}^\mp\rangle_{\tilde{\bf S}{\bf M}}  & = \int\!\frac{{\rm d}^2\bomega_S}{4\pi^2}\,\frac{\tilde{\xi}_{\bf m}(\bomega_S)}{\sqrt{2}}\left[
|\omega_{S_1}\rangle_{\tilde{S}'_{1_V}}|\omega_{S_2}\rangle_{\tilde{S}_{2_V}}|\psi'_1\rangle_{\bf M}\right. \nonumber \\[.05in]
& \,\,\mp \left. |\omega_{S_1}\rangle_{\tilde{S}_{1_V}}|\omega_{S_2}\rangle_{\tilde{S}'_{2_V}}|\psi'_2\rangle_{\bf M}\right],
\end{align}
where 
\begin{align}
|\psi'_1\rangle_{\bf M} &= \left[r_1(\omega_{S_1})|g_1\rangle_{M_1} + r_2(\omega_{S_1})|g_2\rangle_{M_1}\right]\nonumber \\[.05in]
&\,\, \times \left[|g_1\rangle_{M_2} + |g_2\rangle_{M_2}\right]/2, 
\label{psi1prime}
\end{align}
and
\begin{align}
|\psi'_2\rangle_{\bf M} &= \left[|g_1\rangle_{M_1} + |g_2\rangle_{M_1}\right] \nonumber \\[.05in]
&\,\, \times \left[r_1(\omega_{S_2})|g_1\rangle_{M_2} + r_2(\omega_{S_2})|g_2\rangle_{M_2}\right]/2,
\label{psi2prime}
\end{align}
with $r_1(\omega_{S_k})$ and $r_2(\omega_{S_k})$, for $k=1,2$, being the memories' state-dependent field reflectivities at frequency $\omega_{S_k}$.

Similarly, if $|\tilde{\zeta}_{\bf m}^\mp\rangle_{S_{1_n}S_{2_n}}$ has reached Alice and Bob's quantum memories, which are again prepared in the equal superpositions of their ground states, the initial biphoton-memories product state is
\begin{equation}
|\tilde{\zeta}_{{\bf m}_{\rm in}}^\mp\rangle_{\tilde{\bf S}{\bf M}} \equiv |\tilde{\zeta}_{\bf m}^\mp\rangle_{\tilde{S}_1\tilde{S}_2}\otimes|\psi\rangle_{\bf M}.
\end{equation}
Thus, after the state-dependent reflections from the memory cavities, and $\tilde{S}'_{1_H}$ and $\tilde{S}'_{2_H}$'s passage through Fig.~\ref{DuanKimble}'s HWP, $|\tilde{\zeta}_{{\bf m}_{\rm in}}^\mp\rangle_{{\bf S}{\bf M}} $  becomes the  entangled biphoton-memories state
\begin{align}
|\tilde{\zeta}_{{\bf m}_{\rm out}}^\mp\rangle_{\tilde{\bf S}{\bf M}}  & = \int\!\frac{{\rm d}^2\bomega_S}{4\pi^2}\,\frac{\tilde{\zeta}_{\bf m}(\bomega_S)}{\sqrt{2}}\left[
|\omega_{S_1}\rangle_{\tilde{S}'_{1_V}}|\omega_{S_2}\rangle_{\tilde{S}_{2_V}}|\psi'_1\rangle_{\bf M}\right. \nonumber \\[.05in]
& \,\,\pm \left.|\omega_{S_1}\rangle_{\tilde{S}_{1_V}}|\omega_{S_2}\rangle_{\tilde{S}'_{2_V}}|\psi'_2\rangle_{\bf M}\right].
\end{align}

Now, with $|\tilde{\xi}_{{\bf m}_{\rm out}}^\mp\rangle_{\tilde{\bf S}{\bf M}}$ and $|\tilde{\zeta}_{{\bf m}_{\rm out}}^\mp\rangle_{\tilde{\bf S}{\bf M}}$ in hand, we find $\tilde{\rho}^{(\bf s)}_{M_1,M_2\mid \tilde{\xi}^\mp_{\bf m}}$ and $\tilde{\rho}^{(\bf s)}_{M_1,M_2\mid \tilde{\zeta}^\mp_{\bf m}}$ from 
\begin{equation}
\tilde{\rho}^{(\bf s)}_{M_1M_2\mid \tilde{\upsilon}^\mp_{\bf m}} \equiv {\rm Tr}\!\left[\hat{\prod}_{\bf s}|\tilde{\upsilon}_{{\bf m}_{\rm out}}^\mp\rangle_{\tilde{\bf S}{\bf M}}\,{}_{\tilde{\bf S}{\bf M}}\langle \tilde{\upsilon}_{{\bf m}_{\rm out}}^\mp|\right], \mbox{ for $\tilde{\upsilon} = \tilde{\xi}, \tilde{\zeta}$},
\label{upsilon}
\end{equation}
where $\hat{\prod}_{\bf s}$ is the positive operator-valued measurement for photons being detected by Fig.~\ref{DuanKimble}'s SPD$_{S_{s_1}}$ and SPD$_{S_{s_2}}$~\cite{footnote9a}.  Substituting Eq.~(\ref{upsilon}) into Eq.~(\ref{rhoBFs}) for ${\bf s} = (\pm,\pm)$ gives the loaded memories' unnormalized joint density operator as a function of the detectors that registered photon counts.  

The foregoing procedure is carried out in Appendix~\ref{AppendB} with the following results:
\begin{align}
\tilde{\rho}^{(a)}_{M_1,M_2} &= \tilde{\rho}^{(+,+)}_{M_1,M_2\mid \psi^-_n} = \tilde{\rho}^{(-,-)}_{M_1,M_2\mid \psi^-_n}  \nonumber \\[.05in] &
 = \tilde{\rho}^{(+,-)}_{M_1,M_2\mid \psi^+_n} = \tilde{\rho}^{(-,+)}_{M_1,M_2\mid \psi^+_n},
 \label{rhoAusage}
\end{align}
and
\begin{align}
\tilde{\rho}^{(b)}_{M_1,M_2} &= \tilde{\rho}^{(+,-)}_{M_1,M_2\mid \psi^-_n} = \tilde{\rho}^{(-,+)}_{M_1,M_2\mid \psi^-_n}  \nonumber \\[.05in] &
 = \tilde{\rho}^{(+,+)}_{M_1,M_2\mid \psi^+_n} = \tilde{\rho}^{(-,-)}_{M_1,M_2\mid \psi^+_n},
 \label{rhoBusage}
\end{align}
where
\begin{widetext}
\begin{align}
\tilde{\rho}^{(p)}_{M_1,M_2} &= \sum_{\bf m}\mu_{\bf m}\int\!\frac{{\rm d}^2\bomega_S}{4\pi^2}\,\frac{|\tilde{\xi}_{\bf m}(\bomega_S)|^2}{4}\left[|\phi^+_{p_{\mu{\bf m}}}(\bomega_S)\rangle_{M_1M_2}+ |\phi^-_{p_{\mu{\bf m}}}(\bomega_S)\rangle_{M_1M_2} + |\psi^+_{p_{\mu{\bf m}}}(\bomega_S)\rangle_{M_1M_2}+ |\psi^-_{p_{\mu{\bf m}}}(\bomega_S)\rangle_{M_1M_2}\right] \nonumber \\[.05in]
&\,\,\times\, \left[{}_{M_1M_2}\langle \phi^+_{p_{\mu{\bf m}}}(\bomega_S)|+{}_{M_1M_2}\langle\phi^-_{p_{\mu{\bf m}}}(\bomega_S)|+{}_{M_1M_2}\langle \psi^+_{p_{\mu{\bf m}}}(\bomega_S)|+{}_{M_1M_2}\langle\psi^-_{p_{\mu{\bf m}}}(\bomega_S)|\right] \nonumber \\[.05in]
&+\, \sum_{\bf m}\nu_{\bf m}\int\!\frac{{\rm d}^2\bomega_S}{4\pi^2}\,\frac{|\tilde{\zeta}_{\bf m}(\bomega_S)|^2}{4}\left[|\phi^+_{p_{\nu{\bf m}}}(\bomega_S)\rangle_{M_1M_2}+ |\phi^-_{p_{\nu{\bf m}}}(\bomega_S)\rangle_{M_1M_2} + |\psi^+_{p_{\nu{\bf m}}}(\bomega_S)\rangle_{M_1M_2}+ |\psi^-_{p_{\nu{\bf m}}}(\bomega_S)\rangle_{M_1M_2}\right]  \nonumber \\[.05in]
&\,\,\times\, \left[{}_{M_1M_2}\langle \phi^+_{p_{\nu{\bf m}}}(\bomega_S)|+{}_{M_1M_2}\langle\phi^-_{p_{\nu{\bf m}}}(\bomega_S)|+{}_{M_1M_2}\langle \psi^+_{p_{\nu{\bf m}}}(\bomega_S)|+{}_{M_1M_2}\langle\psi^-_{p_{\nu{\bf m}}}(\bomega_S)|\right] , \mbox{ for $p=a,b$},
\label{rhoAB}
\end{align}
with
\begin{align}
|\phi^+_{a_{\mu{\bf m}}}(\bomega_S)\rangle_{M_1M_2} &= 4^{-1}\!\left[e^{i\omega_{S_2}T}r_1(\omega_{S_1}) -e^{i\omega_{S_1}T}r_1(\omega_{S_2})+e^{i\omega_{S_2}T}r_2(\omega_{S_1}) - e^{i\omega_{S_1}T}r_2(\omega_{S_2})\right]|\phi^+\rangle_{M_1M_2} \label{aStates1}\\[.05in]
|\phi^-_{a_{\mu{\bf m}}}(\bomega_S)\rangle_{M_1M_2} &= 4^{-1}\!\left[e^{i\omega_{S_2}T}r_1(\omega_{S_1}) -e^{i\omega_{S_1}T}r_1(\omega_{S_2})-e^{i\omega_{S_2}T}r_2(\omega_{S_1}) + e^{i\omega_{S_1}T}r_2(\omega_{S_2})\right]|\phi^-\rangle_{M_1M_2} \\[.05in]
|\psi^+_{a_{\mu{\bf m}}}(\bomega_S)\rangle_{M_1M_2} &= 4^{-1}\!\left[e^{i\omega_{S_2}T}r_1(\omega_{S_1}) -e^{i\omega_{S_1}T}r_2(\omega_{S_2})+e^{i\omega_{S_2}T}r_2(\omega_{S_1}) - e^{i\omega_{S_1}T}r_1(\omega_{S_2})\right]|\psi^+\rangle_{M_1M_2} \\[.05in]
|\psi^-_{a_{\mu{\bf m}}}(\bomega_S)\rangle_{M_1M_2} &= 4^{-1}\!\left[e^{i\omega_{S_2}T}r_1(\omega_{S_1}) -e^{i\omega_{S_1}T}r_2(\omega_{S_2})-e^{i\omega_{S_2}T}r_2(\omega_{S_1}) + e^{i\omega_{S_1}T}r_1(\omega_{S_2})\right]|\psi^-\rangle_{M_1M_2} \\[.05in]
|\phi^+_{a_{\nu{\bf m}}}(\bomega_S)\rangle_{M_1M_2} &= 4^{-1}\!\left[e^{i\omega_{S_2}T}r_1(\omega_{S_1}) +e^{i\omega_{S_1}T}r_1(\omega_{S_2})+e^{i\omega_{S_2}T}r_2(\omega_{S_1}) + e^{i\omega_{S_1}T}r_2(\omega_{S_2})\right]|\phi^+\rangle_{M_1M_2} \\[.05in]
|\phi^-_{a_{\nu{\bf m}}}(\bomega_S)\rangle_{M_1M_2} &=4^{-1}\!\left[e^{i\omega_{S_2}T}r_1(\omega_{S_1}) +e^{i\omega_{S_1}T}r_1(\omega_{S_2})-e^{i\omega_{S_2}T}r_2(\omega_{S_1}) - e^{i\omega_{S_1}T}r_2(\omega_{S_2})\right]|\phi^-\rangle_{M_1M_2} \\[.05in]
|\psi^+_{a_{\nu{\bf m}}}(\bomega_S)\rangle_{M_1M_2} &= 4^{-1}\!\left[e^{i\omega_{S_2}T}r_1(\omega_{S_1}) +e^{i\omega_{S_1}T}r_2(\omega_{S_2})+e^{i\omega_{S_2}T}r_2(\omega_{S_1}) + e^{i\omega_{S_1}T}r_1(\omega_{S_2})\right]|\psi^+\rangle_{M_1M_2} \\[.05in]
|\psi^-_{a_{\nu{\bf m}}}(\bomega_S)\rangle_{M_1M_2} &= 4^{-1}\!\left[e^{i\omega_{S_2}T}r_1(\omega_{S_1}) +e^{i\omega_{S_1}T}r_2(\omega_{S_2})-e^{i\omega_{S_2}T}r_2(\omega_{S_1}) - e^{i\omega_{S_1}T}r_1(\omega_{S_2})\right]|\psi^-\rangle_{M_1M_2}, 
\label{aStates8}
\end{align}
and
\begin{align}
|\phi^+_{b_{\mu{\bf m}}}(\bomega_S)\rangle_{M_1M_2} &= 4^{-1}\!\left[e^{i\omega_{S_2}T}r_1(\omega_{S_1}) +e^{i\omega_{S_1}T}r_2(\omega_{S_2})+e^{i\omega_{S_2}T}r_2(\omega_{S_1}) +e^{i\omega_{S_2}T} r_1(\omega_{S_1})\right]|\phi^+\rangle_{M_1M_2} \label{bStates1}\\[.05in]
|\phi^-_{b_{\mu{\bf m}}}(\bomega_S)\rangle_{M_1M_2} &= 4^{-1}\!\left[e^{i\omega_{S_2}T}r_1(\omega_{S_1}) +e^{i\omega_{S_1}T}r_2(\omega_{S_2})-e^{i\omega_{S_2}T}r_2(\omega_{S_1}) - e^{i\omega_{S_1}T}r_1(\omega_{S_2})\right]|\phi^-\rangle_{M_1M_2} \\[.05in]
|\psi^+_{b_{\mu{\bf m}}}(\bomega_S)\rangle_{M_1M_2} &= 4^{-1}\!\left[e^{i\omega_{S_2}T}r_1(\omega_{S_1}) +e^{i\omega_{S_1}T}r_1(\omega_{S_2})+e^{i\omega_{S_2}T}r_2(\omega_{S_1}) + e^{i\omega_{S_1}T}r_2(\omega_{S_2})\right]|\psi^+\rangle_{M_1M_2} \\[.05in]
|\psi^-_{b_{\mu{\bf m}}}(\bomega_S)\rangle_{M_1M_2} &= 4^{-1}\!\left[e^{i\omega_{S_2}T}r_1(\omega_{S_1}) +e^{i\omega_{S_1}T}r_1(\omega_{S_2})-e^{i\omega_{S_2}T}r_2(\omega_{S_1}) - e^{i\omega_{S_1}T}r_2(\omega_{S_2})\right]|\psi^-\rangle_{M_1M_2} \\[.05in]
|\phi^+_{b_{\nu{\bf m}}}(\bomega_S)\rangle_{M_1M_2} &= 4^{-1}\!\left[e^{i\omega_{S_2}T}r_1(\omega_{S_1}) -e^{i\omega_{S_1}T}r_2(\omega_{S_2})+e^{i\omega_{S_2}T}r_2(\omega_{S_1}) - e^{i\omega_{S_1}T}r_1(\omega_{S_2})\right]|\phi^+\rangle_{M_1M_2} \\[.05in]
|\phi^-_{b_{\nu{\bf m}}}(\bomega_S)\rangle_{M_1M_2} &= 4^{-1}\!\left[e^{i\omega_{S_2}T}r_1(\omega_{S_1}) -e^{i\omega_{S_1}T}r_2(\omega_{S_2})-e^{i\omega_{S_2}T}r_2(\omega_{S_1}) + e^{i\omega_{S_1}T}r_1(\omega_{S_2})\right]|\phi^-\rangle_{M_1M_2} \\[.05in]
|\psi^+_{b_{\nu{\bf m}}}(\bomega_S)\rangle_{M_1M_2} &= 4^{-1}\!\left[e^{i\omega_{S_2}T}r_1(\omega_{S_1}) -e^{i\omega_{S_1}T}r_1(\omega_{S_2})+e^{i\omega_{S_2}T}r_2(\omega_{S_1}) - e^{i\omega_{S_1}T}r_2(\omega_{S_2})\right]|\psi^+\rangle_{M_1M_2} \\[.05in]
|\psi^-_{b_{\nu{\bf m}}}(\bomega_S)\rangle_{M_1M_2} &=4^{-1}\!\left[e^{i\omega_{S_2}T}r_1(\omega_{S_1}) -e^{i\omega_{S_1}T}r_1(\omega_{S_2})-e^{i\omega_{S_2}T}r_2(\omega_{S_1}) + e^{i\omega_{S_1}T}r_2(\omega_{S_2})\right]|\psi^-\rangle_{M_1M_2}, \label{bStates8}
\end{align}
\end{widetext}
for 
\begin{equation}
|\phi^\mp\rangle_{M_1M_2} \equiv \frac{|g_1\rangle_{M_1}|g_1\rangle_{M_2}\mp |g_2\rangle_{M_1}|g_2\rangle_{M_2}}{\sqrt{2}},
\label{memoriesBells1}
\end{equation}
and 
\begin{equation}
|\psi^\mp\rangle_{M_1M_2} \equiv \frac{|g_1\rangle_{M_1}|g_2\rangle_{M_2}\mp |g_2\rangle_{M_1}|g_1\rangle_{M_2}}{\sqrt{2}},
\label{memoriesBells2}
\end{equation}
being the memories' Bell states.

Note that Appendix~\ref{AppendB} obtained the preceding results using an augmented version of the Duan-Kimble loading protocol, in which Bob applies a $\pi$ pulse to his memory for a $\psi^-_n$ herald from the ZALM QTX if Alice and Bob's detectors record ${\bf s} = (+,-)$ or $(-,+)$ coincidences, and for a $\psi^+_n$ herald from the ZALM QTX if Alice and Bob's detectors record ${\bf s} = (+,+)$ or $(-,-)$ coincidences.  With this conditional $\pi$-pulse usage, Alice and Bob's memories are primed to load a $|\psi^-\rangle_{M_1M_2}$ singlet state, \emph{regardless} of which state the ZALM QTX has heralded, and which combination of Alice and Bob's detectors have registered photon counts.  Indeed, if the ZALM QTX has a separable channelized biphoton wave function, and Alice and Bob's memories are ideal (in the sense defined below), the augmented Duan-Kimble protocol achieves unit-fidelity loading of that singlet state, as will be seen in the next subsection.  Consequently, we \emph{define} $|\psi^-\rangle_{M_1M_2}$ to be the loaded-memories target state, and
\begin{equation}
F_p \equiv {}_{M_1M_2}\langle \psi^-|\hat{\rho}^{(p)}_{M_1M_2}|\psi^-\rangle_{M_1M_2}, \mbox{ for $p=a,b$}
\end{equation}
to be their entangled-state fidelities for type-a and type-b coincidences.

The preceding development applies to \emph{any} quantum memory whose loading relies on state-dependent reflectivities, including, e.g., the original Duan-Kimble protocol---as demonstrated in Refs.~\cite{Nguyen2019,Bersin2024,Knaut2024}---which our companion paper, Raymer~\emph{et al.}~\cite{Raymer2024}, refers to as on-off operation.  Our interest, however, is in push-pull operation, for which we have~\cite{Raymer2024}
\begin{equation}
r_1(\omega_S) =  \frac{(\gamma+i\Delta_{12}/2 -i\Delta\tilde{\omega}_S)(\kappa-\kappa_J+i\Delta\tilde{\omega}_S) - g^2}{(\gamma+i\Delta_{12}/2 -i\Delta\tilde{\omega}_S)(\kappa+\kappa_J-i\Delta\tilde{\omega}_S) + g^2},
\label{r1pushpull}
\end{equation}
and
\begin{equation}
r_2(\omega_S) =  \frac{(\gamma-i\Delta_{12}/2 -i\Delta\tilde{\omega}_S)(\kappa-\kappa_J+i\Delta\tilde{\omega}_S) - g^2}{(\gamma-i\Delta_{12}/2 -i\Delta\tilde{\omega}_S)(\kappa+\kappa_J-i\Delta\tilde{\omega}_S) + g^2}.
\label{r2pushpull}
\end{equation}

Here: $\gamma$ is  the SiV's spontaneous-emission rate; $\Delta\tilde{\omega}_S \equiv \omega_S-\tilde{\omega}_{S_0}$ is the detuning from the photons' center frequency; $\kappa$ is the memory cavities' field-coupling rate from its output coupler to the incoming and outgoing optical fields; $\kappa_J$ is the field-coupling rate for loss due to leakage from the cavities' HR; and $g$ is the SiV-photon  coupling rate, i.e., the single-photon Rabi frequency.  (The above field damping rates equal one-half the associated energy-damping rates.)

The next subsection reports the loaded-memories entangled-state fidelity for:  (A) ideal operation, in which $|r_1(\omega_S)| = |r_2(\omega_S)| = 1$ and $r_2(\omega_S) = -r_1(\omega_S)$ over the biphoton's bandwidth;   and (B) narrowband push-pull operation, in which Eqs.~(\ref{r1pushpull}) and (\ref{r2pushpull}) apply with 
$r_1(\omega_S) \approx r_1(\tilde{\omega}_{S_0}))$ and $r_2(\omega_S) \approx r_2(\tilde{\omega}_{S_0})$ over the  biphoton's bandwidth.  Neither of these cases require solving Eqs.~(\ref{FredholmC}) and (\ref{FredholmE}).  Thus we postpone consideration of broadband push-pull operation, in which  Eqs.~(\ref{r1pushpull}) and (\ref{r2pushpull}) apply but the biphoton's bandwidth no longer permits approximating the reflectivities by their zero-detuning values, to a future paper.  There, because broadband operation's memory-loading evaluation requires knowledge of  the $\{\mu_{\bf m},\xi_{\bf m}(\bomega_S)\}$ and the $\{\nu_{\bf m},\zeta_{\bf m}(\bomega_S)\}$,  Eqs.~({\ref{xiBFm})--(\ref{nuBFm}) and existing numerical techniques for finding SVDs will be key to getting results.   

\subsection{Two examples of loaded-memories entangled-state fidelity}
Here, we present our two examples' general loaded-memories entangled-state fidelity expressions, which can be evaluated for arbitrary channelized biphoton wave functions, and their loaded-memories entangled-state fidelities for the all-Gaussian biphoton wave function with brickwall filtering and parameter values from Table~\ref{singleParams}'s Case~2.  \\[.05in]
\noindent{\emph{Case A: Ideal operation}  When $|r_1(\omega_S)| = |r_2(\omega_S)| = 1$ and $r_2(\omega_S)  = -r_1(\omega_S)$ over the biphoton's bandwidth, Alice and Bob's QRXs can omit Fig.~\ref{DuanKimble}'s time delay because, with $T=0$, Eqs.~(\ref{rhoAB})--(\ref{bStates8}) collapse to
\begin{align}
\tilde{\rho}^{(a)}_{M_1,M_2} &= \tilde{\rho}^{(b)}_{M_1,M_2} \nonumber \\[.05in]
&= \sum_{\bf m}\mu_{\bf m}\,|\psi^-\rangle_{M_1M_2}\,{}_{M_1M_2}\langle \psi^-| \nonumber \\[.05in]
&\,\,+\, \sum_{\bf m}\nu_{\bf m}\,|\phi^-\rangle_{M_1M_2}\,{}_{M_1M_2}\langle \psi^+|.
\end{align}
Ideal operation is indeed ideal; its normalized loaded-memories density operators are
\begin{align}
\hat{\rho}&^{({\bf s})}_{M_1,M_2 \mid \psi^\mp_n} = \sum_{\bf m}\tilde{\mu}_{\bf m}\,|\psi^-\rangle_{M_1M_2}\,{}_{M_1M_2}\langle \psi^-| \nonumber \\[.05in]
&\,\,+\,\sum_{\bf m}\tilde{\nu}_{\bf m}\,|\phi^-\rangle_{M_1M_2}\,{}_{M_1M_2}\langle \phi^-|, 
\mbox{ for ${\bf s} = (\pm,\pm)$,}
\label{IdealMemDensOp}
\end{align} 
which indicates that $\hat{\rho}_{S_{1_n}S_{2_n}\mid \psi^\mp_n}$ from Eq.~(\ref{heraldedDensOp}) has been perfectly transferred to Alice and Bob's quantum memories.  In general, this ideality means that the loaded-memories entangled-state fidelity obeys
\begin{eqnarray}
\lefteqn{F_{\bf s}\equiv{}_{M_1M_2}\langle \psi^-|\hat{\rho}^{({\bf s})}_{M_1,M_2 \mid \psi^\mp_n}|\psi^-\rangle_{M_1M_2}}
\nonumber \\[.05in]
&=& {}_{S_{1_n}S_{2_n}}\!\langle \psi^\mp|\hat{\rho}^\mp_{S_{1_n},S_{2_n} \mid \psi^\mp_n}|\psi^\mp\rangle_{S_{1_n}S_{2_n}}  \\[.05in]
&=& \frac{1}{2}\!\left(1+\sum_{m=1}^\infty\tilde{\lambda}^4_m\right)  \\[.05in]
&=& \frac{\displaystyle \int\!\frac{{\rm d}^2\bomega_S}{4\pi^2}\,K_{S_{1_n}S_{2_n}}^{(c)}(\bomega_S;\bomega_S)}
{\displaystyle \int\!\frac{{\rm d}^2\bomega_S}{4\pi^2}\sum_{v=c,e}K_{S_{1_n}S_{2_n}}^{(v)}(\bomega_S;\bomega_S)},
\label{IdealMemFidelity}                                     
\end{eqnarray}
where the second equality follows from Eq.~(\ref{PrCn}) and the third equality, which follows from Eqs.~(\ref{eigenXi}) and (\ref{eigenZeta}), can be evaluated directly from the Sagnac sources' channelized biphoton wave function.

For the all-Gaussian biphoton wave function with brickwall filtering and the Case~2 parameter values from Table~\ref{singleParams}, Fig.~\ref{BSMerrorprob_fig} shows that ideal operation of our Duan-Kimble quantum memories has error probability $\Pr(e_n) = 0.386\%$, and fidelity $\Pr(c_n) = 1-\Pr(e_n) = 99.6\%$ across all 81 channels.   \\[.05in]
\noindent{\emph{Case B:  Narrowband push-pull operation}
In narrowband push-pull operation we assume that $r_1(\omega_S) \approx r_1(\tilde{\omega}_{S_0})$ and $r_2(\omega_S) \approx r_2(\tilde{\omega}_{S_0})$ over the biphoton's bandwidth.  Equations~(\ref{r1pushpull}) and (\ref{r2pushpull}) show the push-pull reflectivities have Hermitian symmetry about $\tilde{\omega}_{S_0}$, viz., $r_2(\tilde{\omega}_{S_0}+\Delta\tilde{\omega}_S) = r_1^*(\tilde{\omega}_{S_0}-\Delta\tilde{\omega}_S)$, implying that $r_2(\tilde{\omega}_{S_0}) = r_1^*(\tilde{\omega}_{S_0})$. Thus Alice and Bob's QRX's can again omit Fig.~\ref{DuanKimble}'s time delay, because narrowband operation only attenuates and phase shifts incoming wave packets hence it does not significantly delay or reshape them}.  Under these conditions, $\tilde{\rho}_{M_1M_2}^{(a)}$ and $\tilde{\rho}_{M_1M_2}^{(b)}$ from Eqs.~(\ref{rhoAB})--(\ref{bStates8}) reduce to
\begin{widetext}
\begin{align}
\tilde{\rho}_{M_1M_2}^{(a)} &= \sum_{\bf m}\frac{\mu_{\bf m}}{4}\,{\rm Im}^2[r_1(\tilde{\omega}_{S_0})]|\psi^-\rangle_{M_1M_2}\,{}_{M_1M_2}\!\langle \psi^-| \nonumber \\[.05in]
&\,\,+\, 
\sum_{\bf m}\frac{\nu_{\bf m}}{4}\left\{{\rm Re}[r_1(\tilde{\omega}_{S_0})](|\phi^+\rangle_{M_1M_2} + |\psi^+\rangle_{M_1M_2}) +
i\,{\rm Im}[r_1(\tilde{\omega}_{S_0})]|\phi^-\rangle_{M_1M_2}\right\} \nonumber \\[.05in]
&\,\, \times\, \left\{{\rm Re}[r_1(\tilde{\omega}_{S_0})]({}_{M_1M_2}\!\langle \phi^+| + {}_{M_1M_2}\!\langle \psi^+|) -i\,{\rm Im}[r_1(\tilde{\omega}_{S_0})]{}_{M_1M_2}\!\langle \phi^-|\right\},  
\end{align}
and
\begin{align}
\tilde{\rho}_{M_1M_2}^{(b)} &= 
\sum_{\bf m}\frac{\mu_{\bf m}}{4}\left\{{\rm Re}[r_1(\tilde{\omega}_{S_0})](|\phi^+\rangle_{M_1M_2} + |\psi^+\rangle_{M_1M_2}) +
i\,{\rm Im}[r_1(\tilde{\omega}_{S_0})]|\psi^-\rangle_{M_1M_2}\right\} \nonumber \\[.05in]
&\,\,\times \, \left\{{\rm Re}[r_1(\tilde{\omega}_{S_0})]({}_{M_1M_2}\!\langle \phi^+| + {}_{M_1M_2}\!\langle \psi^+|)-i\,{\rm Im}[r_1(\tilde{\omega}_{S_0})]{}_{M_1M_2}\!\langle \psi^-|\right\} \nonumber \\[.05in]
 &\,\,+\, \sum_{\bf m}\frac{\nu_{\bf m}}{4}\,{\rm Im}^2[r_1(\tilde{\omega}_{S_0})]|\phi^-\rangle_{M_1M_2}\,{}_{M_1M_2}\!\langle \phi^-|,
\end{align}
leading to fidelities to the target singlet-state $|\psi^-\rangle_{M_1M_2}$ given by 
\begin{equation}
F_a  = 
\frac{\displaystyle \sum_{\bf m}\mu_{\bf m}\,{\rm Im}^2[r_1(\tilde{\omega}_{S_0})]} {\displaystyle \sum_{\bf m}\,\mu_{\bf m}\,{\rm Im}^2[r_1(\tilde{\omega}_{S_0})]
+ \sum_{\bf m}\nu_{\bf m}\left(2\,{\rm Re}^2[r_1(\tilde{\omega}_{S_0})] + {\rm Im}^2[r_1(\tilde{\omega}_{S_0})]\right)},
\end{equation}
and
\begin{equation}
F_b   = 
\frac{\displaystyle \sum_{\bf m}\mu_{\bf m}\,{\rm Im}^2[r_1(\tilde{\omega}_{S_0})]} {\displaystyle \sum_{\bf m}\mu_{\bf m}\left(2\,{\rm Re}^2[r_1(\tilde{\omega}_{S_0})] + {\rm Im}^2[r_1(\tilde{\omega}_{S_0})]\right)
+ \sum_{\bf m}\nu_{\bf m}\,{\rm Im}^2[r_1(\tilde{\omega}_{S_0})]}.
\end{equation}
\end{widetext}
These fidelities are easily evaluated, given a set of memory parameters, because
\begin{equation}
\sum_{\bf m}\mu_{\bf m} = \int\!\frac{{\rm d}^2\bomega_S}{4\pi^2}\,K^{(c)}_{S_{1_n}S_{2_n}}(\bomega_S;\bomega_S),
\end{equation}
and
\begin{equation}
\sum_{\bf m}\nu_{\bf m} = \int\!\frac{{\rm d}^2\bomega_S}{4\pi^2}\,K^{(e)}_{S_{1_n}S_{2_n}}(\bomega_S;\bomega_S),
\end{equation}
are easily computed for the all-Gaussian case. We choose, however, to postpone consideration of memory parameters until a future paper, where performance evaluation for narrowband push-pull operation will serve as a prelude to the broadband case.  
Note that performance analysis for broadband operation will be doubly complicated in that:  (1) we must deal with the full frequency-dependent complexity of Eqs.~(\ref{rhoAB})--(\ref{bStates8}); and (2) that complexity will require ZALM QRXs to incorporate pre-loading mode shaping, in addition to their frequency conversion and bandwidth compression, to achieve optimum memory-loading performance.  That said, there is something else to be added about narrowband push-pull memory loading:  with the appropriate, potentially accessible, parameter values for the quantum memories, our results further reduce to an attenuated version of ideal operation, as we now show. Thus, it may turn out that the most aggressive bandwidth compression will be best for ZALM performance, but answering that question is a topic for our future work.

In our companion paper~\cite{Raymer2024}, we rewrote the memories' zero-detuning, state-dependent reflectivities for push-pull operation as 
\begin{align}
r_1(\tilde{\omega}_{S_0}) = r_2^*(\tilde{\omega}_{S_0}) = \frac{(1+i\Delta_{12}/2\gamma)(1-\kappa_J/\kappa) -C}{(1+i\Delta_{12}/2\gamma)(1+\kappa_J/\kappa) +C},
\end{align}
where $C\equiv g^2/\kappa\gamma$ is the cavities' cooperativity, with $C > 1$ being  the strong-coupling regime in which the memories are operated. With this rewriting, and assuming $\kappa>\kappa_J$, they then prove that 
\begin{equation}
C = C_\pi \equiv \sqrt{1+(1-\kappa_J^2/\kappa^2)\Delta^2_{12}/4\gamma^2} - \kappa_J/\kappa
\end{equation}
makes the narrowband reflectivities pure-imaginary
\begin{equation}
r_1(\tilde{\omega}_{S_0}) = i\frac{\sqrt{\gamma^2\kappa^2+\Delta^2_{12}(\kappa^2-\kappa^2_J)/4}-\gamma \kappa}{\Delta_{12}(\kappa+\kappa_J)/2},
\end{equation}
giving the desired $\pi$-rad phase shift between the actions of the two color-center states. It immediately follows that the loaded memories' normalized density operators and entangled-state fidelities match those of ideal operation, as given by Eqs.~(\ref{IdealMemDensOp}) and (\ref{IdealMemFidelity}).  So, for the QTX parameters from Table~\ref{singleParams}'s Case~2, the loaded memories have 99.6\% entangled-state fidelity across all 81 channels.

There is, of course, a difference between ideal operation and narrowband push-pull operation with $C= C_\pi$.  It comes as a reduction in the entanglement-distribution rate.  Ideal operation's memories have no intra-cavity losses, so its probability that a particular pump pulse succeeds in loading both Alice and Bob's memories is $\eta_{\rm qtx}\,\eta^2_{\rm prop}\,\eta^2_{\rm qrx}$, where $\eta_{\rm qrx}$ only accounts for \emph{exo-cavity} memory losses.  For narrowband push-pull operation with $C=C_\pi$, however, that success probability gets an additional sub-unit factor of $\eta_{\rm cavity} \equiv |r_1(\tilde{\omega}_{S_0})|^2$, to account for \emph{intra-cavity} memory losses. 

Broaching the issue of ZALM's entanglement-distribution rate is a perfect segue into the next section, where the tradeoff between ZALM's entanglement-distribution rate and its inter-channel interference will be the first topic on the agenda.

\section{Discussion \label{Discuss}}
In this culminating section we will address a variety of additional issues, but before doing so it is worthwhile to appraise what we did accomplish, as opposed to what remains for the future.  We set out to do a deep dive into ZALM's heralded source of polarization-entangled biphotons and their coupling to a pair of intra-cavity color-center quantum memories, and we accomplished those objectives.  In particular, we derived the density operator for the heralded biphoton, which quantifies its heralding performance and fidelity.  We also derived the loaded memories' density operator and its fidelity with the memories' entangled-state target.  Although we have not undertaken a comprehensive exploration of ZALM performance as its source and memory parameters are varied, our theory is set up for just such an endeavor.  Nevertheless, the examples we have evaluated already demonstrate two critical features of Sec.~\ref{ZALM}'s ZALM architecture:  (1) the necessity of achieving a near-separable channelized biphoton wave function to ensure the biphoton sent to Alice and Bob is of high purity; and (2) the premium placed on Alice and Bob's bandwidth compressors' enabling narrowband push-pull memory loading to ensure the arriving biphoton's state is faithfully transferred to the intra-cavity color centers.

We turn now to our first issue for discussion:  ZALM's entanglement-distribution rate, i.e., its per-pump-pulse rate (probability) of Alice and Bob's QRXs sharing the $|\psi^-\rangle_{M_1M_2}$ target state, 
and that rate's tradeoff with inter-channel interference, which occurs when the signal photon from an $n$th-channel herald does not pass through its DWDM$_S$ filter's $n$th-channel passband but a signal photon whose idler companion lies in a different channel does. As detailed below, ZALM's operating with near-deterministic creation of one polarization-entangled pair per pump pulse across its dual-Sagnac source's full phase-matching bandwidth implies a non-trivial probability that there will be a two-pair event over that bandwidth.  But, before quantifying that interference, we address the entanglement-distribution rate.    

Let $N_p$ be the random number of photon pairs produced by the ZALM QTX's dual-Sagnac source from a single pump pulse, and let $\mathbb{E}(N_p) \sim 1$ be its expected value~\cite{footnote11}.  We then have that the ZALM system's entanglement-distribution rate is $R = \sum_{n= -(N-1)/2}^{(N-1)/2}R_n$, where $R_n$, the entanglement-distribution rate for an $n$th-channel herald, is~\cite{footnote12}
\begin{align}
R_n &= \mathbb{E}(N_p)[\eta_{\rm qtx}\Pr(\psi^\mp_n)][\eta^2_{\rm prop}\Pr(S_{1_n},S_{2_n}\mid \psi^\mp_n)] \nonumber \\[.05in]
&\,\,\times\, [\eta^2_{\rm qrx}\eta_{\rm cavity}][(F_a+F_b)/2],  
\label{entangrate}
\end{align}
where $F_a$ and $F_b$ 
are the fidelities for type-a and type-b coincidence detections at Alice and Bob's QRXs, as defined in Eqs.~(\ref{rhoAusage}) and (\ref{rhoBusage}), viz., $(+,+), (-,-),(+,-),$ or $(-,+)$, which are equally likely to occur.  The first bracketed term in Eq.~(\ref{entangrate}) is the probability of a $\psi^\mp_n$ herald, given that the dual-Sagnac source has emitted a biphoton.  The second bracketed term there is the probability that the $n$th-channel signal photons will reach Alice and Bob's QRXs \emph{and} pass through their DWDM$_S$ filters' $n$th-channel passbands, given that there was a $\psi^\mp$ herald. The third bracketed term in Eq.~(\ref{entangrate}) is the probability that Alice and Bob's QRXs register a photon-detection coincidence, given that a pair of signal photons passed through their DWDM$_S$ filters $n$th-channel passbands.  The final bracketed term in that equation is the average $|\psi^-\rangle_{M_1M_2}$ fidelity, given that Alice and Bob's QRXs register a photon-detection coincidence.

The inter-channel interference problem---alluded to above---now becomes clear.  To maximize the generation of heralded polarization-entangled biphotons for transmission to Alice and Bob, the ZALM QTX will operate with $\mathbb{E}(N_p) \sim 1$, $\Omega_{\rm PM} \gg \Delta B$, and $N \sim \Omega_{\rm PM}/\Delta B$, so that each DWDM channel's heralding probability will be low enough, e.g., $\Pr(\psi^\mp_n) \sim 10^{-3}$, that multi-pair events can be ignored on a per-channel basis.  That said, $\mathbb{E}(N_p) \sim 1$ implies that a 2-pair event, over the dual-Sagnac's full phase-matching bandwidth, will occur with $\sim$20\% probability.  Then, because the ZALM QTX must operate at low heralding efficiency to maintain high fidelity, cf.\@ Figs.~\ref{BSMefficiency_fig} and \ref{BSMpurity_fig}, a 2-pair event that heralds on channels $n$ and $m$ can have its $n$th-channel's signal photon appear in the $m$th channel and vice versa causing memory-load errors.  This effect is most severe for $|n-m| = 1$, as illustrated, for the simple case of a single Sagnac-configured source, in Fig.~\ref{crosstalk1_fig}, assuming the all-Gaussian special case with parameter values from Table~\ref{singleParams}.  There we have plotted the normalized $m=n+1$ inter-channel interference, $\chi_1(n)\equiv \Pr(S_{n+1},I_n)/\Pr(S_n,I_n)$ for $n = -40, -39,\ldots,40,$ with 
\begin{equation}
\Pr(S_m,I_n) \equiv \int\!\frac{{\rm d}\omega_S}{2\pi}\int\!\frac{{\rm d}\omega_I}{2\pi}\,|\Psi_{SI_n}(\omega_S,\omega_I)H_{S_m}(\omega_S)|^2
\end{equation}
being the probability of the Sagnac SPDC's biphoton has a signal component that passes through the DWDM$_S$ filter's $m$th channel and an idler component that passes through the DWDM$_I$ filter's $n$th channel.  As expected, Fig.~\ref{crosstalk1_fig} shows that Case~1, with its high heralding efficiency, has very low inter-channel interference, but Case~2, with its high fidelity coming at the cost of low heralding efficiency, has very significant inter-channel interference.  

We can dramatically reduce the inter-channel interference, at the expense of entanglement-distribution rate, by having ZALM's QTX impose artificially enhanced guard bands, e.g., by only allowing heralds from channels spaced two or three apart.  Figure~\ref{crosstalk2_fig} shows the reductions in normalized inter-channel interference offered by
$\chi_2(n)\equiv \Pr(S_{n+2},I_n)/\Pr(S_n,I_n)$ for $n= -40,-38,\ldots,40,$ and $\chi_3(n)\equiv \Pr(S_{n+3},I_n)/\Pr(S_n,I_n)$ for $n = -39,-36,\ldots,39,$ assuming the all-Gaussian special case with Table~\ref{singleParams}'s Case~2 parameter values.  These benefits are accompanied by a reduction in the number of usable channels to 41 for $\chi_2(n)$, and to 27 for $\chi_3(n)$.  A full exploration of the tradeoff between entanglement-distribution rate and inter-channel interference is postponed to a future paper.
\begin{figure}[hbt]
    \centering
    \includegraphics[width=0.45\textwidth]{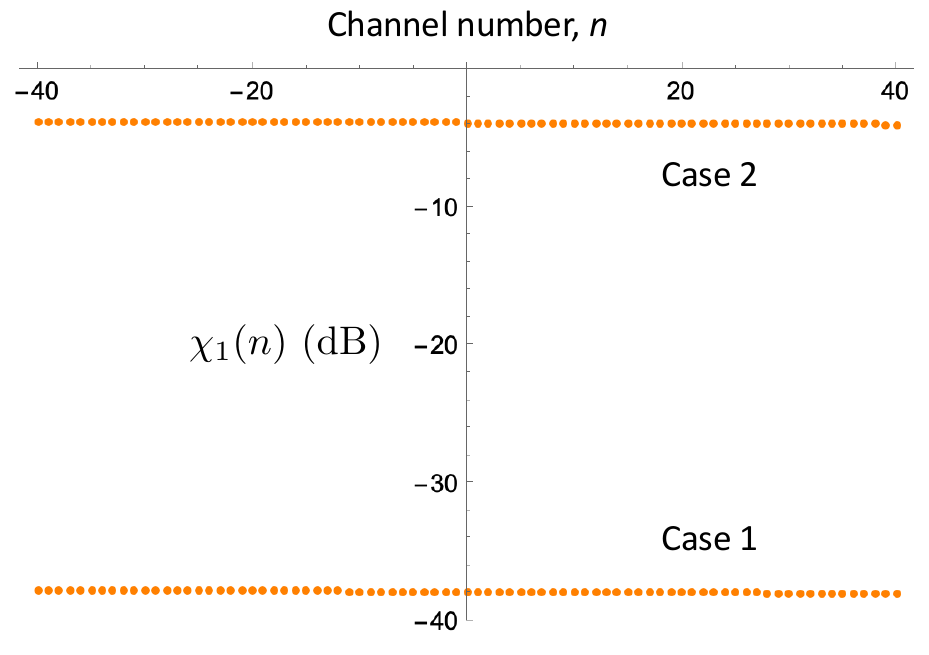}
    \caption{Normalized inter-channel interference, $\chi_1(n)$ for $n = -40,-39,\ldots 40,$ using Eq.~(\ref{allGauss})'s all-Gaussian biphoton wave function with brickwall filtering and parameter values from Table~\ref{singleParams}. \label{crosstalk1_fig}}    
\end{figure}
\begin{figure}
    \centering
    \includegraphics[width=0.45\textwidth]{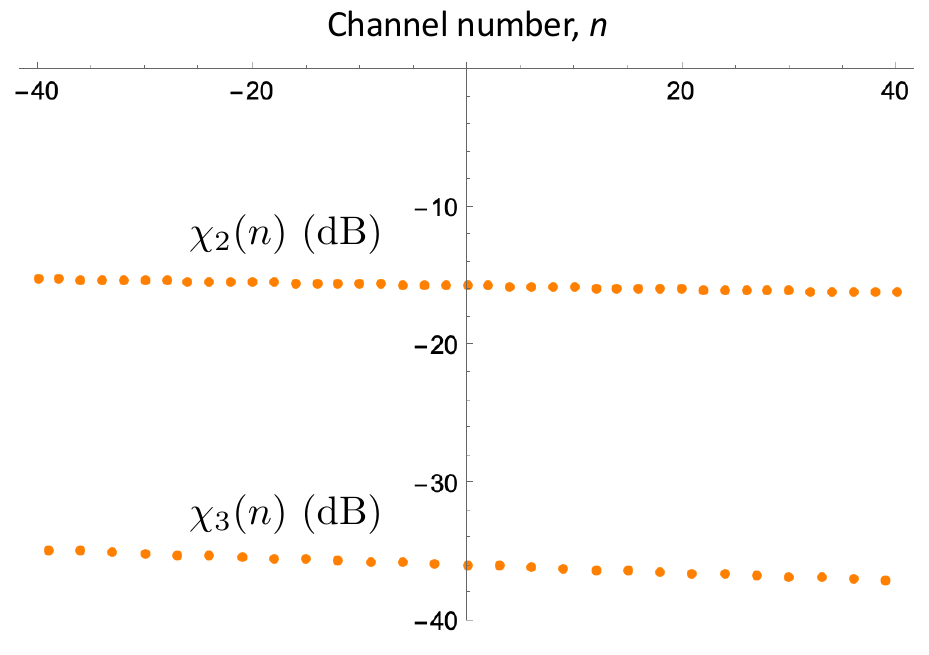}
    \caption{Normalized inter-channel interferences, $\chi_2(n)$ for $n=-40,-38,\ldots,40,$ and $\chi_3(n)$ for $n=39,-36,\ldots,39,$ using Eq.~(\ref{allGauss})'s all-Gaussian biphoton wave function with brickwall filtering and the Case~2 parameter values from Table~\ref{singleParams}. \label{crosstalk2_fig}}    
\end{figure}

In truth, inter-channel interference will not be much of a problem when $\mathbb{E}(N_p) \sim 1$ and $N\gg 1$, because it is unlikely that a two-pair event will herald on adjacent DWDM channels.  Future availability of high-speed, high-quantum-efficiency, number-resolving photodetectors, however, will lead to ZALM's using $\mathbb{E}(N_p) \gg 1$ so as to approach $\Pr(\psi^\mp_n)\sim 0.2$.  In that case, inter-channel interference will be a significant concern.  

Our final comments have to do with extending our analysis to account for equipment nonidealities, and to consider alternative SPDC sources as well as alternative quantum-memory protocols.  With respect to equipment nonidealities, culprits missing from our treatment so far include:  (1) slight asymmetry in the 50--50 beam splitters in the partial BSMs and the Duan-Kimble quantum memories; (2) dark counts in the partial-BSMs and the Duan-Kimble quantum memories' heralding detectors; (3) background-light counts in the Duan-Kimble quantum memories' heralding detectors when ZALM relies on free-space propagation; and (4) imperfect mode conversion in Alice and Bob's QRXs.  Evaluating their impact on ZALM's entanglement-distribution rate and its entangled-state fidelity should be part of an accurate assessment of its utility, and the techniques we have employed here can be extended to account for them.

Turning to the possibility of alternative SPDC sources for use in ZALM, the key issue is to find better ways to approach a separable channelized wave function for the Sagnac sources' polarization-entangled biphotons \emph{without} sacrificing heralding efficiency, which is the bane of our DWDM-filtered sources.  Two possibilities here are intra-cavity dual-SPDC sources and SPDCs whose nonlinear crystals have many phase-matching islands.  Shapiro and Wong~\cite{Shapiro2000} proposed an intra-cavity dual-SPDC source that could directly generate narrowband signal and idler for interfacing to the Rb-atom quantum memories in an early proposal for long-distance qubit teleportation~\cite{Lloyd2001}.  Reference~\cite{Shapiro2000} only considered the output from a single cavity resonance.  For ZALM, however, multiple cavity resonances would be employed to provide the protocol's frequency multiplexing. Cavity losses may be a significant issue for this approach, but a detailed design study for multi-channel, cavity-based sources is certainly warranted and our analysis could be adapted for this purpose and our techniques are capable of addressing the resulting entangled-state fidelity.    

SPDC sources with phase-matching-islands eschew intra-cavity operation.  Instead, they engineer a nonlinear crystal's phase-matching function to produce a joint spectral intensity comprised of 
near-separable, non-overlapping islands well-suited to frequency-multiplexed entanglement distribution.  Reference~\cite{Morrison2022} realized an 8-island demonstration of this technique.  Extending their work to the desired $\sim$100 islands for ZALM may be challenging, but here too further study is needed.   

The final area we will mention for extending our work is consideration of alternative memory protocols, specifically those associated with distributing time-bin entanglement.  The dual-Sagnac source we have assumed affords a large number of frequency-multiplexed channels for high-rate transmission of polarization-entangled biphotons.  Polarization entanglement is the preferred choice for satellite-to-ground entanglement distribution, because atmospheric turbulence does not disturb polarization~\cite{Andrews1998}.  For terrestrial entanglement distribution over optical fiber, however, time-bin entanglement is the way to go, because standard fiber does not preserve polarization and polarization-maintaining fiber is currently too lossy for long-distance operation.  Our dual-Sagnac source's heralded polarization-entangled biphotons can be converted to time-bin entanglement using linear optics, as shown in Fig.~\ref{entconverter}, and then loaded into Duan-Kimble quantum memories set up for time-bin entanglement as in Nguyen \emph{et al}.~\cite{Nguyen2019}.  Raymer~\emph{et al.}'s reflectivity analysis needs only a minor modification, to account for the $\pi$ pulse Ref.~\cite{Nguyen2019} applies to the memories between the arrival of $S_{k_e}$ and $S_{k_\ell}$, after which  the same state-dependent reflectivities as those for Duan-Kimble loading of polarization entanglement are obtained. 
\begin{figure}[h]
    \centering
    \includegraphics[width=0.3\textwidth]{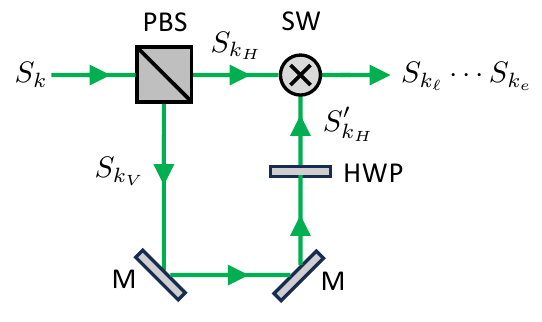}
    \caption{Schematic for converting the polarization-entangled biphoton sent to Alice ($S_1$) and Bob ($S_2$) to time-bin entanglement.  PBS:  polarization beam splitter. HWP:  half-wave plate.  M:  mirror.  SW:  fast switch to first pass $S_{k_H}$ and later pass $S'_{k_H}$ into the output.  $S_{k_\ell}\cdots S_{k_e}$ for $k=1$ and $2$:  the time-bin entangled ($e$ = early, $\ell$ = late) state.  \label{entconverter}}    
\end{figure}

\acknowledgments
This work was supported by the Engineering Research Centers Program of the National Science Foundation
under Grant \#1941583 to the NSF-ERC Center for Quantum Networks. The authors thank Saikat Guha and Dirk Englund for advice and encouragement and Prajit Dhara for valuable discussions.
\appendix

\section{Derivation Details for Sec.~\ref{partial-BSM} \label{AppendA}}
In this appendix we supply derivation details for results that were stated, but not proven, in Sec.~\ref{partial-BSM}. 
\subsection{Derivation of Eq.~(\ref{tilde-rhominusplus})}
We start this derivation by expanding $|\tilde{\psi}\rangle_{{\bf S}_n{\bf I}_n}\,{}_{{\bf S}_n{\bf I}_n}\!\langle \tilde{\psi}|$ in Eq.~(\ref{rho-psiminus})'s first integral using its definition and Eq.~(\ref{beamSplitters})'s beam-splitter relations.  The result so obtained is the enormous expression
\begin{widetext}
\begin{eqnarray}
\lefteqn{\left.\tilde{\rho}_{S_{1_n}S_{2_n}\mid \psi^-_n}\right|_{\mbox{\scriptsize integral 1}} = }\nonumber \\[.05in]
&& \int\!\frac{{\rm d}\omega_{I_+}}{2\pi}\int\!\frac{{\rm d}\omega_{I-}}{2\pi}\int\!\frac{{\rm d}^2\bomega_S}{4\pi^2}\int\!\frac{{\rm d}^2\bomega_I}{4\pi^2}\int\!\frac{{\rm d}^2\bomega_S'}{4\pi^2}\int\!\frac{{\rm d}^2\bomega_I'}{4\pi^2}\,({}_{I_{+H}}\!\langle\omega_{I_+}|\,{}_{I_{-V}}\!\langle \omega_{I_-}|)\Psi_{S_nI_n}(\omega_{S_1},\omega_{I_1})\Psi^*_{S_nI_n}(\omega_{S_1}',\omega_{I_1}') \nonumber \\[.05in]
&&\,\,\times\, \Psi_{S_nI_n}(\omega_{S_2},\omega_{I_2})\Psi^*_{S_nI_n}(\omega_{S_2}',\omega_{I_2}')\left[\frac{|\omega_{S_1}\rangle_{S_{1H}}(|\omega_{I_1}\rangle_{I_{+V}}+|\omega_{I_1}\rangle_{I_{-V}})-|\omega_{S_1}\rangle_{S_{1V}}(|\omega_{I_1}\rangle_{I_{+H}}+|\omega_{I_1}\rangle_{I_{-H}})}{2}\right]  \nonumber \\[.05in]
&&\,\,\times\, \left[\frac{|\omega_{S_2}\rangle_{S_{2H}}(|\omega_{I_2}\rangle_{I_{+V}}-|\omega_{I_2}\rangle_{I_{-V}})-|\omega_{S_2}\rangle_{S_{2V}}(|\omega_{I_2}\rangle_{I_{+H}}-|\omega_{I_2}\rangle_{I_{-H}})}{2}\right]  \nonumber \\[.05in]
&&\,\,\times\,\left[\frac{{}_{S_{2H}}\langle\omega_{S_2}'|({}_{I_{+V}}\langle\omega_{I_2}'|- {}_{I_{-V}}\langle\omega_{I_2}'|)-{}_{S_{2V}}\langle\omega_{S_2}'|({}_{I_{+H}}\langle \omega_{I_2}'|-{}_{I_{-H}}\langle\omega_{I_2}'|)}{2}\right] \nonumber \\[.05in]
&& \,\,\times\,\left[\frac{{}_{S_{1H}}\langle\omega_{S_1}'|({}_{I_{+V}}\langle\omega_{I_1}'|- {}_{I_{-V}}\langle\omega_{I_1}'|)-{}_{S_{1V}}\langle\omega_{S_1}'|({}_{I_{+H}}\langle \omega_{I_1}'|-{}_{I_{-H}}\langle\omega_{I_1}'|)}{2}\right](|\omega_{I_-}\rangle_{I_{-V}}\,|\omega_{I_+}\rangle_{I_{+H}}).
\label{tildeRhoStep2}
\end{eqnarray}
Evaluating the resulting bra-ket inner products gives four impulses that collapse the ${\rm d}^2\bomega_I$ and ${\rm d}^2\bomega_I'$ integrals in Eq.~(\ref{tildeRhoStep2}), reducing it to the slightly less formidable expression 
\begin{eqnarray}
\lefteqn{\left.\tilde{\rho}_{S_{1_n}S_{2_n}\mid \psi^-_n}\right|_{\mbox{\scriptsize integral 1}} = \frac{1}{16}\int\!\frac{{\rm d}\omega_{I_+}}{2\pi}\int\!\frac{{\rm d}\omega_{I-}}{2\pi}\int\!\frac{{\rm d}^2\bomega_S}{4\pi^2}\int\!\frac{{\rm d}^2\bomega_S'}{4\pi^2}}\nonumber \\[.05in]
&& \left\{\Psi_{S_nI_n}(\omega_{S_1},\omega_{I_-})\Psi^*_{S_nI_n}(\omega_{S_1}',\omega_{I_-})\Psi_{S_nI_n}(\omega_{S_2},\omega_{I_+}) \Psi^*_{S_nI_n}(\omega_{S_2}',\omega_{I_+}) |\omega_{S_1}\rangle_{S_{1V}}|\omega_{S_2}\rangle_{S_{2H}}\,{}_{S_{2H}}\langle\omega_{S_2}'|\,{}_{S_{1V}}\langle \omega_{S_1}'|\right. \nonumber\\[.05in]
&& \,\,+\, \Psi_{S_nI_n}(\omega_{S_1},\omega_{I_+})\Psi^*_{S_nI_n}(\omega_{S_1}',\omega_{I_+})\Psi_{S_nI_n}(\omega_{S_2},\omega_{I_-}) \Psi^*_{S_nI_n}(\omega_{S_2}',\omega_{I_-}) |\omega_{S_1}\rangle_{S_{1H}}|\omega_{S_2}\rangle_{S_{2V}}\,{}_{S_{2V}}\langle\omega_{S_2}'|\,{}_{S_{1H}}\langle \omega_{S_1}'| \nonumber \\[.05in]
&&\,\,-\, \Psi_{S_nI_n}(\omega_{S_1},\omega_{I_-})\Psi^*_{S_nI_n}(\omega_{S_1}',\omega_{I_+})\Psi_{S_nI_n}(\omega_{S_2},\omega_{I_+}) \Psi^*_{S_nI_n}(\omega_{S_2}',\omega_{I_-}) |\omega_{S_1}\rangle_{S_{1V}}|\omega_{S_2}\rangle_{S_{2H}}\,{}_{S_{2V}}\langle\omega_{S_2}'|\,{}_{S_{1H}}\langle \omega_{S_1}'| \nonumber \\[.05in]
&&\,\, - \,\left. \Psi_{S_nI_n}(\omega_{S_1},\omega_{I_+})\Psi^*_{S_nI_n}(\omega_{S_1}',\omega_{I_-})\Psi_{S_nI_n}(\omega_{S_2},\omega_{I_-}) \Psi^*_{S_nI_n}(\omega_{S_2}',\omega_{I_+}) |\omega_{S_1}\rangle_{S_{1H}}|\omega_{S_2}\rangle_{S_{2V}}\,{}_{S_{2H}}\langle\omega_{S_2}'|\,{}_{S_{1V}}\langle \omega_{S_1}'| \right\}\!.
\label{tildeRhoStep3}
\end{eqnarray}
Equation~(\ref{tildeRhoStep3}) compacts further, by using Eq.~(\ref{PhiNdefn}) and defining
\begin{equation}
K_{S_{1n}S_{2n}}^{(1)}(\bomega_S;\bomega_S') \equiv \Phi_n(\omega_{S_1},\omega'_{S_1}) \Phi_n(\omega_{S_2},\omega'_{S_2})
\mbox{ and }
K_{S_{1n}S_{2n}}^{(2)}(\bomega_S;\bomega_S') \equiv \Phi_n(\omega_{S_1},\omega'_{S_2}) \Phi_n(\omega_{S_2},\omega'_{S_1}), 
\end{equation}
which gives us
\begin{eqnarray}
\lefteqn{\left.\tilde{\rho}_{S_{1_n}S_{2_n}\mid \psi^-_n}\right|_{\mbox{\scriptsize integral 1}} = \frac{1}{16}\int\!\frac{{\rm d}^2\bomega_S}{4\pi^2}\int\!\frac{{\rm d}^2\bomega_S'}{4\pi^2}}\nonumber \\[.05in]
&& \left\{K_{S_{1n}S_{2n}}^{(1)}(\bomega_S;\bomega_S')(|\omega_{S_1}\rangle_{S_{1V}}|\omega_{S_2}\rangle_{S_{2H}}\,{}_{S_{2H}}\!\langle\omega_{S_2}'|\,{}_{S_{1V}}\!\langle \omega_{S_1}'|+|\omega_{S_1}\rangle_{S_{1H}}|\omega_{S_2}\rangle_{S_{2V}}\,{}_{S_{2V}}\!\langle\omega_{S_2}'|\,{}_{S_{1H}}\langle \omega_{S_1}'| )\right. \nonumber\\[.05in]
&&\,\,-\, \left.K_{S_{1n}S_{2n}}^{(2)}(\bomega_S;\bomega_S')(|\omega_{S_1}\rangle_{S_{1V}}|\omega_{S_2}\rangle_{S_{2H}}\,{}_{S_{2V}}\!\langle\omega_{S_2}'|\,{}_{S_{1H}}\!\langle \omega_{S_1}'| + |\omega_{S_1}\rangle_{S_{1H}}|\omega_{S_2}\rangle_{S_{2V}}\,{}_{S_{2H}}\langle\omega_{S_2}'|\,{}_{S_{1V}}\langle \omega_{S_1}'|) \right\}.
\label{tildeRhoKernel1}
\end{eqnarray}
Now, recognizing that
\begin{equation}
K^{(1)}_{S_{1n}S_{2n}}(\bomega_S;\bomega'_S) = \frac{K^{(c)}_{S_{1n}S_{2n}}(\bomega_S;\bomega'_S)+K^{(e)}_{S_{1n}S_{2n}}(\bomega_S;\bomega'_S)}{2},
\end{equation}
and 
\begin{equation}
K^{(2)}_{S_{1n}S_{2n}}(\bomega_S;\bomega'_S) = \frac{K^{(c)}_{S_{1n}S_{2n}}(\bomega_S;\bomega'_S)- K^{(e)}_{S_{1n}S_{2n}}(\bomega_S;\bomega'_S)}{2}, 
\end{equation}
we arrive at
\begin{align}
\left.\tilde{\rho}_{S_{1_n}S_{2_n}\mid \psi^-_n}\right|_{\mbox{\scriptsize integral 1}}&= \frac{1}{16}\int\!\frac{{\rm d}^2\bomega_S}{4\pi^2}\int\!\frac{{\rm d}^2\bomega'_S}{4\pi^2}\,K^{(c)}_{S_{1n}S_{2n}}(\bomega_S;\bomega'_S)|\psi^-(\bomega_S)\rangle_{S_1S_2}\,{}_{S_1S_2}\langle \psi^-(\bomega'_S)|  \nonumber \\[.05in]
&\,\,+\,\frac{1}{16}\int\!\frac{{\rm d}^2\bomega_S}{4\pi^2}\int\!\frac{{\rm d}^2\bomega'_S}{4\pi^2}\,K^{(e)}_{S_{1n}S_{2n}}(\bomega_S;\bomega'_S)|\psi^+(\bomega_S)\rangle_{S_1S_2}\,{}_{S_1S_2}\langle \psi^+(\bomega'_S)|.
\end{align}
\end{widetext}
The same steps can be followed to show that
\begin{equation}
\left.\tilde{\rho}_{S_{1_n}S_{2_n}\mid \psi^-_n}\right|_{\mbox{\scriptsize integral 2}} =
\left.\tilde{\rho}_{S_{1_n}S_{2_n}\mid \psi^-_n}\right|_{\mbox{\scriptsize integral 1}},
\end{equation}
completing the derivation of the $\tilde{\rho}_{S_{1_n}S_{2_n}\mid \psi^-_n}$ part of Eq.~(\ref{tilde-rhominusplus}).  
That equation's $\tilde{\rho}_{S_{1_n}S_{2_n}\mid \psi^+_n}$ part is found by repeating for Eq.~(\ref{rho-psiplus}) what we have just done for Eq.~(\ref{rho-psiminus}).  

\subsection{Verification of Eqs.~(\ref{xiBFm})--(\ref{nuBFm})}
Substituting Eqs.~(\ref{KcSVD}) and (\ref{xiBFm}) into the left-hand side of Eq.~(\ref{FredholmC}), we have that
\begin{widetext}
\begin{align}
\int\!&\frac{{\rm d}^2\bomega'_S}{4\pi^2}\,K^{(c)}_{S_{1_n}S_{2_n}}(\bomega_S;\bomega'_S)\xi_{\bf m}(\bomega'_S) \nonumber \\[.05in]
&=  \sum_{\ell =1}^\infty\sum_{\ell'=1}^\infty \int\!\frac{{\rm d}^2\bomega'_S}{4\pi^2}\,\lambda^2_\ell \lambda_{\ell'}^2 [\phi_\ell(\omega_{S_1})\phi^*_\ell(\omega'_{S_1})\phi_{\ell'}(\omega_{S_2})\phi^*_{\ell'}(\omega'_{S_2}) + \phi_\ell(\omega_{S_1})\phi^*_\ell(\omega'_{S_2})\phi_{\ell'}(\omega_{S_2})\phi^*_{\ell'}(\omega'_{S_1})]\phi_m(\omega'_{S_1}) \phi_m(\omega'_{S_2}) \nonumber \\[.05in]
&= \sum_{m=1}^\infty 2\lambda_m^4 \,\phi_m(\omega_{S_1})\phi_m(\omega_{S_2}), \mbox{ for $m =1,2,\ldots,$}
\end{align}
and
\begin{align}
\int\!&\frac{{\rm d}^2\bomega'_S}{4\pi^2}\,K^{(c)}_{S_{1_n}S_{2_n}}(\bomega_S;\bomega'_S)\xi_{\bf m}(\bomega'_S) \nonumber \\[.05in]
&=  \sum_{\ell =1}^\infty\sum_{\ell'=1}^\infty \int\!\frac{{\rm d}^2\bomega'_S}{4\pi^2}\,\lambda^2_\ell \lambda_{\ell'}^2 [\phi_\ell(\omega_{S_1})\phi^*_\ell(\omega'_{S_1})\phi_{\ell'}(\omega_{S_2})\phi^*_{\ell'}(\omega'_{S_2}) + \phi_\ell(\omega_{S_1})\phi^*_\ell(\omega'_{S_2})\phi_{\ell'}(\omega_{S_2})\phi^*_{\ell'}(\omega'_{S_1})]\nonumber \\[.05in]
& \,\,\times\, [\phi_{m_1}(\omega'_{S_1})\phi_{m_2}(\omega'_{S_2})+ \phi_{m_1}(\omega'_{S_2})\phi_{m_2}(\omega'_{S_1})]/\sqrt{2}, \mbox{ for $m_1>m_2 = 1,2,\ldots, $} \nonumber \\[.05in]
&= \sum_{m_1=2}^\infty\sum_{m_2 =1}^{m_1-1} 2\lambda_{m_1}^2\lambda_{m_2}^2 \,[\phi_{m_1}(\omega_{S_1})\phi_{m_2}(\omega_{S_2})+ \phi_{m_1}(\omega_{S_2})\phi_{m_2}(\omega_{S_1})]/\sqrt{2},
\end{align}
\end{widetext}
and our verification is complete for Eqs.~(\ref{xiBFm}) and (\ref{muBFm})'s being the eigenfunctions and eigenvalues of $K^{(c)}_{S_{1_n}S_{2_n}}(\bomega_S;\bomega'_{S})$. 

Following the same approach, starting with substitution of Eqs.~(\ref{KeSVD}) and (\ref{zetaBFm}) into Eq.~(\ref{FredholmE}) will verify Eqs.~(\ref{zetaBFm}) and (\ref{nuBFm})'s being the eigenfunctions and eigenvalues of $K^{(e)}_{S_{1_n}S_{2_n}}(\bomega_S;\bomega'_{S})$. 

\section{Derivation Details for Sec.~\ref{memory}\label{AppendB}}
Here we provide derivation details for the $\tilde{\rho}^{(p)}_{M_1,M_2}$ results presented in Sec.~\ref{memory}.

Suppose Alice and Bob's QRXs are illuminated by the $|\tilde{\xi}^-_{\bf m}\rangle_{\tilde{S}_1\tilde{S}_2}$ state from Eq.~(\ref{tildeXiState}), and that they record photon counts from their $\tilde{S}_{1+}$ and $\tilde{S}_{2+}$ detectors.  Their quantum memories are then left in the state whose unnormalized density operator is
\begin{widetext}
\begin{align}
\tilde{\rho}^{(+,+)}_{M_1,M_2\mid \tilde{\xi}^-_{\bf m}} &= \int\!\frac{{\rm d}^2\bomega_+}{4\pi^2}\int\!\frac{{\rm d}^2\bomega_S}{4\pi^2}\int\!\frac{{\rm d}^2\bomega'_S}{4\pi^2}\,\frac{\tilde{\xi}_{\bf m}(\bomega_S)\tilde{\xi}^*_{\bf m}(\bomega'_S)}{8}\,
{}_{\tilde{S}'_{1+}}\!\langle \omega_{1+}|\,\,{}_{\tilde{S}'_{2+}}\!\langle \omega_{2+}|\nonumber \\[.05in]
&\,\,\times \left\{|\omega_{S_1}\rangle_{\tilde{S}'_{1_V}}|\omega_{S_2}\rangle_{\tilde{S}_{2_V}}\left[r_1(\omega_{S_1})|g_1\rangle_{M_1} + r_2(\omega_{S_1})|g_2\rangle_{M_1}\right]\left[|g_1\rangle_{M_2}+|g_2\rangle_{M_2}\right] \right. \nonumber \\[.05in]
& \,\, - \left.|\omega_{S_1}\rangle_{\tilde{S}_{1_V}}|\omega_{S_2}\rangle_{\tilde{S}'_{2_V}}\left[|g_1\rangle_{M_1} + |g_2\rangle_{M_1}\right]\left[r_1(\omega_{S_2})|g_1\rangle_{M_2} + r_2(\omega_{S_2})|g_2\rangle_{M_2}\right]
\right\} \nonumber \\[.05in]
&\,\,\times \left\{\left[r^*_1(\omega'_{S_1})\,{}_{M_1}\!\langle g_1| + r^*_2(\omega'_{S_1})\,{}_{M_1}\!\langle g_2|\right]\left[{}_{M_2}\!\langle g_1|+{}_{M_2}\!\langle g_2|\right] {}_{\tilde{S}_{2_V}}\!\!\langle \omega'_{S_2}|\,{}_{\tilde{S}'_{S_{1_V}}}\!\!\langle \omega'_{S_1}|\right. \nonumber \\[.05in]
& \,\, - \left.\left[{}_{M_1}\!\langle g_1| + {}_{M_2}\!\langle g_2|\right]\left[r^*_1(\omega'_{S_2})\,{}_{M_2}\langle g_1| + r^*_2(\omega'_{S_2})\,{}_{M_2}\!\langle g_2|\right]\,{}_{\tilde{S}'_{S_{2_V}}}\!\!\langle \omega'_{S_2}|\, {}_{\tilde{S}_{1_V}}\!\!\langle \omega'_{S_1}|
\right\} |\omega_{2+}\rangle_{\tilde{S}'_{2+}}\,|\omega_{1+}\rangle_{\tilde{S}'_{1+}}.
\label{rhoMemStep1}
\end{align}
\end{widetext}
Using the beam-splitter relations,
\begin{align}
|\omega_{S_k}\rangle_{\tilde{S}'_{k_V}} &= \frac{|\omega_{S_k}\rangle_{\tilde{S}'_{k+}} +|\omega_{S_k}\rangle_{\tilde{S}'_{k-}}}{\sqrt{2}},  
\label{newBS1}\\[.05in]
|\omega_{S_k}\rangle_{\tilde{S}_{k_V}} &= \frac{e^{i\omega_{S_k}T}(|\omega_{S_k}\rangle_{\tilde{S}'_{k+}} - |\omega_{S_k}\rangle_{\tilde{S}'_{k-}})}{\sqrt{2}},
\mbox{ for $k=1,2$},
\label{newBS2}
\end{align}
in Eq.~(\ref{rhoMemStep1}), evaluating the ensuing bra-ket inner products, and integrating over $
\bomega_+$, we get
\begin{widetext}
\begin{align}
\tilde{\rho}&^{(+,+)}_{M_1,M_2\mid \tilde{\xi}^-_{\bf m}} = \int\!\frac{{\rm d}^2\bomega_S}{4\pi^2}\,\frac{|\tilde{\xi}_{\bf m}(\bomega_S)|^2}{32}\,\nonumber \\[.05in]
&\,\, \times \left\{\left[e^{i\omega_{S_2}T}r_1(\omega_{S_1})-e^{i\omega_{S_1}T}r_1(\omega_{S_2})\right]|g_1\rangle_{M_1}|g_1\rangle_{M_2}+\left[e^{i\omega_{S_2}T}r_2(\omega_{S_1})-e^{i\omega_{S_1}T}r_2(\omega_{S_2})\right]|g_2\rangle_{M_1}|g_2\rangle_{M_2} \right. \nonumber \\[.05in]
&\,\, +\, \left.\left[e^{i\omega_{S_2}T}r_1(\omega_{S_1})-e^{i\omega_{S_1}T}r_2(\omega_{S_2})\right]|g_1\rangle_{M_1}|g_2\rangle_{M_2} + 
\left[e^{i\omega_{S_2}T}r_2(\omega_{S_1})-e^{i\omega_{S_1}T}r_1(\omega_{S_2})\right]|g_2\rangle_{M_1}|g_1\rangle_{M_2}\right\} \nonumber \\[.05in]
&\,\,\times\, \left\{{}_{M_2}\langle g_1|\,{}_{M_1}\langle g_1|\left[e^{-i\omega_{S_2}T}r^*_1(\omega_{S_1})-e^{-i\omega_{S_1}T}r^*_1(\omega_{S_2})\right]+{}_{M_2}\langle g_2|\,{}_{M_1}\langle g_2|\left[e^{-i\omega_{S_2}T}r^*_2(\omega_{S_1})-e^{-i\omega_{S_1}T}r^*_2(\omega_{S_2})\right] \right. \nonumber \\[.05in]
&\,\, +\, \left.{}_{M_2}\langle g_2|\,{}_{M_1}\langle g_1|\left[e^{-i\omega_{S_2}T}r^*_1(\omega_{S_1})-e^{-i\omega_{S_1}T}r^*_2(\omega_{S_2})\right] + 
{}_{M_2}\langle g_1|\,{}_{M_1}\langle g_2|\left[e^{-i\omega_{S_2}T}r^*_2(\omega_{S_1})-e^{-i\omega_{S_2}T}r^*_1(\omega_{S_2})\right]|\right\}
\label{rhoMemStep2}.
\end{align}
\end{widetext}
Rewriting this density operator in terms of the quantum memories' Bell states from Eq.~(\ref{memoriesBells1})
and (\ref{memoriesBells2}), and then inserting the result in Eq.~(\ref{rhoBFs}), we arrive at the $\sum_{\bf m}\mu_{\bf m}$, $p=a$ term in Eq.~(\ref{rhoAB}) that applies to $\tilde{\rho}^{(+,+)}_{M_1,M_2\mid \psi^-_n}$ from Eq.~(\ref{rhoAusage}) with the memories' frequency-dependent Bell states being given by Eqs.~(\ref{aStates1})--(\ref{aStates8}).

To get the $\sum_{\bf m}\nu_{\bf m}$, $p=a$ term in Eq.~(\ref{rhoAB}), we parallel what we have just done, starting from Alice and Bob's memories being illuminated by the $|\tilde{\zeta}^-_{\bf m}\rangle_{\tilde{S}_1\tilde{S}_2}$ state from Eq.~(\ref{tildeZetaState}) and assuming Alice and Bob's QRXs  record photon counts from their $\tilde{S}_{1+}$ and $\tilde{S}_{2+}$ detectors.  After using the beam-splitter relations from Eqs.~(\ref{newBS1}) and (\ref{newBS2}), evaluating the ensuing bra-ket inner products, 
and then integrating over $\bomega_+$, we find that 
\begin{widetext}
\begin{align}
\tilde{\rho}&^{(+,+)}_{M_1,M_2\mid \tilde{\zeta}^-_{\bf m}} = \int\!\frac{{\rm d}^2\bomega_S}{4\pi^2}\,\frac{|\tilde{\zeta}_{\bf m}(\bomega_S)|^2}{32}\,\nonumber \\[.05in]
&\,\, \times \left\{\left[e^{i\omega_{S_2}T}r_1(\omega_{S_1})+e^{i\omega_{S_1}T}r_1(\omega_{S_2})\right]|g_1\rangle_{M_1}|g_1\rangle_{M_2}+\left[e^{i\omega_{S_2}T}r_2(\omega_{S_1})+ e^{i\omega_{S_1}T}r_2(\omega_{S_2})\right]|g_2\rangle_{M_1}|g_2\rangle_{M_2} \right. \nonumber \\[.05in]
&\,\, +\, \left.\left[e^{i\omega_{S_2}T}r_1(\omega_{S_1})+e^{i\omega_{S_1}T}r_2(\omega_{S_2})\right]|g_1\rangle_{M_1}|g_2\rangle_{M_2} + 
\left[e^{i\omega_{S_2}T}r_2(\omega_{S_1})+e^{i\omega_{S_1}T}r_1(\omega_{S_2})\right]|g_2\rangle_{M_1}|g_1\rangle_{M_2}\right\} \nonumber \\[.05in]
&\,\,\times\, \left\{{}_{M_2}\langle g_1|\,{}_{M_1}\langle g_1|\left[e^{-i\omega_{S_2}T}r^*_1(\omega_{S_1})+ e^{-i\omega_{S_1}T}r^*_1(\omega_{S_2})\right]+{}_{M_2}\langle g_2|\,{}_{M_1}\langle g_2|\left[e^{-i\omega_{S_2}T}r^*_2(\omega_{S_1})+ e^{-i\omega_{S_1}T}r^*_2(\omega_{S_2})\right] \right. \nonumber \\[.05in]
&\,\, +\, \left.{}_{M_2}\langle g_2|\,{}_{M_1}\langle g_1|\left[e^{-i\omega_{S_2}T}r^*_1(\omega_{S_1})+ e^{-i\omega_{S_1}T}r^*_2(\omega_{S_2})\right] + 
{}_{M_2}\langle g_1|\,{}_{M_1}\langle g_2|\left[e^{-i\omega_{S_2}T}r^*_2(\omega_{S_1})+ e^{-i\omega_{S_2}T}r^*_1(\omega_{S_2})\right]|\right\}
\label{rhoMemStep3}.
\end{align}
\end{widetext}
Rewriting this expression in terms of the quantum memories' Bell states, and inserting the result in Eq.~(\ref{rhoBFs}), we recover the $\sum_{\bf m}\nu_{\bf m}$, $p=a$ term in Eq.~(\ref{rhoAB}) that applies to $\tilde{\rho}^{(+,+)}_{M_1,M_2\mid \psi^-_n}$ from Eq.~(\ref{rhoAusage}) with the memories' frequency-dependent Bell states being given by Eqs.~(\ref{aStates1})--(\ref{aStates8}).  This completes the derivation of $\tilde{\rho}^{(+,+)}_{M_1,M_2 \mid\psi^-_n}$.  Repeating this procedure three more times will show that the $\tilde{\rho}^{(a)}_{M_1,M_2}$ we have just verified as applying to $\psi^-_n$ heralds with ${\bf s} = (+,+)$ detections also applies to $\psi^-_n$ heralds with ${\bf s} = (-,-)$ detections, as well as to $\psi^+_n$ heralds with ${\bf s} = (+,-)$ or ${\bf s} = (-,+)$ detections.  We omit the details.   

At this point, it only remains for us to demonstrate that $\tilde{\rho}^{(b)}_{M_1,M_2}$ from Eq.~(\ref{rhoAB}) characterizes the density operators specified for it in Eq.~(\ref{rhoBusage}).  So, let us assume that Alice and Bob's QRXs are illuminated by the $|\tilde{\xi}^-_{\bf m}\rangle_{\tilde{S}_1\tilde{S}_2}$ state or the $|\tilde{\zeta}^-_{\bf m}\rangle_{\tilde{S}_1\tilde{S}_2}$ state, and that in either case they record photon counts from their $S_{1+}$ and $S_{2-}$ detectors.  Repeating the procedure we have been using throughout this subsection results in the following expressions, respectively, for $\tilde{\rho}^{(+,-)}_{M_1,M_2\mid \tilde{\xi}^-_{\bf m}}$ and $\tilde{\rho}^{(+,-)}_{M_1,M_2\mid \tilde{\zeta}^-_{\bf m}}$ \emph{after} Bob has applied a $\pi$ pulse to his memory:
\begin{widetext}
\begin{align}
\tilde{\rho}^{(+,-)}_{M_1M_2\mid \tilde{\xi}^-_{\bf m}} &= \int\!\frac{{\rm d}^2\bomega_S}{4\pi^2}\,\frac{|\tilde{\xi}_{\bf m}(\bomega_S)|^2}{4}\left[|\phi^+_{b_{\mu{\bf m}}}(\bomega_S)\rangle_{M_1M_2}+ |\phi^-_{b_{\mu{\bf m}}}(\bomega_S)\rangle_{M_1M_2} + |\psi^+_{b_{\mu{\bf m}}}(\bomega_S)\rangle_{M_1M_2}+ |\psi^-_{b_{\mu{\bf m}}}(\bomega_S)\rangle_{M_1M_2}\right] \nonumber \\[.05in]
&\,\,\times\, \left[{}_{M_1M_2}\langle \phi^+_{b_{\mu{\bf m}}}(\bomega_S)|+{}_{M_1M_2}\langle\phi^-_{b_{\mu{\bf m}}}(\bomega_S)|+{}_{M_1M_2}\langle \psi^+_{b_{\mu{\bf m}}}(\bomega_S)|+{}_{M_1M_2}\langle\psi^-_{b_{\mu{\bf m}}}(\bomega_S)|\right],
\label{rhoABxi}
\end{align}
and
\begin{align}
\tilde{\rho}^{(+,-)}_{M_1M_2\mid \tilde{\zeta}^-_{\bf m}}&= \int\!\frac{{\rm d}^2\bomega_S}{4\pi^2}\,\frac{|\tilde{\zeta}_{\bf m}(\bomega_S)|^2}{4}\left[|\phi^+_{b_{\nu{\bf m}}}(\bomega_S)\rangle_{M_1M_2}+ |\phi^-_{b_{\nu{\bf m}}}(\bomega_S)\rangle_{M_1M_2} + |\psi^+_{b_{\nu{\bf m}}}(\bomega_S)\rangle_{M_1M_2}+ |\psi^-_{b_{\nu{\bf m}}}(\bomega_S)\rangle_{M_1M_2}\right]  \nonumber \\[.05in]
&\,\,\times\, \left[{}_{M_1M_2}\langle \phi^+_{b_{\nu{\bf m}}}(\bomega_S)|+{}_{M_1M_2}\langle\phi^-_{b_{\nu{\bf m}}}(\bomega_S)|+{}_{M_1M_2}\langle \psi^+_{b_{\nu{\bf m}}}(\bomega_S)|+{}_{M_1M_2}\langle\psi^-_{b_{\nu{\bf m}}}(\bomega_S)|\right].
\label{rhoABzeta}
\end{align}
\end{widetext}
Inserting these equations into Eq.~(\ref{rhoBFs}) we verify that $\tilde{\rho}^{(b)}_{M_1,M_2}$ does equal $\tilde{\rho}^{(+,-)}_{M_1,M_2\mid \psi^-_n}$.  Similar calculations, \emph{including} Bob's application of a $\pi$ pulse to his memory after its interaction with the biphoton, will verify that $\tilde{\rho}^{(b)}_{M_1,M_2}$ equals $\tilde{\rho}^{(-,+)}_{M_1,M_2\mid \psi^-_n}$, $\tilde{\rho}^{(+,+)}_{M_1,M_2\mid \psi^+_n}$, and $\tilde{\rho}^{(-,-)}_{M_1,M_2\mid \psi^+_n}$.  Again we omit the details.

Strictly speaking, Bob's use of a $\pi$ pulse is not necessary for Duan-Kimble memory loading:  the herald Alice and Bob receive from the ZALM QTX and knowing each other's detector that registered a photon will tell them whether to expect loading of a $|\psi^-\rangle_{M_1M_2}$ or a $|\phi^-\rangle_{M_1M_2}$ state.  We have chosen to augment the Duan-Kimble protocol as we have to ensure they will always expect loading of a $|\psi^-\rangle_{M_1M_2}$ state when their QRXs register a coincidence. Under completely ideal conditions---i.e., a separable channelized biphoton wave function and an ideal quantum memory---the $|\psi^-\rangle_{M_1M_2}$ singlet will then be loaded into Alice and Bob's memories with probability~1.


\begin{thebibliography}{99}
\bibitem{Einstein1935}A. Einstein, B. Podolsky, and N. Rosen, Can quantum-mechanical
description of physical reality be considered complete?, Phys. Rev. 
{\bf 47}, 777--780 (1935).

\bibitem{Bennett1993}C. H. Bennett, G. Brassard, C. Cr\'{e}peau, R. Jozsa, A.
Peres, and W. K. Wootters, Teleporting an unknown
quantum state via dual classical and
Einstein-Podolsky-Rosen channels, Phys. Rev. Lett. {\bf 70}, 1895--1899 (1993).

\bibitem{Lloyd2004}S. Lloyd, J. H. Shapiro, F. N. C. Wong, P. Kumar, S.M. Shahriar, and H. P. Yuen, ,
Infrastructure for the quantum internet, Comput. Commun. Rev. {\bf 34}, 9--20 (2004).

\bibitem{Kimble2008}H. J. Kimble, The quantum internet, Nature {\bf 453}, 1023--1030 (2008).  

\bibitem{Wehner2018}S. Wehner, D. Elkouss, and R. Hanson, Quantum internet:  A vision for the road ahead, Science {\bf 362}, eaam9288 (2018).

\bibitem{Azuma2023}K. Azuma, S. E. Economou, D. Elkouss, P. Hilaire, L. Jiang, H.-K. Lo, and I. Tzitrin, Quantum repeaters: From quantum networks to the quantum internet, Rev. Mod. Phys. {\bf 95}, 045006 (2023).

\bibitem{Neumann2022}S. P. Neumann, A. Buchner, L. Bulla, M. Bohmann, and R. Ursin, Continuous entanglement distribution over a transnational 248 km fiber link, Nat. Commun. {\bf 13}, 6134 (2022).

\bibitem{Sangouard2011}N. Sangouard C. Simon, H. de Riedmatten, and N. Gisin, Quantum repeaters based on atomic ensembles and linear optics, Rev. Mod. Phys. {\bf 83}, 33-80 (2011).

\bibitem{Muralidharan2015}S. Muralidharan, L. Li, J. Kim, N. L\"{u}kenhaus, M. D. Lukin, and L. Jiang, Optimal architectures for long distance quantum communication, Sci. Rep. {\bf 6}, 20463 (2015).

\bibitem{Yin2017}J. Yin, Y. Cao, Y.-H. Li, S.-K. Liao, L. Zhang,
J.-G. Ren, W.-Q. Cai, W.-Y. Liu, B. Li, H. Dai, \emph{et al}., Satellite-based entanglement distribution over 1200 kilometers, Science {\bf 356}, 1140--1144 (2017).

\bibitem{Liao2017}S.-K. Liao, Y. Cao, Y.-H. Li, S.-K. Liao, L. Zhang, 
J.-G. Ren, W.-Q. Cai, W.-Y. Liu, B. Li, H. Dai, \emph{et al}., Satellite-to-ground quantum key distribution, Nature {\bf 549}, 43--47 (2017).

\bibitem{Chen2023}K. C. Chen, P. Dhara, M. Heuck, Y. Lee, W. Dai, S. Guha, and D. Englund, Zero-added-loss entangled-photon multiplexing for ground- and space-based quantum networks, Phys. Rev. Appl. {\bf 19}, 054029 (2023).

\bibitem{Braunstein1995}S. L. Braunstein and A. Mann, Measurement of the Bell operator and quantum teleportation, Phys. Rev. A {\bf 51}, R1727--R1730 (1995).

\bibitem{Kumar1990}P. Kumar, Quantum frequency conversion, Opt.
Lett. {\bf 15}, 1476--1478 (1990)

\bibitem{Huang1992}J. M. Huang and P. Kumar, Observation of quantum frequency
conversion, Phys. Rev. Lett. {\bf 68}, 2153--2156 (1992).

\bibitem{Karpinski2017}M. Karpinski, M. Jachura, L. J. Wright, and B. J. Smith,
Bandwidth manipulation of quantum light by an electro-optic time lens, Nat. Photon. {\bf 11}, 53--57 (2017).

\bibitem{Mower2011}J. Mower and D. Englund, Efficient generation of single and entangled photons on a silicon photonic integrated chip, Phys. Rev. A {\bf 84}, 052326 (2011).

\bibitem{Dhara2022}P. Dhara, S. J. Johnson, C. N. Gagatsos, P. G. Kwiat, and S. Guha, Heralded-multiplexed high-efficiency cascaded source of dual-rail polarization-entangled photon pairs using spontaneous parametric down conversion, Phys. Rev. Appl. {\bf 17}, 034071 (2022). 

\bibitem{Jones2016}C. Jones, D. Kim, M. T. Rakher, P. G. Kwiat, and T. D. Ladd, Design and analysis of communication protocols for quantum repeater networks, New J. Phys. {\bf 18}, 083015 (2016).

\bibitem{Duan2004}L.-M. Duan and H. J. Kimble, Scalable quantum computation through cavity-assisted interactions, Phys. Rev. Lett. {\bf 92}, 127902 (2004).

\bibitem{Raymer2024}M. G. Raymer, C. Embleton, and J. H. Shapiro, The Duan-Kimble cavity-atom quantum memory loading scheme revisited, to appear in Phys. Rev. Appl.; arXiv:2406.12201 [quant-ph].

\bibitem{Wong2006}F. N. C. Wong, J. H. Shapiro, and T. Kim, Efficient generation of polarization-entangled photons in a nonlinear crystal, Laser Phys. {\bf 16}, 1517--1524 (2006).

\bibitem{footnote1}We assume our ZALM QTX's Sagnac SPDC sources use PPLN crystals because of their high nonlinearity and $\sim$10\,THz phase-matching bandwidth.

\bibitem{footnote1a}The $T$-sec-long time delay is chosen to maximize the mode overlap at the beam splitter when the cavity contains an SiV color center, i.e., the mode overlap is optimized with $T=0$ in the absence of that color center.

\bibitem{footnote2}Alternatively, the single-ended Fabry-P\'{e}rot cavity could be replaced with a ring cavity, obviating the need for a circulator.

\bibitem{footnote3}Duan-Kimble quantum memories can also be constructed using the tin-vacancy (SnV) color center.  

\bibitem{Nguyen2019}C. T. Nguyen, D. D. Sukachev, M. K. Bhaskar, B. Machielse, D. S. Levonian, E. N. Knall, P. Stroganov, R. Riedinger, H. Park, M. Lon\u{c}ar, \emph{et al}., Quantum network nodes based on diamond qubits with an efficient nanophotonic interface, Phys. Rev. Lett. {\bf 123}, 183602 (2019).

\bibitem{Bersin2024}E. Bersin, M. Sutula, Y. Q. Huan, A. Suleymanzade, D. R. Assumpcao, Y-C. Wei, P.-J, Stas, C. M. Knaut, E. N. Knall, C. Langrock, \emph{et al}., Telecom networking with a diamond quantum memory, PRX Quantum {\bf 5}, 010303 (2024).

\bibitem{Knaut2024}C. M. Knaut, A. Suleymanzade, Y.-C. Wei, D. R. Assumpcao, P.-J. Stas, Y. Q. Huan, B. Machielse, E. N. Knall, M. Sutula, G. Baranes, \emph{et al}., Entanglement of nanophotonic quantum
memory nodes in a telecom network, Nature, {\bf 629}, 573--578 (2024).

\bibitem{Ou1999}Z. Y. Ou, J.-K. Rhee, and L. J. Wang, Photon bunching and multiphoton interference in parametric down-conversion, Phys. Rev. A {\bf 60}, 593--604 (1999).

\bibitem{footnote4}These single-photon states have ${}_{K_P}\langle \omega_K|\omega'_K\rangle_{K_P} = 2\pi \delta(\omega_K-\omega'_K)$ inner product, where $\delta(\cdot)$ is the unit impulse function.

\bibitem{U'Ren2005}A. B. U'Ren, C. Silberhorn, R. Erdmann, K. Banaszek, W. P. Grice, I. A. Walmsley, and M. G. Raymer. Generation of pure-state single-photon wavepackets by conditional preparation based on spontaneous parametric downconversion, Laser Phys. {\bf 15}, 146--161] (2005).

\bibitem{Zhang2014}Z. Zhang, J. Mower, D. Englund, F. N. C. Wong, and J. H. Shapiro, Unconditional security of time-energy entanglement quantum key distribution using dual-basis interferometry, Phys. Rev. Lett. {\bf 112}, 120506 (2014).

\bibitem{Dixon2013}P. B. Dixon, J. H. Shapiro, and F. N. C. Wong, Spectral engineering by Gaussian phase-matching for quantum photonics, Opt. Express {\bf 21}, 5879--5890 (2013).

\bibitem{Chen2017}C. Chen, C. Bo, M. Y. Niu, F. Xu, Z. Zhang, J. H. Shapiro, and F. N. C. Wong, Efficient generation and characterization of spectrally factorable biphotons, Opt. Express {\bf 25}, 7300-7312 (2017).

\bibitem{Chen2019}C. Chen, J. E. Heyes, K.-H. Hong, M. Y. Niu, A. E. Lita, T. Gerrits, S. W. Nam, J. H. Shapiro, and F. N. C. Wong,  Indistinguishable single-mode photons from spectrally-engineered biphotons, Opt. Express {\bf 27}, 11626--11634 (2019).

\bibitem{footnote6}The SVD's $\{\lambda_\ell\}$, $\{\phi_\ell(\omega_S)\}$, and $\{\psi_\ell(\omega_I)\}$ have implicit dependence on the channel number $n$.

\bibitem{footnote7}Sub-unit efficiencies will appear in Sec.~\ref{Discuss}, where we discuss ZALM's entanglement-distribution rate.

\bibitem{footnote8}These kernels' eigenvalues and eigenfunctions have implicit dependence on the channel number $n$.  

\bibitem{footnote9}Below, sums of $\mu_{\bf m}$ and $\nu_{\bf m}$ only include the ${\bf m} = (m_1,m_2)$ indices given in Eqs.~(\ref{muBFm}) and (\ref{nuBFm}), respectively.  The same will apply later to sums of $\tilde{\mu}_{\bf m}$, $\tilde{\nu}_{\bf m}$, $\tilde{\mu}^2_{\bf m}$, and $\tilde{\nu}^2_{\bf m}$.  

\bibitem{footnote9a}For example, when ${\bf s} = (+,+)$, as in Eq.~(\ref{rhoMemStep1}), $\hat{\prod}_{\bf s}$ becomes
\begin{displaymath}
\hat{\prod}_{(+,+)} \equiv \int\!\frac{{\rm d}^2\bomega_{++}}{4\pi^2}\,|\omega_{1+}\rangle_{\tilde{S}'_{1+}}|\omega_{2+}\rangle_{\tilde{S}'_{2+}}\,{}_{\tilde{S}'_{2+}}\!\langle \omega_{2+}|\,{}_{\tilde{S}'_{1+}}\!\langle \omega_{1+}|,
\end{displaymath}
where $\bomega_{++} \equiv (\omega_{1+},\omega_{2+})$ and ${\rm d}^2\bomega_{++} \equiv {\rm d}\omega_{1+}\,{\rm d}\omega_{2+}$.

\bibitem{footnote11}The astute reader will note that $\mathbb{E}(N_p) \sim 1$ contradicts the perturbative assumption made in Sec.~\ref{singlePerturbation}, seemingly undercutting all of our subsequent analysis.  Such is not the case.  All of our principal results rely on the channelized Sagnac-configured source having a perturbative representation, i.e., its $n$th channel produces biphotons with channelized wave function $\Psi_{S_nI_n}(\omega_S,\omega_I)$.  In the absence of pump depletion, SPDC produces signal and idler that are in a squeezed state, i.e., a zero-mean, jointly-Gaussian state that is completely characterized by its normally-ordered and phase-sensitive correlation functions~\cite{Wong2006}.  After channelization, to the point that the average signal and idler photon numbers are $\sim$$10^{-2}$ or less, the $n$th-channel signal and idler can be approximated as being in a biphoton state whose $\Psi_{S_nI_n}(\omega_S,\omega_I)$ is proportional to the channelized version of the phase-sensitive cross correlation~\cite{Wong2006}.  Inasmuch as the phase-sensitive cross correlation can be found from SPDC theory that goes beyond the biphoton~\cite{Quesada2022}, our theory is on sound footing, albeit with the all-Gaussian special case becoming a somewhat more arbitrary, rather than fully-justified, example. 

\bibitem{Quesada2022}N. Quesada, L. G. Helt, M. Menotti, M. Liscidini, and J. E. Sipe, Beyond photon pairs---nonlinear quantum photonics in the high-gain regime: a tutorial, Adv. Opt. Photon. {\bf 14}, 291--403 (2022).

\bibitem{footnote12}Here, $0 < \eta_{\rm cavity} <1$ is an efficiency factor accounting, in the general case, for intra-cavity losses in the two memories.  For narrowband operation with $C = C_\pi$ we have $\eta_{\rm cavity} = |r_1(\tilde{\omega}_{S_0})|^2.$

\bibitem{Shapiro2000}J. H. Shapiro and N. C. Wong, An ultrabright narrowband source of polarization-entangled photon pairs,  J. Opt. B: Quantum and Semiclass. Opt. {\bf 2}, L1--L4 (2000).

\bibitem{Lloyd2001}S. Lloyd, M. S. Shahriar, J. H. Shapiro, and P. R. Hemmer, Long distance, unconditional teleportation of atomic states via complete Bell state measurements, Phys. Rev. Lett. {\bf 87}, 167903 (2001).

\bibitem{Morrison2022}C. L. Morrison, F. Graffitti, P. Barrow, A. Pickston, J. Ho, and A. Fedrizzi, Frequency-bin entanglement from domain-engineered down-conversion, APL Photon. {\bf 7}, 066102 (2022).

\bibitem{Andrews1998}L. C. Andrews and R. L. Phillips, \emph{Laser Beam Propagation through Random Media} (SPIE, Bellingham, 1998).

\end{thebibliography}
\end{document}